\pdfoutput=1
\documentclass[cernpreprint, atlasdraft=false, UKenglish, texmf, orcidlogo]{atlasdoc}
\usepackage[full]{atlaspackage}
\usepackage{atlasbiblatex}

\usepackage{atlasphysics}
\graphicspath{{logos/}{figures/}}

\usepackage{ANA-TOPQ-2019-03-PAPER-defs}
\usepackage{multirow}
\usepackage{tabularx,booktabs}
\newcolumntype{Y}{>{\centering\arraybackslash}X}
\AtlasTitle{Measurement of $t\bar{t}$ production in association with additional $b$-jets in the $e\mu$ final state in proton--proton collisions at $\sqrt{s}=13$~\ensuremath{\text{TeV}}\xspace with the ATLAS detector}

\AtlasAbstract{%
This paper presents measurements of top-antitop quark pair (\ensuremath{t\bar{t}}\xspace) production in association with additional $b$-jets.  The analysis utilises 140~fb\ensuremath{^{-1}} of proton--proton collision data collected with the ATLAS detector at the Large Hadron Collider at a centre-of-mass energy of 13~\ensuremath{\text{TeV}}\xspace.
Fiducial cross-sections are extracted in a final state featuring one electron and one muon, with at least three or four \ensuremath{b\text{-jets}}\xspace. Results are presented at the particle level for both integrated cross-sections and normalised differential cross-sections, as functions of global event properties, jet kinematics, and \ensuremath{b\text{-jet}}\xspace pair  properties. Observable quantities characterising \ensuremath{b\text{-jets}}\xspace originating from the top quark decay and additional \ensuremath{b\text{-jets}}\xspace are also measured at the particle level, after correcting for detector effects. The measured integrated fiducial cross-sections are consistent with \ensuremath{t\bar{t}b\bar{b}}\xspace predictions from various next-to-leading-order matrix element calculations matched to a parton shower within the uncertainties of the predictions. State-of-the-art theoretical predictions are compared with the differential measurements; none of them simultaneously describes all observables.  Differences between any two predictions are smaller than the measurement uncertainties for most observables.
}

\AtlasRefCode{TOPQ-2019-03}

\PreprintIdNumber{CERN-EP-2024-191}

\AtlasJournal{JHEP}
\AtlasJournalRef{JHEP 01 (2025) 068}
\AtlasDOI{10.1007/JHEP01(2025)068}

\hypersetup{pdftitle={ATLAS document},pdfauthor={The ATLAS Collaboration}}

\begin{document}

\maketitle

\tableofcontents

\clearpage
\section{Introduction}
%

\begin{figure}[tbh!]
\centering
\subfloat[]{\includegraphics[width=0.35\textwidth]{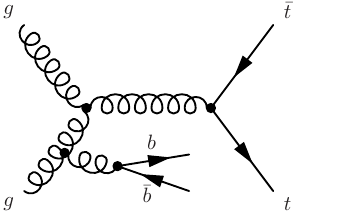}\label{fig:ttbb}}
\subfloat[]{\includegraphics[width=0.35\textwidth]{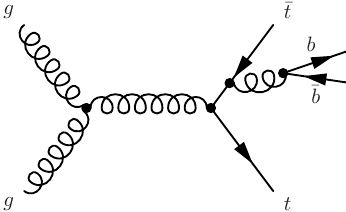}\label{fig:ttgluonbb}}\\
\subfloat[]{\includegraphics[width=0.35\textwidth]{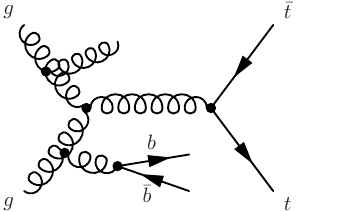}\label{fig:ttbbInitg}}
\subfloat[]{\includegraphics[width=0.35\textwidth]{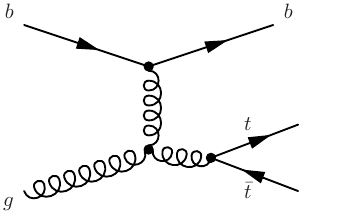}\label{fig:ttb}}
\caption{Example Feynman diagrams of QCD processes leading to a \ttbj final state:
\protect \subref{fig:ttbb} with additional $b$-quarks produced from initial-state gluon radiation, \protect \subref{fig:ttgluonbb} with additional $b$-quarks produced from final-state gluon radiation, \protect \subref{fig:ttbbInitg} the process with more than one gluon emission, where one of them produces a pair of $b$-quarks, and \protect \subref{fig:ttb} the process initiated with a $b$-quark in the initial state.}
\label{fig:introd:FD}
\end{figure}

Measurements of the production cross-section of a top--antitop quark pair (\ttbar) with jets emanating from gluon radiation (additional jets) at the Large Hadron Collider (LHC) provide essential tests of the predictions of quantum chromodynamics (QCD).
Among these, the process of \ttbar events produced in association with jets originating from $b$-quarks (\bjets), as shown in Figure~\ref{fig:introd:FD}, is of special interest, as it is difficult to predict due to the scale hierarchy between \ttbar production and $b\bar{b}$ production from gluon emission and the non-negligible mass of the $b$-quark.  There can be a single gluon emission in the initial or final state producing a pair of $b$-quarks as shown in Figure~\ref{fig:ttbb} and Figure~\ref{fig:ttgluonbb}, or multiple gluon emissions where at least one splits into a pair of $b$-quarks as shown in Figure~\ref{fig:ttbbInitg}, or a $b$-quark participates in initiating the process leading to a \ttbar in association with a final state $b$-quark as shown in Figure~\ref{fig:ttb}.

\begin{figure}[tbh!]
\centering
\subfloat[]{\includegraphics[width=0.35\textwidth]{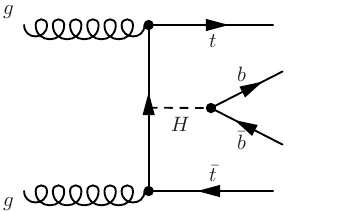}\label{fig:ttH}}
\subfloat[]{\includegraphics[width=0.35\textwidth]{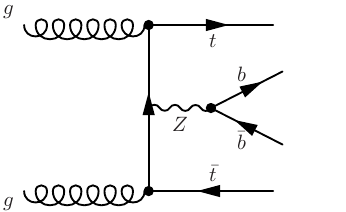}\label{fig:ttZ}}
\caption{Example Feynman diagrams of electroweak processes leading to a \ttbj final state,
\protect \subref{fig:ttH} the \ttH($H\to b\bar{b}$) process and \protect \subref{fig:ttZ} the \ttZ($Z\to b\bar{b}$) process.}
\end{figure}

Since the discovery of the Higgs boson~\cite{HIGG-2012-27,CMS-HIG-12-028}, the measurement of the Higgs boson coupling to the heaviest elementary particle, the top quark, has become an essential test of the Standard Model (SM).
Direct measurements of the top quark Yukawa coupling are only possible in events where a Higgs boson is produced in association with one or two top quarks, $tH$ or \ttH~\cite{HIGG-2018-13,CMS-HIG-17-035}. The production of \ttH with a subsequent decay of $H\rightarrow b\bar{b}$ is shown in Figure~\ref{fig:ttH}.
While $H\rightarrow b\bar{b}$ has the largest Higgs boson branching fraction, its detection suffers from a large irreducible background from QCD \ttbj production~\cite{CMS-HIG-18-016,HIGG-2020-23}.
Similarly, the production of \ttZ with $Z \rightarrow b\bar{b}$ as shown in Figure~\ref{fig:ttZ} faces significant challenges due to large backgrounds, making it crucial to enhance our understanding of QCD \ttbj production.
A better understanding of the QCD production of \ttbj as predicted by the SM and improved Monte Carlo (MC) modelling would greatly benefit the \ttH($H\to b\bar{b}$) measurements and other analyses with similar final states.

MC simulations of the \ttbj process differ in the treatment of the $b$-quark mass, the calculation of the $b$-quark production in the matrix element at next-to-leading order (NLO) or parton shower, and the merging of $b$-quark production into the inclusive \ttbar prediction~\cite{Frixione:2007nw,Nason:2004rx,Frixione:2007vw,Alioli:2010xd,Bothmann:2019yzt,Alwall:2014hca,Jezo:2018yaf,FabioPhilippPozzoriniSiegert:2014,SherpaFusion:2024}. The predictions vary in the angular distribution of the produced jets, their transverse momenta (\pt), mass and energy, requiring a large number of observables to comprehensively probe the various aspects of the \ttbj process.
Comparing the predictions with both integrated and differential cross-section measurements of \ttbar production with additional $b$-jets is crucial for improving the MC modelling.
State-of-the-art QCD calculations give predictions for the \ttbar production cross-section with up to two additional massless partons at NLO
in perturbation theory matched to a parton shower algorithm~\cite{HoechePSmatch:2014}. QCD predictions of \ttbb are also calculated at NLO matched to a parton shower algorithm~\cite{Jezo:2018yaf,FabioPhilippPozzoriniSiegert:2014,SherpaFusion:2024}.

The ATLAS Collaboration has measured \ttj production with $7$~\TeV~\cite{TOPQ-2011-21, TOPQ-2012-03}, $8$~\TeV~\cite{TOPQ-2015-04}, and early $13$~\TeV\ data~\cite{TOPQ-2017-01, TOPQ-2015-17}, as well as \ttbj with $7$~\TeV~\cite{TOPQ-2012-16}, $8$~\TeV~\cite{TOPQ-2014-10}, and partial $13$~\TeV\ data~\cite{TOPQ-2017-12}. The data were found to be consistently higher than the MC predictions in the previous \ttbj fiducial inclusive measurements, while the sensitivity of fiducial differential measurements to predictions suffered from a limited amount of data.  The CMS Collaboration has also reported measurements of \ttj, \ttbj and \ttcj production using a similar amount of data collected at different centre-of-mass energies~\cite{CMS-TOP-12-018,CMS-TOP-12-041,CMS-TOP-17-002,CMS-TOP-18-002,CMS-TOP-20-003}. A new set of \ttbb measurements in the semileptonic \ttbar decay channel was recently reported using the full 13 \TeV\ data sample by the CMS Collaboration~\cite{CMS-TOP-22-009}, where the data exceed the predictions from several MC configurations.

In this paper, measurements of QCD \ttbj fiducial cross-sections are presented using data recorded with the ATLAS detector during $2015-2018$ in proton--proton ($pp$) collisions at a centre-of-mass energy $\sqrt{s} = 13$~\TeV, corresponding to an integrated luminosity of \lumi.  Various fiducial regions are defined at the stable particle level with the criteria as close as possible to the detector-level event selections to minimise the MC modelling uncertainties due to extrapolation to the full phase space.  Additionally, fiducial differential measurements are presented as a function of several observables.
Since the top quark decays into a $b$-quark and $W$ boson nearly 100\% of the time, \ttbar events are usually classified according to how the two $W$ bosons decay.
Only the $e\mu$ channel, in which both the $W$ bosons decay leptonically, is considered.
This includes events where one $W$ boson decays into an electron and electron neutrino and the other into a muon and muon neutrino, as well as those where one or both $W$ bosons produce a leptonically decaying $\tau$-lepton.  %
The dilepton channel has the advantage of having no contamination from jets coming from the hadronic decay of the $W$ boson, as compared to the single-lepton or all-hadronic channels. Requiring one electron and one muon makes the background composition much simpler than other dilepton channels. The observables measured in \ttbj events in the fiducial phase space include: \bjet multiplicity, \pt spectra of the leading four \pt-ordered \bjets, \pt spectra of the two \bjets originating from the top quark decay, \pt spectra of the two highest-\pt additional \bjets, dijet mass and angular distribution of the first two additional \bjets ordered in \pt.
A variable characterising the \ttbar invariant mass, i.e. the invariant mass of the electron, the muon, and the two $b$-jets presumably from the top quark decay, is also measured.  The fiducial phase space definitions considered in these differential measurements extend those presented in the previous ATLAS results~\cite{TOPQ-2017-12}.

This paper is organised as follows. The ATLAS detector is described in Section~\ref{sec:detector}. The simulation of signal and background processes and the event reconstruction and selection are discussed in Sections~\ref{sec:mcsim} and~\ref{sec:recosel}, respectively.
The definition of the fiducial phase space is presented in Section~\ref{sec:fiducial}.
A summary of all measured observables in each fiducial phase space region is detailed in Section~\ref{subsec:observables}.
The estimate of the background processes is summarised in Section~\ref{sec:bkgd}.
The extraction of fiducial cross-sections is detailed in Section~\ref{sec:fidxsec}.
The sources of systematic uncertainties considered in the measurements are described in Section~\ref{sec:systs}.
The results of the inclusive and differential cross-section measurements are presented in Section~\ref{sec:result}.
Finally, a summary of the results is given in Section~\ref{sec:conclusion}.


%

%
\section{ATLAS detector}
\label{sec:detector}

The ATLAS detector~\cite{PERF-2007-01} at the LHC covers nearly the entire solid angle around the collision point~\footnote{ATLAS uses a right-handed coordinate system with its origin at the nominal interaction point (IP) in the centre of the detector and the \(z\)-axis along the beam pipe. The \(x\)-axis points from the IP to the centre of the LHC ring, and the \(y\)-axis points upwards. Polar coordinates \((r,\phi)\) are used in the transverse plane, \(\phi\) being the azimuthal angle around the \(z\)-axis. The pseudorapidity is defined in terms of the polar angle \(\theta\) as \(\eta = -\ln \tan(\theta/2)\) and is equal to the rapidity $ y = \frac{1}{2} \ln \left( \frac{E + p_z c}{E - p_z c} \right) $ in the relativistic limit. Angular distance is measured in units of \(\Delta R \equiv \sqrt{(\Delta y)^{2} + (\Delta\phi)^{2}}\).}.
It consists of an inner tracking detector surrounded by a thin superconducting solenoid, electromagnetic and hadronic calorimeters,
and a muon spectrometer incorporating three large superconducting air-core toroidal magnets.

The inner-detector system (ID) is immersed in a \qty{2}{\tesla} axial magnetic field
and provides charged-particle tracking in the range \(|\eta| < 2.5\).
The high-granularity silicon pixel detector covers the vertex region and typically provides four measurements per track,
the first hit generally being in the insertable B-layer (IBL) installed before Run~2~\cite{ATLAS-TDR-19,PIX-2018-001}.
It is followed by the SemiConductor Tracker (SCT), which usually provides eight measurements per track.
These silicon detectors are complemented by the transition radiation tracker (TRT),
which enables radially extended track reconstruction up to \(|\eta| = 2.0\).
The TRT also provides electron identification information
based on the fraction of hits (typically 30 in total) above a higher energy-deposit threshold corresponding to transition radiation.

The calorimeter system covers the pseudorapidity range \(|\eta| < 4.9\).
Within the region \(|\eta|< 3.2\), electromagnetic calorimetry is provided by barrel and
endcap high-granularity lead/liquid-argon (LAr) calorimeters,
with an additional thin LAr presampler covering \(|\eta| < 1.8\)
to correct for energy loss in material upstream of the calorimeters.
Hadronic calorimetry is provided by the steel/scintillator-tile calorimeter,
segmented into three barrel structures within \(|\eta| < 1.7\), and two copper/LAr hadronic endcap calorimeters.
The solid angle coverage is completed with forward copper/LAr and tungsten/LAr calorimeter modules
optimised for electromagnetic and hadronic energy measurements respectively.

The muon spectrometer (MS) comprises separate trigger and
high-precision tracking chambers measuring the deflection of muons in a magnetic field generated by the superconducting air-core toroidal magnets.
The field integral of the toroids ranges between \num{2.0} and \qty{6.0}{\tesla\metre}
across most of the detector.
Three layers of precision chambers, each consisting of layers of monitored drift tubes, cover the region \(|\eta| < 2.7\),
complemented by cathode-strip chambers in the forward region, where the background is highest.
The muon trigger system covers the range \(|\eta| < 2.4\) with resistive-plate chambers in the barrel, and thin-gap chambers in the endcap regions.

The luminosity is measured mainly by the LUCID--2~\cite{LUCID2} detector that records Cherenkov light produced in the quartz windows of photomultipliers located close to the beampipe.

Events are selected by the first-level trigger system implemented in custom hardware,
followed by selections made by algorithms implemented in software in the high-level trigger~\cite{TRIG-2016-01}.
The first-level trigger accepts events from the \qty{40}{\MHz} bunch crossings at a rate below \qty{100}{\kHz},
which the high-level trigger further reduces in order to record complete events to disk at about \qty{1}{\kHz}.

A software suite~\cite{SOFT-2022-02} is used in data simulation, in the reconstruction
and analysis of real and simulated data, in detector operations, and in the trigger and data acquisition
systems of the experiment.

%
%
%
%
%
%
%
%
%
%
%
%
%
%
%
%


%

%
\section{Signal and background simulations}
\label{sec:mcsim}

Simulated events were produced using various MC algorithms to estimate contributions from \ttbb signal and background processes, determine detector resolution and acceptance correction factors, and evaluate systematic uncertainties.
Additionally, comparisons between unfolded data and theoretical predictions were facilitated using MC generators.
All samples were normalised based on the best available theory predictions as indicated in the text below.

The response of the ATLAS detector and trigger were simulated based on the detailed model implemented in the \textsc{GEANT4} program~\cite{SOFT-2010-01,Agostinelli:2002hh} or, for the estimation of some of the systematic uncertainties, using a faster approach employing parameterised showers in the calorimeter~\cite{ATL-PHYS-PUB-2010-013}.
The effect of multiple $pp$ interactions in each bunch crossing (pile-up) was modelled by overlaying each hard-scattering event with inelastic $pp$ collisions generated using \PYTHIA[8.186]~\cite{Sjostrand:2007gs} with the \texttt{NNPDF2.3LO} set of parton distribution functions (PDFs)~\cite{Ball:2012cx} and the A3 set of tuned parameters~\cite{ATL-PHYS-PUB-2016-017}.
For all samples of simulated events, except those generated using \SHERPA, the \EVTGEN%
program~\cite{Lange:2001uf} was used to model the decays of bottom and charm hadrons.
The top quark mass $m_{\text{top}}$ was set to $172.5$~\GeV\ and the Higgs boson mass was set to $125$~\GeV.
Simulated events were reconstructed and analysed using the same software as the data.

\subsection{\ttbar\ samples}
The \ttbar\ event production was simulated using the \powhegbox~v2 heavy-quark (hvq)~\cite{Frixione:2007nw,Nason:2004rx,Frixione:2007vw,Alioli:2010xd} event generator, employing matrix elements calculated at NLO precision in QCD with the \texttt{NNPDF3.0NLO}~\cite{Ball:2014uwa} PDF set in the five-flavour scheme (5FS).
These events were then interfaced to \PYTHIA[8.230]~\cite{Sjostrand:2014zea} to simulate the processes for parton showering, hadronisation and the underlying event.
Top quark decays, including spin correlations, were modelled at leading-order (LO) precision in QCD\@.
To regulate the \pt\ of the first additional emission beyond the Born configuration, the \hdamp\ parameter was set to 1.5\mtop~\cite{ATL-PHYS-PUB-2016-020}.
Additionally, the renormalisation and factorisation scales defined for the Born process were set to the default scale $\sqrt{m_{\text{top}}^2 + p_{\text{T,top}}^2}$, derived from the top quark mass, $m_{\text{top}}$, and the transverse momentum of the top quark before radiation, $p_{\text{T,top}}$.  The $p_{\mathrm{T}}^{\mathrm{hard}}$ parameter~\cite{Mrenna:2022hmt}, which determines the region of phase space vetoed during showering when matched to a parton shower, was set to zero by default.
This sample used a recoil scheme where partons recoil against a $b$-quark.
This recoil scheme changes the modelling of second and subsequent gluon emissions from the $b$-quark produced by the top quark decay, and therefore affects how the momentum is rearranged between the $W$ boson and the $b$-quark.
This sample is referred to as \powpy.  All samples using \PYTHIA[8] for the modelling of parton showering, hadronisation and underlying event used the A14 set of tuned parameters~\cite{ATL-PHYS-PUB-2014-021} and the \texttt{NNPDF2.3LO} set of PDFs. All \ttbar samples were normalised to the cross-section prediction at next-to-next-to-leading order (NNLO)
in QCD including the resummation of next-to-next-to-leading logarithmic (NNLL) soft-gluon terms calculated using
\TOPpp[2.0]~\cite{Beneke:2011mq,Cacciari:2011hy,Baernreuther:2012ws,Czakon:2012zr,Czakon:2012pz,Czakon:2013goa,Czakon:2011xx}.
In $pp$ collisions at a centre-of-mass energy of \(\rts = \qty{13}{\TeV}\), this cross-section corresponds to
\(\sigma(\ttbar)_{\text{NNLO+NNLL}} = 832 \pm 51\,\unit{\pb}\) using a top quark mass of \(\mtop = \qty{172.5}{\GeV}\).
The uncertainties in the cross-section due to the PDF and the strong coupling constant \alphas were calculated using the \PDFforLHC[15] prescription~\cite{Butterworth:2015oua}
with the \texttt{MSTW2008NNLO}~\cite{Martin:2009iq,Martin:2009bu}, \texttt{CT10NNLO}~\cite{Lai:2010vv,Gao:2013xoa}
and \texttt{NNPDF3.0NNLO} PDF sets in the 5FS, and were added in quadrature to the scale uncertainty.

Various alternative \ttbar samples were generated to assess the uncertainties arising from QCD MC model choices and to compare with unfolded data.
Uncertainties due to initial- and final-state radiation were assessed by using event weights to vary the QCD renormalisation and factorisation scales in the
matrix element independently by factors of $2.0$ and $0.5$ from their default values, using the \texttt{Var3c} variation of the A14 tune, and by changing the renormalisation and factorisation scales used in the parton shower by factors of $2.0$ and $0.625$.  A higher scaling factor than $0.5$ was used for the downward variation to keep the variations of the kinematic variables in the upward and downward directions symmetrical.

Another sample was generated mirroring the nominal \ttbar settings, but with the $h_{\mathrm{damp}}$ parameter set to $3 m_{\text{top}}$, enabling the assessment of the impact of variations in the first QCD emission scale. This sample is referred to as \powpy\ $h_{\mathrm{damp}}$.
Furthermore, the uncertainty in the matrix element and parton shower matching is evaluated by utilising an alternative \powpy\ sample.
The $p_{\mathrm{T}}^{\mathrm{hard}}$ parameter of \PYTHIA[8] was set to 1 instead of the default value of 0, altering the overlap in the phase space filled by \pow\ and \PYTHIA[8]. The sample is referred to as \powpy\ $p_{\mathrm{T}}^{\mathrm{hard}}$.

Another sample was generated with identical settings to the nominal \ttbar sample, except for a modified recoil scheme.
Here, the default gluon recoil, typically recoiling against the $b$-quark, was replaced by the top quark itself serving as the recoil for the second and subsequent gluon emissions from the $b$-quark. This sample is referred to as \powpy RecoilToTop. A \powpy\ sample, with the same settings as the nominal but replacing the recoil scheme of the initial-state emission by an alternative scheme, was used only for comparison with the measurement.  In the alternative scheme, only one final state parton takes the recoil of an emission in contrast to the global recoil.  This sample is referred to as \powpy\ dipole.

Additionally, further \ttbar samples were generated
by replacing \PYTHIA[8] with \HERWIG[7.1.3]~\cite{Bellm:2017jjp} to model the parton showering, hadronisation and underlying event, and are used to evaluate uncertainties.
This alternative approach used the \HERWIG[7.1] default set of tuned parameters~\cite{Bellm:2017jjp}
and the \texttt{MMHT2014LO}~\cite{Harland-Lang:2014zoa} PDF set.

Further predictions were computed for comparison with data, including alternative NLO samples of \ttbar events produced with \MGNLO v2.6.0 ~\cite{Alwall:2014hca} interfaced to either \PYTHIA[8.230] or \HERWIG[7.1.3].
These samples are referred to as \amcnlopyeight and \mcnlohwseven, respectively. %
Additionally, samples were generated using the \SHERPA[2.2.12] generator~\cite{Bothmann:2019yzt} and \OPENLOOPS~\cite{Buccioni:2019sur,Cascioli:2011va} using NLO accurate matrix elements for up to one additional parton and LO accurate matrix elements for up to four additional partons. The additional parton emissions were matched and merged with the \SHERPA parton shower based on Catani--Seymour dipole factorisation~\cite{Gleisberg:2008fv,Schumann:2007mg} using the \textsc{MePs@Nlo} prescription~\cite{Hoeche:2011fd,Catani:2001cc}. Samples were generated using the \texttt{NNPDF3.0NNLO} PDF set, along with a dedicated set of tuned parton shower parameters developed by the \SHERPA authors.  Additional $b$-quarks were treated as massless. The \SHERPA samples include approximate NLO EW$_{\mathrm{virtual}}$ corrections ~\cite{Kallweit:2014xda,Kallweit:2015dum,ChristianLindertMarek:2018}. The scale used for the multijet merging was set to 30~\GeV. This sample is referred to as \sherpa \ttbar.
These samples are summarised in Table~\ref{tab:mc_signal}.

\subsection{$t\bar{t}b\bar{b}$ samples}
Dedicated samples focusing on $t\bar{t}b\bar{b}$ events were produced to capture additional $b$-quarks calculated by the matrix element, ensuring NLO QCD accuracy in the four-flavour scheme (4FS), and interfaced to \PYTHIA[8.230]. %
Using the \POWHEGBOXRES~\cite{Jezo:2018yaf} generator alongside \OPENLOOPS%
, and the four-flavour \texttt{NNPDF3.0NLO NF4}~\cite{Ball:2014uwa} PDF set, this set-up incorporated a specialised implementation to take the scale hierarchy into account~\cite{Jezo:2018yaf,ttbbTopLHCWG:2023}.  This includes setting the factorisation scale to $\frac{1}{2}\Sigma_{i=t,\bar{t},b,\bar{b},j}m_{\mathrm{T},i}$ (where $j$ stands for extra partons), and the renormalisation scale to $\sqrt[4]{\prod_{i=t,\bar{t},b,\bar{b}} m_{\mathrm{T},i}}$, where
$m_{\mathrm{T},i}=\sqrt{m_i^2 +p_{\mathrm{T},i}^2}$ is the transverse mass of a given parton.
The \hdamp\ parameter was set to $\frac{1}{2}\Sigma_{i=t,\bar{t},b,\bar{b}}m_{\mathrm{T},i}$, and the $p_{\mathrm{T}}^{\mathrm{hard}}$ parameter was set to zero.
The global recoil scheme, as implemented by default in \PYTHIA[8], was used such that the recoil of the initial-state emission is taken by the whole final state.
The parameter $h_{\mathrm{bzd}}$ that controls the enhancement in the Born cross-section due to soft and collinear NLO radiative contributions from light jets~\cite{Jezo:2018yaf} was set to 5, and the \POWHEGBOXRES hardness criterion (\texttt{POWHEG:pTdef}) used in the matching was set to 1 following studies reported in Ref.~\cite{ATL-PHYS-PUB-2022-006}.
This sample is referred to as \powpy\ \ttbb.
Further variations were explored through three additional \ttbb samples.
One sample had $h_{\mathrm{bzd}}$ set to 2, the second sample had $p_{\mathrm{T}}^{\mathrm{hard}}$ set to 1, and the third sample used a different scheme for the recoil of the initial-state emission such that only one final-state parton took the recoil of an emission and had $h_{\mathrm{bzd}}$ set to 2, all other settings were the same as the nominal \powpy\ \ttbb sample. These three samples are referred to as \powpy\ \ttbb $h_{\mathrm{bzd}}$, \powpy\ \ttbb $p_{\mathrm{T}}^{\mathrm{hard}}$, and \powpy\ \ttbb dipole, respectively.

An additional \ttbb (4FS) sample was generated, preserving the \POWHEGBOXRES%
generator settings while substituting \PYTHIA[8] with \HERWIG[7.1.6].
This sample, referred to as \phwsevenone \ttbb, used the \HERWIG[7.1] set of tuned parameters and the \texttt{MMHT2014LO} PDF set.

Furthermore, a dedicated sample of \ttbb events was generated using \SHERPA[2.2.10] in conjunction with \OPENLOOPS. %
Matrix element calculations, performed at NLO accuracy using the \textsc{Comix}~\cite{Gleisberg:2008fv} and \OPENLOOPS %
generators, were merged with the \SHERPA parton shower~\cite{Schumann:2007mg}. The renormalisation scale was set to $\sqrt[4]{\prod_{i=t,\bar{t},b,\bar{b}} m_{\mathrm{T},i}}$.
The factorisation and resummation scales were both set to $\frac{1}{2}\Sigma_{i=t,\bar{t},b,\bar{b},j}m_{\mathrm{T},i}$ where $j$ refers to extra partons.
The \texttt{NNPDF3.0NNLO NF4} PDF set was used. This sample is referred to as \sherpa \ttbb.
In all \ttbb (4FS) samples, the mass of the $b$-quark was set to 4.75~\GeV\ in the matrix element.
The MC yields of all \ttbb samples were normalised to the cross-sections computed by the corresponding matrix element generators.

Additionally, the results featuring at least four \bjets were compared to the NLO computation of the $pp\rightarrow e\nu_e \mu\nu_{\mu} b\bar{b} b\bar{b}$ process, encompassing resonant and non-resonant top quark contributions, and all interference effects among them, along with the non-resonant and off-shell effects of the $W$ boson.
These calculations were performed using the \textsc{Helac-NLO} framework~\cite{HelacNLO} by the authors of Refs.~\cite{malgorzata_ttbb_2021} and~\cite{malgorzata_ttbb_2022}.
The renormalisation and factorisation scales were set to $\mu_{\mathrm{R}} = \mu_{\mathrm{F}} = \mu_{\mathrm{0}} = H_{\mathrm{T}} / 3$, and employed the \texttt{NNPDF3.1NLO}~\cite{Ball:2017uwa} PDF sets.
Here, $H_{\mathrm{T}}$ represents the scalar sum of the \pt of all four \bjets, electron, muon, and the missing transverse momentum from the two neutrinos.
The predictions were corrected to account for the effects of non-perturbative contributions such as multiple parton interactions and hadronisation, and the leptonic decays of at least one $\tau$-lepton in the dileptonic \ttbar channel yielding the prompt $e\mu$ final state. The \ppyeight \ttbb MC samples, as discussed above, were used to derive the corrections.
These predictions are denoted by \textsc{Helac-NLO} (off-shell) $e\mu+4b$ and are summarised in Table~\ref{tab:mc_signal}.

\subsection{Background samples}
Single-top-quark production processes, namely $tW$ associated production, $t$-channel, and $s$-channel production, were simulated using the \POWHEGBOX[v2]~\cite{Re:2010bp,Nason:2004rx,Frixione:2007vw,Alioli:2010xd,Alioli:2009je,Frederix:2012dh} generator at NLO in QCD interfaced to \PYTHIA[8] for the parton showering, hadronisation and the underlying event, using the A14 set of tuned parameters and the \texttt{NNPDF2.3LO} set of PDFs.
The $t$-channel process was generated in the 4FS with the \texttt{NNPDF3.0NLO NF4}~\cite{Ball:2014uwa} PDF set, while the 5FS was used for $tW$ and $s$-channel processes.
To handle the interference between $tW$ and \ttbar\ production, the diagram removal scheme~\cite{Frixione:2008yi} was employed.
An alternative $tW$ sample was generated with the same settings as the nominal one except that the diagram subtraction scheme~\cite{Frixione:2008yi} was used to evaluate the uncertainty due to the treatment of the interference with \ttbar production.
Additionally, one sample was simulated for $tW$ using the same set-up as the nominal one, but replacing \PYTHIA[8] with \HERWIG[7.0.4], and another sample was generated at NLO in QCD with \MGNLO interfaced to \PYTHIA[8].  The single-top MC samples for the $t$- and $s$-channels were normalised to cross-sections from NLO predictions~\cite{Aliev:2010zk,Kant:2014oha}, while the $tW$-channel MC sample was normalised to NLO+NNLL~\cite{Kidonakis:2010ux}.

The production of \ttH\ events was simulated using the \POWHEGBOX generator at NLO in QCD with the \hdamp parameter set to $0.75\times(2\mtop+m_H)$, and interfaced to \PYTHIA[8].
All possible SM decay modes of the Higgs boson were included and simulated using \PYTHIA[8].
Alternative \ttH samples were generated using \POWHEGBOX with the same settings as the nominal ones, but replacing the parton shower model with \HERWIG[7].
Another sample was generated with \MGNLO and interfaced to \PYTHIA[8].

Events for \ttZ, \ttW, \tWZ, \tWH, \tHbj, \tZ and \tttt were generated with \MGNLO at NLO in QCD\@.
The \ttZ and \ttW background processes are collectively labelled as \ttV.
The MC yields were normalised to the corresponding cross-sections from the \MGNLO computation. %

All of these nominal background events generated using \POWHEGBOX or \MGNLO, as mentioned above, were interfaced to \PYTHIA[8.230]~\cite{Sjostrand:2014zea} for the parton showering, hadronisation and the underlying event, using the A14 set of tuned parameters and the \texttt{NNPDF2.3LO} set of PDFs.

For $V+$jets events (where $V=W$ or $Z$), simulation was carried out using the \SHERPA[2.2.1] generator~\cite{Bothmann:2019yzt} %
with NLO-accurate matrix elements for up to two partons and LO-accurate matrix elements for up to four partons. The MC yields were normalised to the NNLO cross-sections, computed using \textsc{Fewz}~\cite{Anastasiou:2003ds} with
the \texttt{MSTW2008NNLO} PDF set.
Additionally, samples of diboson final states were simulated using the \SHERPA[2.2.2] generator~\cite{Bothmann:2019yzt} %
at NLO in QCD for up to one additional parton and at LO accuracy for up to three additional partons. The samples were normalised according to the cross-sections computed by the generator.

\begin{table}
\caption{Summary of the MC sample set-ups for modelling the signal processes for the data analysis and for comparison with the measured cross-sections and differential distributions. The different blocks indicate, from top to bottom, the samples used as the nominal MC (nom.), systematic variations (syst.) and for comparison only (comp.).  Dashes (--) imply that the corresponding settings are not applicable for the fixed-ordered calculations. The $t\bar{t}b\bar{b}$ simulations are performed in the four-flavour scheme (4FS), where the $b$-quark is treated as a massive particle in the matrix element. Other simulations employ the five-flavor scheme (5FS), where the $b$-quark is treated as massless within the matrix element. However, all parton shower models consider the $b$- and $c$-quarks to be massive.
For details see Section~\ref{sec:mcsim}.}
\footnotesize
\centering
\begin{scalebox}{0.76}{
\begin{tabular}{llllllll}
\toprule
MC sample                                  & Generator     &  Process        & Flavour   & Parton shower   & Matching/                            & Tune          & Use  \\
&               &                 & Scheme       &                 & Parton shower                        &               &   \\
&               &                 & (4FS/5FS)                 &                 & settings                             &               &   \\
\midrule
\ppyeight                                  & \POWHEGBOX[v2] & $t\bar{t}$ NLO & 5FS              & \PYTHIA[8.230] & \POWHEG                              & A14           & nom. \\
&                &                &                  &                & \hdamp=1.5\mtop                      &               &       \\
&               &                 &                  &                & $p_{\mathrm{T}}^{\mathrm{hard}}=0$   &               &       \\
&               &                 &                  &                & \texttt{globalRecoil}                &               &       \\
&               &                 &                  &                & \texttt{recoilToColoured}=ON         &               &        \\
\midrule
\ppyeighthdamp                             & \POWHEGBOX[v2] & $t\bar{t}$ NLO & 5FS              & \PYTHIA[8.230] & \textsc{Powheg}                      & A14           & syst. \\
&               &                 &                  &                & \hdamp=3\mtop                        &               &       \\
\ppyeight $p_{\mathrm{T}}^{\mathrm{hard}}$ & \POWHEGBOX[v2] & $t\bar{t}$ NLO & 5FS              & \PYTHIA[8.230] & \textsc{Powheg}                      & A14           & syst. \\
&               &                 &                  &                & $p_{\mathrm{T}}^{\mathrm{hard}}=1$                    &               &       \\
\ppyeight RecoilToTop                      & \POWHEGBOX[v2] & $t\bar{t}$ NLO & 5FS              & \PYTHIA[8.230] & \textsc{Powheg}                      & A14           & syst. \\
&               &                 &                  &                & \texttt{recoilToTop}                 &               &       \\
\phwsevenone                               & \POWHEGBOX[v2] & $t\bar{t}$ NLO & 5FS              & \HERWIG[7.1.3] & \textsc{Powheg}                      & H7.1-Default & syst. \\
\midrule
\ppyeight dipole                           & \POWHEGBOX[v2] & $t\bar{t}$ NLO & 5FS              & \PYTHIA[8.230] & \textsc{Powheg}                      & A14           & comp. \\
&                &                &                  &                & \texttt{dipoleRecoil} on             &               &       \\
\amcnlopyeight                             & \MG            & $t\bar{t}$ NLO &  5FS             & \PYTHIA[8.230] & \textsc{MC@NLO}                      & A14           & comp. \\
& \amNLO[v2.6.0] &                &                  &                &                                      &               &  \\

\mcnlohwseven                              & \MG            & $t\bar{t}$ NLO  &  5FS            & \HERWIG[7.1.3] & \textsc{MC@NLO}                      & H7.1-Default & comp. \\
& \amNLO[v2.6.0] &                 &                 &                &                                      &               &        \\
\sherpat                                   & \SHERPA[2.2.12] & \ttbar+0,1 @NLO & 5FS            & \SHERPA       & \textsc{MePs@Nlo}                     & Author's tune       & comp. \\
&                 & ~+2,3,4 @LO    &               &                                                         &               &       \\
\ppyeight \ttbb                            & \POWHEGBOXRES   & $t\bar{t}b\bar{b}$ NLO & 4FS   & \PYTHIA[8.230] & \POWHEGBOXRES                          & A14           & comp. \\
&                 &                        &       &                & $h_{\mathrm{bzd}}$=5                   &               &       \\
&                 &                        &       &                & $p_{\mathrm{T}}^{\mathrm{hard}}=0$     &               &       \\
&                 &                        &       &                & \texttt{globalRecoil}                  &               &       \\
\ppyeight \ttbb $p_{\mathrm{T}}^{\mathrm{hard}}$ & \POWHEGBOXRES   & $t\bar{t}b\bar{b}$ NLO & 4FS &  \PYTHIA[8.230] & \POWHEGBOXRES                     & A14           & comp. \\
&                 &                        &     &                 &  $p_{\mathrm{T}}^{\mathrm{hard}}=1$   &               &       \\
\ppyeight \ttbb $h_{\mathrm{bzd}}$         & \POWHEGBOXRES   & $t\bar{t}b\bar{b}$ NLO       & 4FS &  \PYTHIA[8.230] & \POWHEGBOXRES                                      & A14           & comp. \\
&                 &                              &     &                 & $h_{\mathrm{bzd}}$=2                               &               &       \\
\ppyeight \ttbb dipole                     & \POWHEGBOXRES   & $t\bar{t}b\bar{b}$ NLO & 4FS & \PYTHIA[8.230] & \POWHEGBOXRES                             & A14           & comp. \\
&                 &                        &     &                & $h_{\mathrm{bzd}}$=2                      &               &       \\
&                 &                        &     &                & \texttt{dipoleRecoil} on                  &               &       \\
\phwsevenone \ttbb                         & \POWHEGBOXRES   & $t\bar{t}b\bar{b}$ NLO & 4FS & \HERWIG[7.1.6] & \POWHEGBOXRES                             & H7.1-Default & comp. \\
\sherpat \ttbb                             & \SHERPA[2.2.10] & $t\bar{t}b\bar{b}$ NLO & 4FS & \SHERPA         & \textsc{MePs@Nlo}                        & Author's tune       & comp. \\
&                 &      &          &                                          &        &  \\
\textsc{Helac-NLO} (off-shell)             & \textsc{Helac-NLO}       & $e\mu+4b$ NLO & 5FS &                             & --                  & --            & comp. \\

\bottomrule
\end{tabular}
}\end{scalebox}

\label{tab:mc_signal}
\end{table}

%
%
%

%
%
%
%
%
%
%
%
%
%
%
%
%
%
%
%
%
%
%
%
%
%
%
%
%
%
%
%
%
%
%
%
%
%
%
%


%

%
%

%
\section{Detector-level event reconstruction and selection}
\label{sec:recosel}

%
\subsection{Detector-level object reconstruction}
\label{subsec:detobjs}
Interaction vertices resulting from $pp$ collisions are reconstructed based on a minimum of two tracks with $\pt>500~\mev$, demonstrating consistency with originating from the beam collision region in the $x$--$y$ plane. Of all primary vertex candidates, the one associated with the highest sum of squared $\pt$ of its associated tracks is designated as the hard-scatter primary vertex~\cite{ATL-PHYS-PUB-2015-026}.

Electron candidates are reconstructed by using energy clusters in the EM calorimeter, which are then matched to a track in the ID~\cite{EGAM-2018-01}. These candidates must satisfy $\pt>25~\gev$ and pseudorapidity $|\eta_{\mathrm{cluster}}|<2.47$, excluding the transition region between the endcap and barrel calorimeters ($1.37<|\eta_{\mathrm{cluster}}|<1.52$).
The selection criteria use tight electron identification working point, characterised by a likelihood discriminant utilising calorimeter, tracking, and combined variables to effectively discriminate between electrons and jets~\cite{EGAM-2018-01}.

Muon candidates are reconstructed by combining tracks in the ID with tracks in the MS~\cite{MUON-2018-03}.
The resulting muon candidates undergo a refitting process using comprehensive track information from both detector systems.
Selection criteria ensure that these candidates have $\pt > 25~\gev$ and $|\eta| < 2.5$ and satisfy the medium muon identification working point (\textit{Medium})~\cite{MUON-2018-03}.

The association of electron (muon) candidate tracks with the primary vertex is achieved by requiring that the significance of their transverse impact parameter, $d_0$, satisfies $|d_0/\sigma(d_0)|<5\ (3)$, where $\sigma(d_0)$ denotes the measured uncertainty in $d_0$.
Furthermore, the longitudinal impact parameter is required to satisfy $|z_0 \sin\theta|<0.5$~mm. %

To mitigate the presence of non-prompt leptons and jets misidentified as leptons, stringent isolation criteria are imposed on lepton candidates in both the tracker and calorimeter.
For track-based lepton isolation, the quantity $I_R = \sum \pt^{\textrm{trk}}$ is computed, summing the transverse momenta of all tracks (excluding the lepton candidate itself) within a cone of size $R_{\textrm{cut}}$ around the lepton's direction. Here, $R_{\textrm{cut}}$ is set to the minimum of $r_{\textrm{min}}$ and $10~\gev/\pt^\ell$, with $r_{\textrm{min}}$ taking the values 0.2~(0.3) for electron (muon) candidates, and $\pt^\ell$ representing the lepton's transverse momentum.
The electron selection uses \pt and $\eta$-dependent thresholds on the $I_R/\pt^\ell$ ratio, while all muon candidates must satisfy $I_R/\pt^\ell < 0.15$.

Moreover, both electrons and muons must satisfy a calorimeter-based isolation requirement that calculates the sum of transverse energy within a cone of $R_{\textrm{cut}}=0.2$ around the lepton, after subtracting contributions from pile-up and the lepton's own energy deposition.
This sum is required to be below 30\% of the muon's transverse momentum, while electron candidates are subject to \pt and $\eta$-dependent thresholds for this quantity.

Reconstruction, identification and isolation efficiencies of electrons (muons) are corrected to match those observed in data using $Z\rightarrow e^+ e^-(\mu^+ \mu^-)$ events, while the position and width of the observed $Z$ peak is used to calibrate the electron (muon) energy (momentum) scale and resolution~\cite{EGAM-2018-01,MUON-2018-03,MUON-2022-01}.

Jet reconstruction relies on identifying constituents through a particle flow (PFlow) algorithm that combines measurements from both the ID and the calorimeter~\cite{PERF-2015-09}.
Subsequently, jet candidates are formed from these PFlow objects using the anti-$k_t$ algorithm with a radius parameter of $R=0.4$~\cite{Cacciari:2008gp,Fastjet}.
These candidates undergo calibration using simulation, with corrections derived from in situ techniques applied to data~\cite{JETM-2018-05}.
Only jet candidates with $\pT > 25$~\GeV\ and $|\eta|<2.5$ are retained.
To mitigate pile-up effects, jets with $\pT<60$~\GeV\ and $|\eta|<2.4$ are required to satisfy the `\textit{Tight}' working point of the jet vertex tagger (JVT) criteria, ensuring an efficiency of 95\%--97\% depending on the jet \pt, which identifies jets as originating from the selected primary vertex~\cite{PERF-2014-03}.
Additionally, a set of quality criteria is applied to discard events containing jets originating from non-collision sources or detector noise~\cite{ATLAS-CONF-2015-029}.

Jets likely to contain $b$-hadrons, i.e. weakly decaying hadrons composed of a $b$-quark, are identified as \bjets using the DL1r algorithm~\cite{FTAG-2019-07}, which employs a deep neural network. These identified jets are known as $b$-tagged jets.  The network utilises distinctive features of $b$-hadrons, such as the impact parameters of tracks and the displaced vertices reconstructed in the ID\@.
Additionally, discriminating variables constructed by a recurrent neural network are included as input, exploiting spatial and kinematic correlations between tracks originating from the same $b$-hadron. For each jet, a multivariate $b$-tagging discriminant is computed, and the jet is considered $b$-tagged if this value exceeds a predefined threshold.

The baseline working point, chosen to define signal regions, corresponds to an approximate $77\%$ tagging efficiency for $b$-jets in simulated \ttbar events. However, the $b$-tagging efficiency varies with $\pT$ and $\eta$ and is measured using collision data~\cite{FTAG-2018-01} with uncertainties ranging from $1\%$ to $8\%$ depending on the jet $\pT$.  %
The tagging algorithm achieves a rejection factor of about 170 against light-flavoured jets ($u$-, $d$-, $s$-quark, or gluon) and approximately 5 against jets originating from $c$-quarks.
The selected working point is optimised to maximise the signal-to-background ratio, while keeping the number of signal events high.

Ambiguities between selected leptons and jets are resolved using an overlap removal procedure.
If a muon shares a track with an electron, indicating possible bremsstrahlung, the electron is not selected.
To avoid double-counting electron energy deposits as jets, the closest jet within $\Delta R=0.2$ of a selected electron is removed.
Subsequently, if the nearest surviving jet is within $\Delta R = 0.4$ of the electron, the electron is discarded to reduce non-prompt background contributions.
Muons are removed if they are within $\Delta R = 0.4$ of the nearest jet, which helps mitigate background from heavy-flavour decays inside jets.
However, if this jet is associated with fewer than three tracks, the muon is retained, and the jet is removed instead.
This approach prevents inefficiencies caused by high-energy muons experiencing significant energy loss in the calorimeter.

The missing transverse energy (denoted by $E_{\mathrm{T}}^{\mathrm{miss}}$) is defined as the magnitude of the negative vector sum of the $\pt$ of all selected and calibrated objects in the event.
This includes a term to accommodate the momentum from soft particles in the event that are not associated with any of the selected objects~\cite{JETM-2020-03}.
The soft term is computed from ID tracks matched to the selected primary vertex, enhancing its resilience to contamination from pile-up interactions.

\subsection{Event selection}
\label{subsec:evtsel}
A data sample of $pp$ collisions at a centre-of-mass energy of $\sqrt{s}=13$~\TeV, collected with the ATLAS detector during the period of 2015--2018, is used in this analysis.
This data sample corresponds to an integrated luminosity of \lumi~\cite{DAPR-2021-01}.
Events were only recorded under stable beam conditions and when all components of the ATLAS detector were confirmed to be fully functioning.

The data were collected using a combination of single-electron and single-muon triggers, with requirements on the identification, isolation, and \pt of the leptons to maintain high efficiency across the full momentum range while controlling the trigger rates~\cite{TRIG-2018-05,TRIG-2018-01}.
For electrons the trigger thresholds were \pt = 26, 60 and 140~$\gev$, whereas for muons the thresholds were \pt = 26 and 50~$\gev$.\footnote{Lower \pt thresholds of 24~$\gev$ and 120~$\gev$ for electrons and 20~$\gev$ for muons were applied for 2015 data.}
Isolation requirements were applied to the triggers with the lowest \pt thresholds of 26~$\gev$ for both electrons and muons~\cite{ATL-DAQ-PUB-2016-001,ATL-DAQ-PUB-2017-001,ATL-DAQ-PUB-2018-002,ATL-DAQ-PUB-2019-001}.

To ensure events originate from the hard-scattering process, they are required to contain at least one primary vertex candidate.
A sample enriched in \ttbar events is pre-selected by requiring events to contain exactly one electron and one muon with a minimum \pt of $28$~\GeV, along with at least two jets tagged as containing $b$-hadrons using the DL1r $b$-tagging algorithm at the $77\%$ efficiency working point.
The selected leptons are required to have opposite charges and to match the corresponding lepton reconstructed by the single-lepton trigger, with a \pt exceeding the trigger \pt threshold by up to 2~$\gev$ to ensure they remain at the trigger efficiency plateau.
Additionally, the invariant mass of the electron and muon pair is required to be larger than 15~\GeV\ to remove contributions from low-mass $\tau\tau$ Drell--Yan processes with only minimal signal efficiency loss. The \ttbj signal events are selected by requiring at least three $b$-tagged jets at the $77\%$ efficiency working point.


%

%
%
%

%
\section{Definitions of fiducial phase space}
\label{sec:fiducial}

%
\subsection{Particle-level object selections}
\label{subsec:partleveldef}
Particle-level objects are selected in simulated events using definitions that mirror closely the detector-level objects outlined in Section~\ref{subsec:detobjs}.
These particle-level objects are defined based on stable particles with proper lifetimes greater than 30 ps.

Only prompt electrons and muons that are produced directly from the primary vertex, including those from $\tau$-lepton decays, are considered.
To account for final-state photon radiation, the four-momentum of each lepton is adjusted by adding the four-momenta of all photons, not emanating from hadrons, within a $\Delta R=0.1$ cone around the lepton.
Electrons and muons are required to have  $\pt > 25~\GeV$ and $|\eta|< 2.5$.

Jet clustering is performed using the anti-$k_t$ algorithm with a radius parameter of $R=0.4$.
All stable particles are considered for jet clustering except for prompt electrons and muons, along with the associated photons.
Neutrinos from $W$ boson decays (prompt neutrinos) are excluded from jet clustering, but neutrinos from hadron decays are included.
These jets exclude particles from pile-up events but incorporate those from the underlying event. The decay products of hadronically decaying $\tau$-leptons are included in the jet clustering.
Jets are required to have $\pt > 25~\GeV$ and $|\eta| < 2.5$.

Jets are defined as $b$-jets, if at least one $b$-hadron with $\pt > 5~\GeV$  is associated with the jet via ghost association~\cite{Cacciari:2008gn}.
In this ghost-association procedure, $b$-hadrons are included in the jet clustering while their $\pt$ is scaled to a negligible value.
A similar approach is adopted to define $c$-jets, with the $b$-jet definition taking precedence.
In other words, a jet containing both a $b$-hadron and a $c$-hadron is classified as a $b$-jet.
Jets lacking both $b$-hadrons and $c$-hadrons are categorised as light-flavour jets.

Electrons and muons, satisfying the selection criteria outlined above, must have a separation from selected jets of $\Delta R(\text{lepton}, \text{jet}) > 0.4$.
This criterion ensures consistency with the detector-level selection described in Section~\ref{subsec:detobjs}.

\subsection{Phase-space definitions}
\label{subsec:fiddef}
The fiducial phase spaces are defined using particle-level objects with kinematic requirements similar to those applied to the reconstructed objects in the event selections.
Fiducial integrated cross-sections and fiducial normalised differential cross-sections are determined by requiring the presence of exactly one electron and one muon, each with $\pt \geq 28$~\GeV\ and opposite charges at the particle level, and with two different criteria for the minimum number of \bjets with \pt $>25$~\GeV.
Detailed kinematic requirements can be found in Section~\ref{subsec:partleveldef}.
One set of measurements are performed in the particle-level phase space with three or more \bjets accounting for at least one \bjet in addition to the two from the top quark decays, and another set with the requirement of four or more \bjets accounting for at least two additional \bjets.  The normalised fiducial differential cross-sections as a function of \bjets multiplicity are measured with two different criteria for the minimum number of \bjets: (i) at least two \bjets, and (ii) at least three \bjets. A few percent of events may contain at least one \bjet from a different parton interaction than the one producing the \ttbar pair (referred to as `multiple parton interactions'), with reduced contribution as the \bjet multiplicity requirement increases.  Such \bjets are included in the particle-level event selection.

\section{Observables}
\label{subsec:observables}
The normalised differential distributions of various physics quantities are measured in the fiducial phase space as outlined in Table~\ref{tab:obs_list}.
These distributions fall into two categories of particle-level observables closely corresponding to their detector-level counterparts.
In the first category, \bjets are ordered by \pt, while in the second category, \bjets are classified as originating from the top quark decay or additional partons using an algorithm based on the angular distances as described in Section~\ref{subsec:bjetclass}.
For each category, the observables are defined independently of any specific generator, probing the kinematics of \bjets produced from the top quark decay, dynamics of additional \bjets and non-\bjets, and overall event characteristics within each region of phase space.

Observables such as the number $N_{b \text{-jets}}$ of \bjets in the event, the scalar sum $H_{\mathrm{T}}^{\mathrm{had}}$ of the \pt of each jet, and the scalar sum $H_{\mathrm{T}}^{\mathrm{all}}$ of the \pt of each lepton, jet and the $E_{\mathrm{T}}^{\mathrm{miss}}$ represent the overall event properties. These observables are sensitive to matrix elements and parton shower modelling.
Additionally, variables such as the average angular distance $\Delta R_{\text{avg}}^{bb}$ between any two \bjets and the maximum separation $\Delta \eta_{\text{max}}^{jj}$ in $\eta$ between any two jets are indicative of additional QCD emissions.  Further measurements include the \pt and $\eta$ observables of up to four leading \bjets, as listed in Table~\ref{tab:obs_list}, where the \bjets are labelled as $b_1$, $b_2$, $b_3$, and $b_4$ according to their rank in descending order of their \pt.  These variables encapsulate contributions from both the top quark decay products and the additional jets, with the leading \bjets kinematics dominantly reflecting the top quark. The invariant mass $m(e\mu b_1b_2)$ of the system of electron, muon, and two leading \bjets, as well as the invariant mass $m(b_1b_2)$, transverse momentum $\pt(b_1b_2)$, and angular distance $\Delta R(b_1,b_2)$ of the system of two leading \bjets have sensitivity to the dynamics of top quark decay products.  Moreover, the additional \bjet activity is explored through distributions of the angular distance min$\Delta R(bb)$, invariant mass $m (bb^{\text{min}\Delta R})$, and transverse momentum $p_{\text{T}}(bb^{\text{min}\Delta R})$ of the two closest \bjets. The $N_{b \text{-jets}}$ distributions are measured separately in phase spaces with $\geq 2b$ and $\geq 3b$, while all other differential observables are measured in $\geq 3b$ and $\geq 4b$ phase spaces, where $2b$, $3b$ and $4b$ refer to the number of \bjets.

In the second category, \pt and $\eta$ distributions are obtained after assigning \bjets as originating from top quarks ($b_1^{\text{top}}$ and $b_2^{\text{top}}$) using the reconstruction algorithm to better probe the kinematics of the top quark, where the subscripts $1$ and $2$ indicate the leading- and sub-leading \bjets assigned to the top quarks, respectively.
Similarly, the \pt and $\eta$ distributions of the leading additional \bjet ($b_1^{\text{add}}$) assigned using the reconstruction algorithm are measured.
The invariant mass $m (e\mu bb^{\text{top}})$ of the system of electron, muon, and two \bjets from top quarks, and the invariant mass $m (bb^{\text{top}})$ and transverse momentum $\pt(bb^{\text{top}})$ of the system of two \bjets assigned to top quarks are studied.  Another interesting observable is the angular distance between the leading additional \bjet and the system of electron, muon, and two \bjets from the top quarks, which is sensitive to the recoil momentum distribution shared by final-state partons.  These observables are measured in $\geq 3b$ and $\geq 4b$ phase spaces.
Additionally, the \pt and $\eta$ distributions of the sub-leading additional \bjet ($b_2^{\text{add}}$), invariant mass $m (bb^{\text{add}})$, and transverse momentum $\pt(bb^{\text{add}})$ of two additional \bjets system are measured in the $\geq 4b$ phase space.

Additional non-\bjets (denoted by $l/c$-jets), defined as those not containing any ghost-matched $b$-hadron, can occur in the dileptonic events only via QCD radiation.  This is only partially predicted by the matrix element with limited accuracy in QCD\@. The production rate of additional $l/c$-jets are sensitive to both the matrix element and the parton shower models. The activity of those jets is measured in terms of the multiplicity $N_{l/c\text{-jets}}$ of additional $l/c$-jets in the $\geq 3b$ and $\geq 4b$ phase spaces, the $\pt$ and $\eta$ variables of the leading additional $l/c$-jet ($l/c\text{-jet}_1$), the scalar difference $\pt (l/c\text{-jet}_1) - \pt(b_1^{\text{add}})$ in the transverse momenta of the leading additional $l/c$-jet and the leading additional \bjet, and the angular separation $\Delta R (e\mu bb^{\text{top}}, l/c\text{-jet}_1)$ between the leading $l/c$-jet and the direction of the system of electron, muon and \bjets assigned to the top quarks in the $\geq 3b$ $\geq 1 l/c$ and $\geq 4b$ $\geq 1 l/c$ phase spaces, as listed in Table~\ref{tab:obs_list}.

All of these observables are also constructed in a similar way at the detector level, where the $b$-tagged jets are considered as a proxy for the \bjets.  The $b$-tagged jets are ordered in \pt at the detector-level unless mentioned otherwise.  Similary, the additional $l/c$-jets at the detector-level refer to all those reconstructed jets that are not $b$-tagged.

%

\begin{table}
\caption{Summary of all measured observables in each fiducial phase space region.  The availability of an observable is indicated by $\checkmark$ in the last five columns. The first half of the table lists the global event variables and the kinematics of \bjets ordered in \pt, whereas the second half lists the quantities specific to the \bjets assigned to top quark decays or to extra $b\bar{b}$ system as well as those related to the additional light- or $c$-jets ($l/c$-jets).} %
\begin{center}
\resizebox{0.95\textwidth}{!}{
\begin{tabular}{c|l|ccccc}
\toprule
Observable  & Description & \multicolumn{5}{c}{Phase spaces}  \\
&             & \multicolumn{1}{c|}{$\geq 2b$} & \multicolumn{1}{c|}{$\geq 3b$} & \multicolumn{1}{c|}{$\geq 3b$} & \multicolumn{1}{c|}{$\geq 4b$} & \multicolumn{1}{c}{$\geq 4b$} \\
&             & \multicolumn{1}{c|}{}          & \multicolumn{1}{c|}{}          & \multicolumn{1}{c|}{$\geq 1 l/c$} & \multicolumn{1}{c|}{} & \multicolumn{1}{c}{$\geq 1 l/c$} \\
\midrule
$\sigma^{\mathrm{fid}}$ &  Fiducial total cross-section & & $\checkmark$ & $\checkmark$ & $\checkmark$ & $\checkmark$ \\
$N_{b \text{-jets}}$ & Number of \bjets & $\checkmark$ & $\checkmark$ &  &  &\\
$N_{l/c\text{-jets}}$ & Number of light- or $c$-jets & & $\checkmark$ &  & $\checkmark$ \\
$H_{\mathrm{T}}^{\mathrm{had}}$ & Scalar sum of \pt of all jets & & $\checkmark$ & & $\checkmark$ & \\
$H_{\mathrm{T}}^{\mathrm{all}}$ & Scalar sum of \pt of charged leptons, jet and missing \et & & $\checkmark$ & & $\checkmark$ & \\
$\Delta R_{\text{avg}}^{bb}$ &  Average angular distance in $\Delta R$ of \bjet pairs & & $\checkmark$ & & $\checkmark$ & \\
$\Delta \eta_{\text{max}}^{jj}$ &  Maximum absolute difference in $\eta$ between any pair of jets & &  $\checkmark$ & & $\checkmark$ & \\
$\pt(b_1)$ & \pt of the hardest \bjet & & $\checkmark$ & & $\checkmark$ & \\
$\pt(b_2)$ & \pt of second-hardest \bjet & & $\checkmark$ & & $\checkmark$ & \\
$\pt(b_3)$ & \pt of third-hardest \bjet & & $\checkmark$ & & $\checkmark$ & \\
$\pt(b_4)$ & \pt of fourth-hardest \bjet & & & & $\checkmark$ &  \\
$\eta(b_1)$ & $\eta$ of hardest \bjet & & $\checkmark$ & & $\checkmark$ & \\
$\eta(b_2)$ & $\eta$ of second-hardest \bjet & & $\checkmark$ & & $\checkmark$ & \\
$\eta(b_3)$ & $\eta$ of third-hardest \bjet & & $\checkmark$ & & $\checkmark$ & \\
$\eta(b_4)$ & $\eta$ of fourth-hardest \bjet & & & & $\checkmark$ &  \\
$\pt(l/c\text{-jet}_1)$ & \pt of the hardest light- or $c$-jet & & & $\checkmark$ & & $\checkmark$  \\
$\eta(l/c\text{-jet}_1)$ & $\eta$ of the hardest light- or $c$-jet & & & $\checkmark$ & & $\checkmark$ \\
$m(b_1b_2)$ & Invariant mass of two hardest \bjets in \pt & & $\checkmark$  & & $\checkmark$ & \\
$\Delta R(b_1,b_2)$ & $\Delta R$ between two hardest \bjets & & $\checkmark$ & & $\checkmark$ &\\
$\pt(b_1b_2)$ &  \pt of two hardest \bjets & & $\checkmark$ & & $\checkmark$ & \\
$m (bb^{\text{min}\Delta R})$ & Invariant mass of two closest \bjets in $\Delta R$ & &  & & $\checkmark$ & \\
$p_{\text{T}}(bb^{\text{min}\Delta R})$ & \pt of the closest \bjets pair & &  & & $\checkmark$ & \\
min$\Delta R(bb)$ & Closest angular distance in $\Delta R$ among \bjets & & & & $\checkmark$ & \\
$m(e\mu b_1b_2)$ & Invariant mass of electron, muon and two hardest \bjets & & $\checkmark$ & & $\checkmark$ & \\
\midrule
$\pt(b_1^{\text{top}})$ & \pt of the hardest \bjet assigned to top quark & & $\checkmark$ & & $\checkmark$ & \\
$\pt(b_2^{\text{top}})$ & \pt of the second-hardest \bjet assigned to top quark & & $\checkmark$ & & $\checkmark$ & \\
$\pt(b_1^{\text{add}})$ & \pt of the hardest additional \bjet & & $\checkmark$ & & $\checkmark$ & \\
$\pt(b_2^{\text{add}})$ & \pt of the second-hardest additional \bjet & &  & & $\checkmark$ & \\
$\eta (b_1^{\text{top}})$ & $\eta$ of the hardest \bjet assigned to top quark & & $\checkmark$ & & $\checkmark$ & \\
$\eta (b_2^{\text{top}})$ & $\eta$ of the second-hardest \bjet assigned to top quark & & $\checkmark$ & & $\checkmark$ & \\
$\eta (b_1^{\text{add}})$ & $\eta$ of the hardest additional \bjet & & $\checkmark$ & & $\checkmark$ & \\
$\eta (b_2^{\text{add}})$ & $\eta$ of the second-hardest additional \bjet & &  & & $\checkmark$ & \\
$m (bb^{\text{top}})$ & Invariant mass of a pair of \bjets assigned to top quarks & & $\checkmark$ & & $\checkmark$ & \\
$\pt(bb^{\text{top}})$ & \pt of a pair of \bjets assigned to top quarks & & $\checkmark$ & & $\checkmark$ & \\
$m (bb^{\text{add}})$ & Invariant mass of a pair of additional \bjets & & & & $\checkmark$ & \\
$\pt (bb^{\text{add}})$ & \pt of a pair of additional \bjets & & & & $\checkmark$ & \\
$m (e\mu bb^{\text{top}})$ &  Invariant mass of $e\mu$ and the \bjets pair assigned to top quarks & & $\checkmark$ & & $\checkmark$ & \\
$\Delta R(e\mu bb^{\mathrm{top}}, b_1^{\mathrm{add}})$ & $\Delta R$ between the direction of the system of $e\mu$ & & $\checkmark$ & & $\checkmark$ & \\
& and \bjet pair assigned to top and the direction of the hardest additional \bjet & & & & & \\
$\Delta R (e\mu bb^{\text{top}}, l/c\text{-jet}_1)$ &  $\Delta R$ between the direction of the system of $e\mu$ \\
& and \bjet pair assigned to top and the direction of the hardest light- or $c$-jet & & & $\checkmark$ & & $\checkmark$ \\
$\pt (l/c\text{-jet}_1) - \pt(b_1^{\text{add}})$  & Difference in \pt between the hardest $l/c$-jet and the additional \bjet & & & $\checkmark$ & & $\checkmark$ \\
\bottomrule
\end{tabular}
}
\end{center}
\label{tab:obs_list}
\end{table}


\subsection{Classification of \bjets}
\label{subsec:bjetclass}
In events with three or more \bjets, ambiguity arises regarding whether these \bjets originate from top quark decays or gluon radiation.
An empirical algorithm is developed to better identify the origins of representative \bjets and compare their kinematics with particle-level predictions from various MC generators.
This approach aims to improve the understanding of \bjets sources and enhance the accuracy of kinematic measurements.

The discriminating kinematic variables are investigated in the simulated events after attaching `truth' labels to \bjets to indicate their true origin in the MC.  The particle-level \bjets originating from top quark decay are classified based on the origin of hadrons, a method used in this study as described in Ref.~\cite{HIGG-2020-23}.
This approach involves mapping the original heavy hadrons to the $b$-quarks that are closest in $\Delta R$.
Subsequent heavy hadrons from decay chains are excluded from this mapping.
Those hadrons associated with a $b$-quark originating from the top quark decay are denoted by $b$-hadrons derived from a top quark.
Correspondingly, the selected particle-level \bjet that is ghost-associated with such $b$-hadrons is designated as a \bjet from top, while the remaining \bjets are labelled as additional \bjets.

\begin{figure}[htbp]
\centering
\subfloat[]{\includegraphics[width=0.33\textwidth]{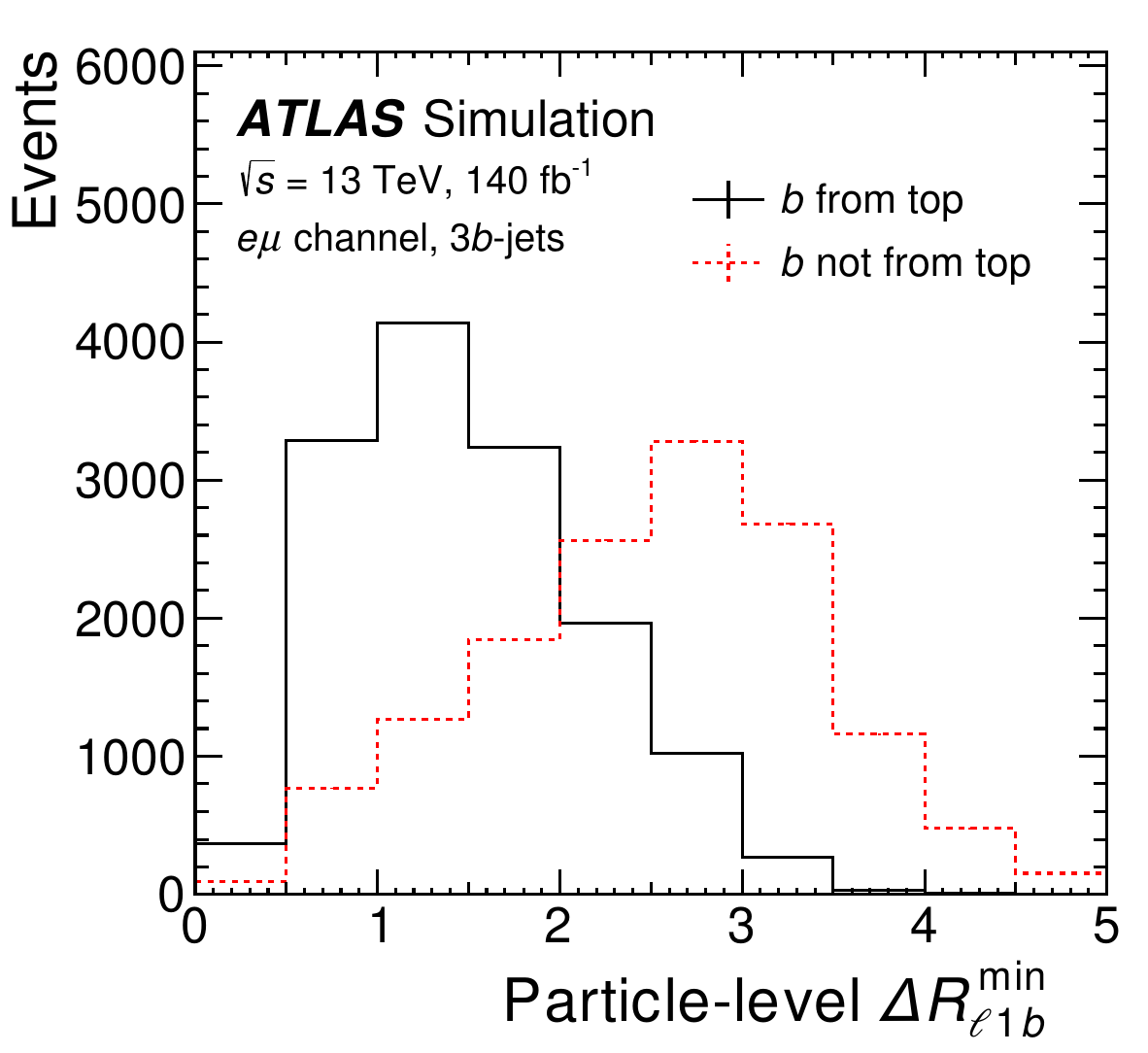}\label{fig:discrim3b:dRl1b}}
\subfloat[]{\includegraphics[width=0.33\textwidth]{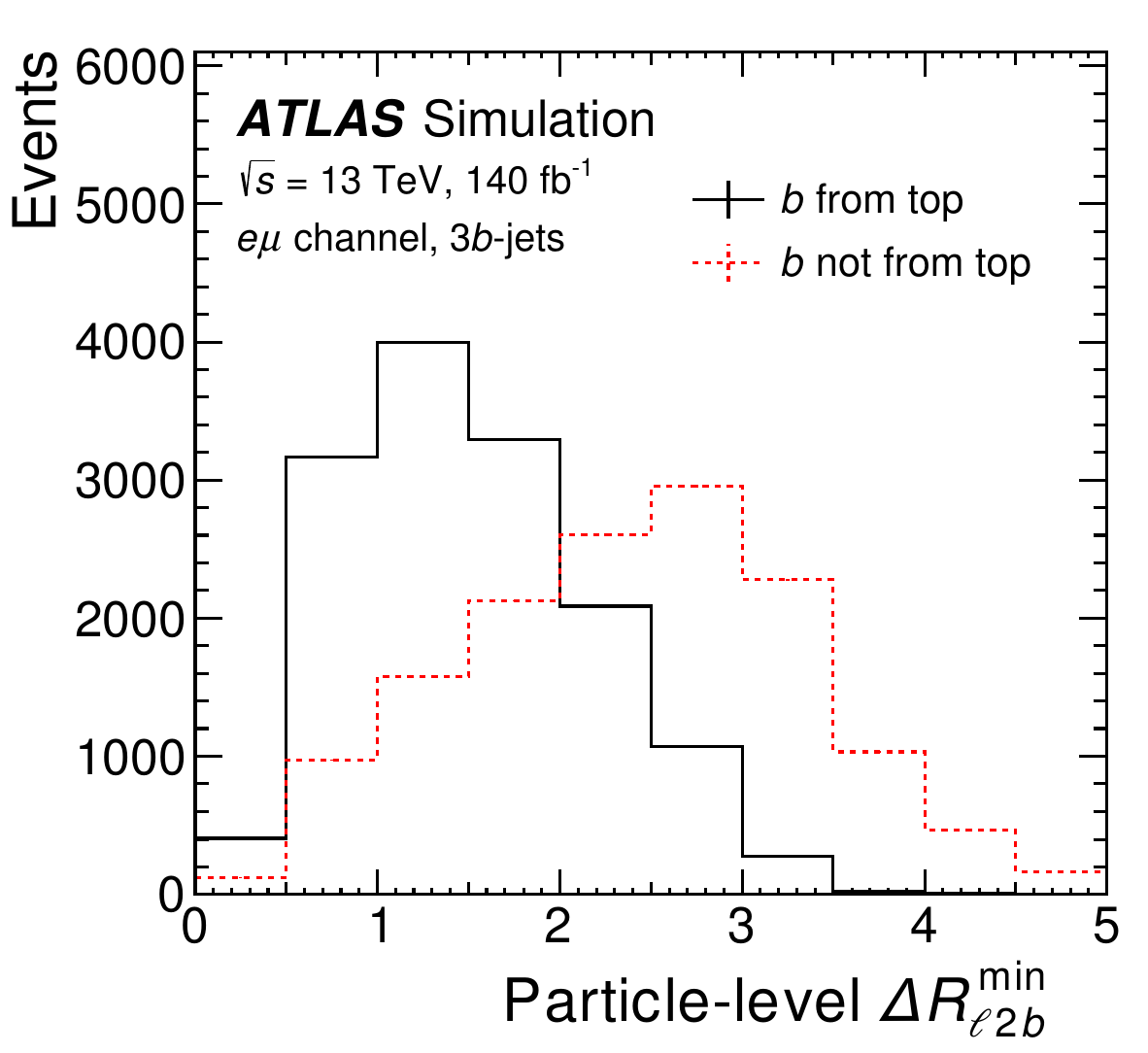}\label{fig:discrim3b:dRl2b}} \\
\subfloat[]{\includegraphics[width=0.345\textwidth]{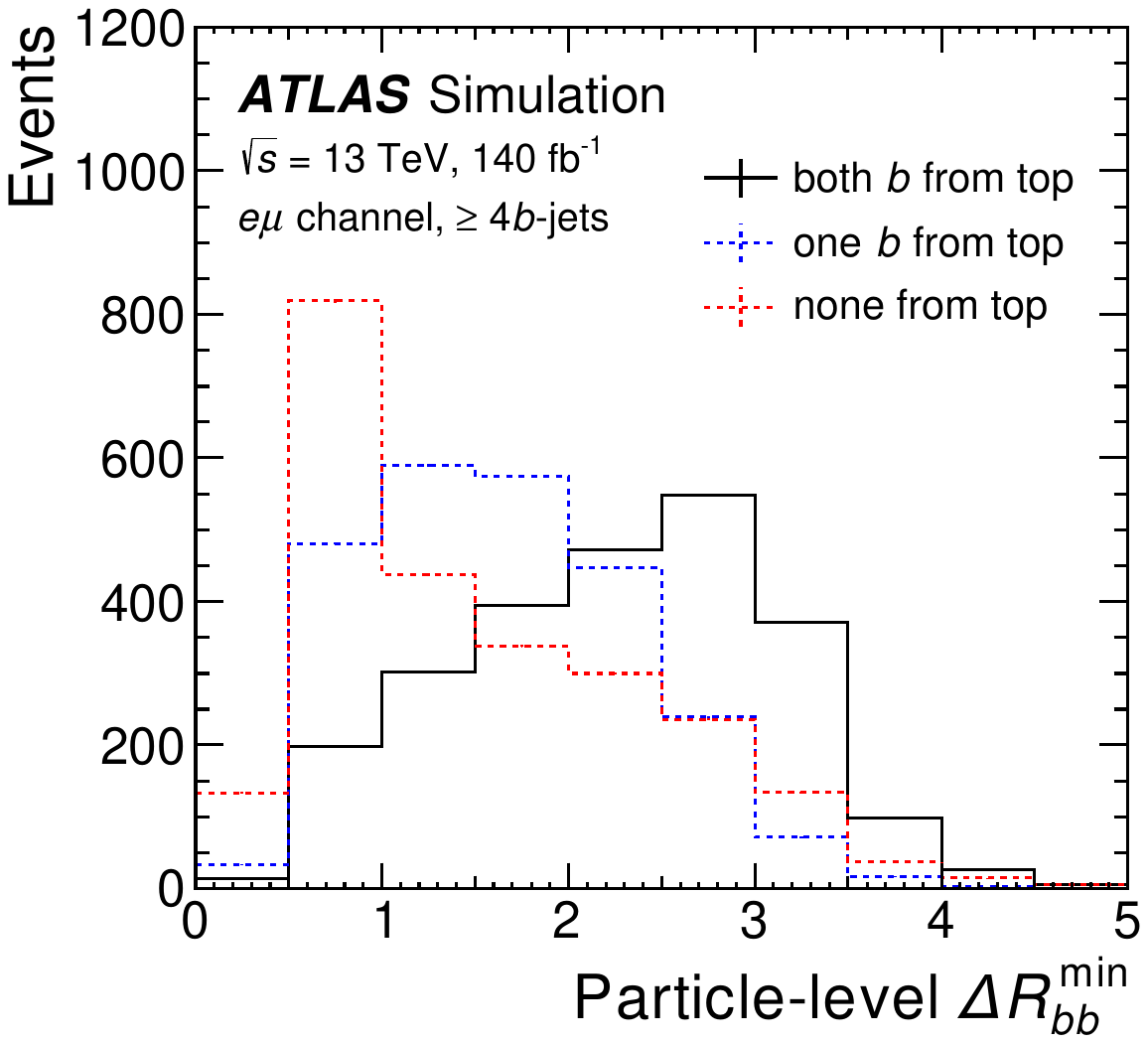}\label{fig:discrim4b:dRlbbmin}}
\subfloat[]{\includegraphics[width=0.33\textwidth]{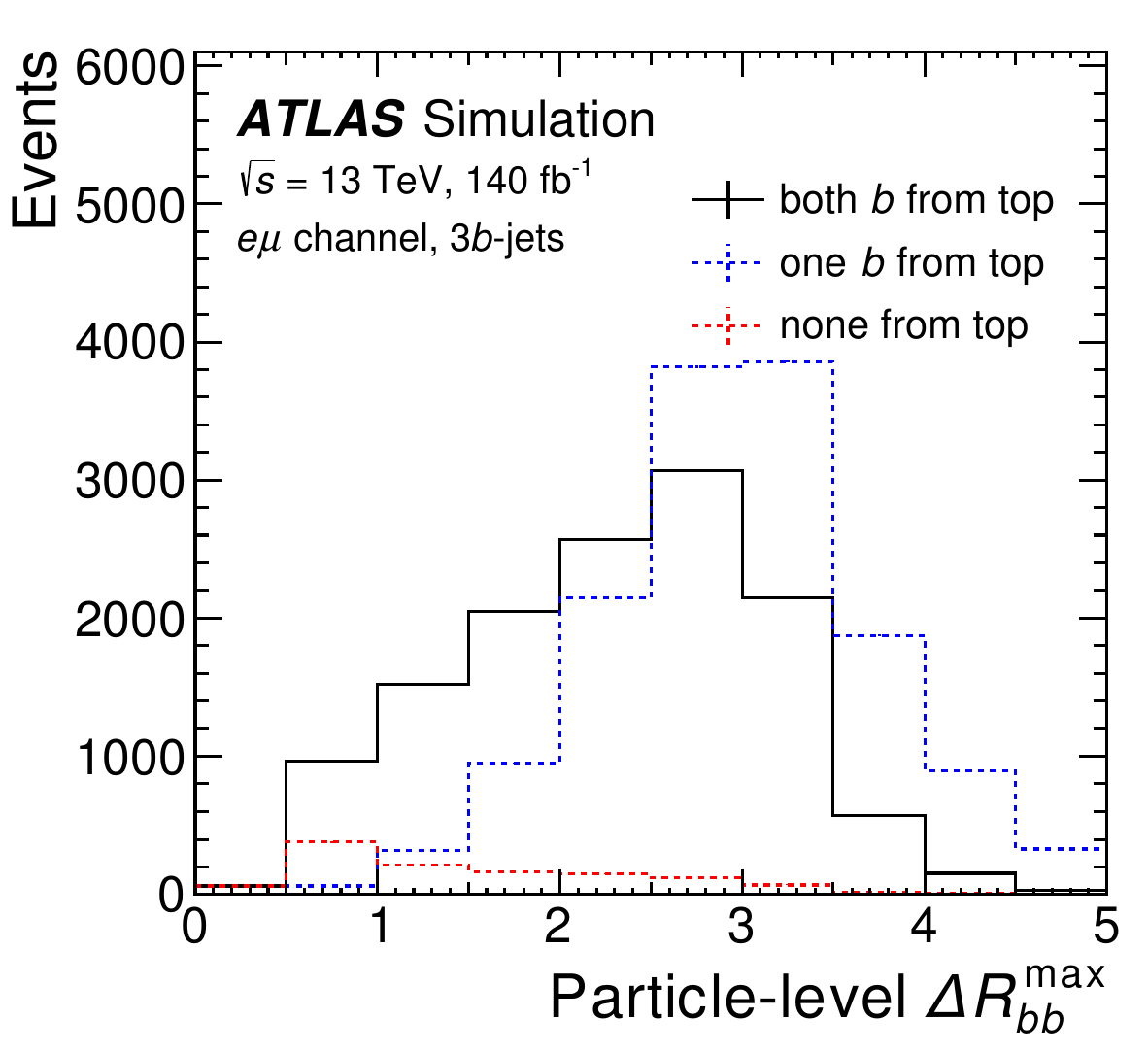}\label{fig:discrim3b:dRbbmax}}

\caption{Distributions of discriminating variables at the particle-level used in the reconstruction algorithm are compared for the \bjets originating from top quark ($b$ from top) and those originating from the gluon system ($b$ not from top):
(a) $\Delta R^{\mathrm{min}}_{\ell 1 b}$,
(b) $\Delta R^{\mathrm{min}}_{\ell 2 b}$ in the fiducial phase space with one electron, one muon, and exactly three \bjets;
(c) $\Delta R^{\mathrm{min}}_{bb}$ in the fiducial phase space with one electron, one muon, and at least four \bjets; and
(d) $\Delta R^{\mathrm{max}}_{bb}$ in the fiducial phase space with one electron, one muon, and exactly three \bjets.
}
\label{fig:discrim:dRs}
\end{figure}

The algorithm uses angular separations, as defined in Section~\ref{sec:detector}, between two physics objects and is applied to both detector-level and particle-level events.
Key metrics such as the minimum distance $\Delta R^{\mathrm{min}}_{\ell 1 b}$ between the leading lepton and any \bjet, the minimum distance $\Delta R^{\mathrm{min}}_{\ell 2 b}$ between the sub-leading lepton and any \bjet, the maximum distance $\Delta R^{\mathrm{max}}_{bb}$ between two \bjets, and the minimum distance $\Delta R^{\mathrm{min}}_{bb}$ between two \bjets, are essential for discriminating between \bjets from top quark decays and additional \bjets, as illustrated in Figure~\ref{fig:discrim:dRs}.  In most of the \ttbj MC events, leptons are closer to the \bjet originating from top quark decays, as shown in Figures~\ref{fig:discrim3b:dRl1b} and~\ref{fig:discrim3b:dRl2b}. Additionally, two \bjets not originating from top quark decays often show the smallest angular distance among any \bjet pairs as shown in Figure~\ref{fig:discrim4b:dRlbbmin}.
In instances where one of the \bjets fails the fiducial requirement, the angular separation between a \bjet from a top quark and an additional \bjet may be larger than that of any other \bjet pair combination, as shown in Figure~\ref{fig:discrim3b:dRbbmax}.

At the particle level, the algorithm is restricted to four input \bjets ordered in \pt. If four \bjets are not available, it uses only three \bjets per event.
Events featuring fewer than three \bjets are not classified relative to the \bjets.
The \bjets are arranged into a number of permuted sets depending on the \bjet multiplicity, with the first two \bjet positions in the set representing the \bjets from top quarks, and the remaining \bjets representing the additional \bjets. The chosen permutation among these sets is the one that minimises $-\ln w$, defined in terms of the sum of squares of the differences between the $\Delta R$ values of the lepton-\bjet pairs relative to the global quantities $\Delta R^{\mathrm{min}}_{\ell 1 b}$ or $\Delta R^{\mathrm{min}}_{\ell 2 b}$, and of the differences between the $\Delta R$ values of \bjet pairs in a given permutation relative to $\Delta R^{\mathrm{max}}_{bb}$ or $\Delta R^{\mathrm{min}}_{bb}$. The full expression is:
{\small{
\begin{equation*}
-\ln w = \left\{ \begin{array}{ll} \left(\Delta R_{\ell 1 b1}  - \Delta R^{\mathrm{min}}_{\ell 1 b}\right)^2 + \left(\Delta R_{\ell 2 b2}  - \Delta R^{\mathrm{min}}_{\ell 2 b}\right)^2 + \left(\mathrm{max}(\Delta R_{b1 b3}, \Delta R_{b2 b3})  - \Delta R^{\mathrm{max}}_{bb}\right)^2 &  \mathrm{if}\  N_{b\text{-jets}} = 3, \\
\left(\Delta R_{\ell 1 b1}  - \Delta R^{\mathrm{min}}_{\ell 1 b}\right)^2 + \left(\Delta R_{\ell 2 b2}  - \Delta R^{\mathrm{min}}_{\ell 2 b}\right)^2 + \left(\Delta R_{b3 b4} - \Delta R^{\mathrm{min}}_{bb}\right)^2  &  \mathrm{if}\  N_{b\text{-jets}} \geq 4, \end{array} \right.
\end{equation*}
}}

where $\Delta R_{\ell 1 b1}$ ($\Delta R_{\ell 2 b2}$) represents the angular distance between the leading (sub-leading) lepton in the event and the first (second) \bjet of a given set, $\Delta R_{b1 b3}$ ( $\Delta R_{b2 b3}$ ) refers to the angular distance between first (second) and third \bjet of the set, and $\Delta R_{b3 b4}$ represents the angular distance between the third and fourth \bjet in a given permutation. Each variable entering in the squared sum in the expression is assigned the same weight for simplicity.  The $N_{b\text{-jets}}$ is the number of \bjets in the event.  The set that gives the smallest value of $-\ln w$ is finally chosen to classify its first two \bjets as belonging to the top quark, and the remaining \bjets are assigned as additional \bjets.
At the detector level, the same algorithm is applied to reconstructed leptons and $b$-tagged jets, after taking only up to four $b$-tagged jets into account. If there are more than four $b$-tagged jets present, then the first four ordered in \pt are taken.

The purities of \bjet classifications according to this algorithm are evaluated using `truth' labels to \bjets in simulated events, as discussed above.  This is done solely to assess the performance of the algorithm, it is not integral to the definitions of the measured observables.
According to this definition of `truth' \bjet origin, the typical fraction of QCD \ttbj simulated events wherein the two highest-\pt particle-level \bjets originate from top quarks amounts to $42\%$ ($27\%$) in events containing at least three (four) \bjets.
The fractions of particle-level events where two \bjets are correctly assigned to top quarks according to this technique are found to be $53\%$ ($56\%$) in the selected sample of simulated \ttbj events containing three (four) \bjets.  Slightly lower (by 1\%--2\%) purities are obtained with the corresponding reconstructed events where the $b$-tagged jets are geometrically matched to the particle-level \bjets.   The probabilities of correct assignment of \bjet (two or more \bjets in case of kinematic observable defined with multiple \bjets) in a given bin of the measured observable ranges from $50\%$ to $85\%$ ($40\%$ to $75\%$).  The reconstruction algorithm is not optimal for signal events that have at least one of the \bjets from \ttbar decay outside the fiducial volume ($<10\%$ fraction of events with three or more \bjets), and hence the correct assignments of \bjets cannot be made to both top quarks.

%
%

%


%

%
\section{Background estimation}
\label{sec:bkgd}

%
%
The \ttj candidate events are pre-selected with a prompt electron and a prompt muon of opposite charge and two or more $b$-tagged jets ($\geq 2b@77\%$), as detailed in Section~\ref{sec:recosel}.  The \ttbj signal events consist of a prompt electron and a prompt muon from $W$ decays, have three or more $b$-tagged jets ($\geq 3b@77\%$), and at least three or more particle-level \bjets. The preselection of events with two or more $b$-tagged jets yields a sample primarily dominated by \ttbar production with minimal contributions from other processes.
The background events in this sample can be categorised into two main types.
Firstly, there are events with genuine prompt leptons originating from top, $W$, and $Z$ decays, including those arising from leptonic $\tau$-lepton decays, which are elaborated upon in Section~\ref{subsec:promptbkg}.
Secondly, there are events where at least one reconstructed lepton candidate is non-prompt or misidentified as a different object. These can be a non-prompt lepton from the decay of a $b$- or $c$-hadron, an electron from a photon conversion, hadronic jet activity misidentified as an electron, or a muon from an in-flight decay of a pion or kaon.
The estimation of this second type of background involves a combination of data-driven and simulation-based approach, detailed in Section~\ref{subsec:fakelepbkg}.

The signal event category \ttbj (also denoted as \ttb) refer to events in the \ttbar\ MC sample consisting of three or more particle-level \bjets.  Events that contain less than three \bjets but have at least one \cjet are categorised as \ttcj (\ttc), and all remaining events are defined as \ttlj (\ttl).  The events with three or more $b$-tagged jets contain substantial background from \ttl or \ttc where \cjets or light jets are mis-identified as \bjets by the $b$-tagging algorithm.  The MC yields of these background processes before any corrections are given in Table~\ref{tab:yields}.  These backgrounds are estimated simultaneously with the extraction of fiducial cross-sections, as discussed in Section~\ref{sec:fidxsec}.

\subsection{Backgrounds with prompt leptons}
\label{subsec:promptbkg}
The estimation of various background processes with the presence of oppositely charged prompt electrons and muons in the final state is primarily performed using the MC simulation.  The \ttH process constitutes the primary irreducible background, contributing approximately $1\%$ to $6\%$ of the total expected event yields in the $3b@77\%$ and $\geq 4b@77\%$ regions, with its contribution increasing as the number of required $b$-tagged jets increases. The \ttV production yields a similar contribution as \ttH production in the $3b@77\%$ region, and nearly half the size of \ttH in the $\geq 4b@77\%$ region.  The \ttZ production dominates the \ttV process; the \ttW background is about $30\%$ ($1\%$) of the total \ttV yield in $3b@77\%$ ($\geq 4b@77\%$) regions. %

The single-top-quark production contributes approximately $3\%$ to the expected event yields, decreasing with the number of selected $b$-tagged jets.
Contributions from $Z/\gamma^{*}$+jets and diboson processes are found to be minimal.
Other rare SM processes such as $t\bar{t}t\bar{t}$, $tZ$, $tWZ$, $tWH$, and $tHbj$ are also included in the total background estimate, all of which are estimated from MC simulations.

\subsection{Background with non-prompt or misidentified leptons}
\label{subsec:fakelepbkg}
The contribution of non-prompt or misidentified leptons (referred to as `non-prompt lepton background') is estimated using a data-driven approach, relying on the data events where the electron and muon have the same electric charges.  This method, detailed in Ref.~\cite{TOPQ-2015-09}, is based on the assumption that jets misidentified as electrons or muons yield a comparable number of events with opposite-charge (opposite-sign, OS) and same-charge (same-sign, SS) $e\mu$ pairs in the \ttbar sample, particularly in instances where at least one $W$ boson decays hadronically.  This estimate includes the non-prompt and misidentified leptons produced not only in the \ttbar events, but also in other processes such as \Wjets.  Events are pre-selected with the same baseline criteria for the electron and muon as for the signal except for the requirements on their charges, and that they contain at least two or more $b$-tagged jets. Known sources of same-sign prompt leptons contribution estimated from the signal and background MC simulation are subtracted from the data distribution of the same-sign events, and the non-prompt lepton background in each bin is extracted by scaling the remaining data events by the lepton-\pt dependent and $b$-tagged jets multiplicity dependent asymmetry factor $R$.

\begin{equation*}
N_{i,\mathrm{non-prompt}} = R \cdot (N_{i,\mathrm{SS}}^{\mathrm{Data}} - N_{i,\mathrm{SS-prompt}}^{\mathrm{MC}}),
\end{equation*}
where $N_{i,\mathrm{SS}}^{\mathrm{Data}}$ is the number of observed data events with same-sign $e\mu$ pair in the $i$th bin of the observable, and $N_{i,\mathrm{SS-prompt}}^{\mathrm{MC}}$ is the MC simulated yield of the prompt same-sign events in that bin. The factor $R$, which scales the non-prompt lepton background from the same-sign region to the opposite-sign signal region, is determined using \powpy\ \ttbar events where at least one of the leptons originates from the hadronically decaying $W$ boson and is hence misidentified. It is defined as:
\begin{equation*}
R = \frac{N_{\mathrm{OS-non-prompt}}^{\mathrm{MC}}}{N_{\mathrm{SS-non-prompt}}^{\mathrm{MC}}},
\end{equation*}
where $N_{\mathrm{OS-non-prompt}}^{\mathrm{MC}}$ and $N_{\mathrm{SS-non-prompt}}^{\mathrm{MC}}$ are the non-prompt lepton background event yields estimated from the MC simulation in the opposite-sign and same-sign events, respectively.  It is determined as a function of lepton \pt in the $2b@77\%$ events, while it is evaluated inclusively in lepton \pt in the $\geq 3b@77\%$ events due to the very low background and limited MC statistics. The values of $R$ range from $1.98\pm0.52$ to $2.38\pm0.74$ across the lepton-\pt bins in $2b@77\%$ events, while it is $1.65\pm0.45$ for inclusive $\geq 3b@77\%$ events, where the quoted error represents the uncertainty due to limited MC statistics and the MC modelling uncertainty, with the latter being the dominant component.  The MC modelling uncertainty includes the uncertainty in the matrix element and the uncertainty in the modelling of the parton shower, hadronisation and underlying event.  Table~\ref{tab:yields} presents the total estimated background yields in various bins of $b$-tagged jet multiplicity.  %

%

%
%
%
%
%
%

%
%
%
%
%
%
%
%
%
%
%
%

%
%
\begin{table}
\centering
\caption{Predicted and observed event yields for the events pre-selected with two or more reconstructed jets ($\geq 2j$) and further categorised in exactly $2b@77\%$,
exactly $3b@77\%$ and $\geq 4b@77\%$ selections before any flavour composition scale factors are applied to \ttbar events.  Predicted events in the \ttbar sample are split into three categories, with \ttbar+\bjets referring to the signal events consisting of at least three particle-level \bjets, and \ttc (\ttl) referring to background events consisting of less than three particle-level \bjets but more than one (no) \cjets. The background with non-prompt or misidentified leptons is denoted by non-prompt lepton. The quoted errors are symmetrised and indicate total statistical and systematic uncertainties in the predictions due to experimental and theoretical sources. %
}
\label{tab:yields}
\begin{tabular}{l
r@{$\,\pm\,$}l
r@{$\,\pm\,$}l
r@{$\,\pm\,$}l
}
\toprule
Process     & \multicolumn{2}{c}{$\geq 2j$, $2b@77\%$}  & \multicolumn{2}{c}{$\geq 3j$, $3b@77\%$}  & \multicolumn{2}{c}{$\geq 4j$, $\geq 4b@77\%$}   \\
\midrule

\ttbar+\bjets  &  4100 & 790 & ~~~3550 &  650 & ~~~~~~474 & 99 \\
\midrule
\ttc  & 11600  & 2200 & ~~~2190 & 430 & ~~~~~~57 & 15  \\
\ttl  & 263000 & 33000 & ~~~2080 & 440 & ~~~~~~25 & 15 \\
\midrule
\hspace{0mm} $Wt$                      & 9100 & 1800 & ~~~227 & 94 &  ~~~~~~14 &  11  \\
\hspace{0mm} $t\bar{t}V$            & 740 & 230  &  ~~~94 & 30 &  ~~~~~~16.3  &  5.1   \\
\hspace{0mm} $t\bar{t}H$            & 180 & 22 & ~~~108 & 13 &  ~~~~~~37.2  &  5.3   \\
\hspace{0mm} Non-prompt lepton            & 340 & 210 &  ~~~37 &  20 &  ~~~~~~10.9 &  6.1  \\
\hspace{0mm} $Z/\gamma^*$+jets       &  96 & 38 &   ~~~3.4 &  1.4 &   ~~~~~~0.15 &  0.09   \\
\hspace{0mm} Diboson                &  85 &  43 &   ~~~3.0 &  1.5 &   ~~~~~~0.11 &  0.07  \\
\hspace{0mm} Others                  &  41 &  20 &  ~~~16.4 &  8.2 &   ~~~~~~6.4 &  2.9  \\
\midrule
Total predicted   & 290000 & 35000 & ~~~8300 & 1300 & ~~~~~~640 & 120  \\
Observed     & \multicolumn{2}{c}{281213}  & \multicolumn{2}{c}{10235} & \multicolumn{2}{c}{798}   \\
\bottomrule
\end{tabular}
\end{table}

%
%
%
%
%
%
%
%
%
%
%
%
%
%
%
%
%
%
%
%
%
%
%
%
%
%
%
%
%
%
%
%
%
%
%
%
%
%
%
%
%
%
%
%
%
%
%
%
%
%
%
%

%
%
%
%
%
%
%
%
%
%
%
%
%
%
%
%
%
%
%
%
%
%
%
%
%
%
%
%
%
%
%
%
%
%
%
%
%
%
%
%
%
%
%
%
%
%
%
%
%
%



%

%
\section{Extraction of fiducial cross-sections}
\label{sec:fidxsec}

\begin{figure}[htbp]
\centering
\subfloat[]{\includegraphics[width=0.44\textwidth]{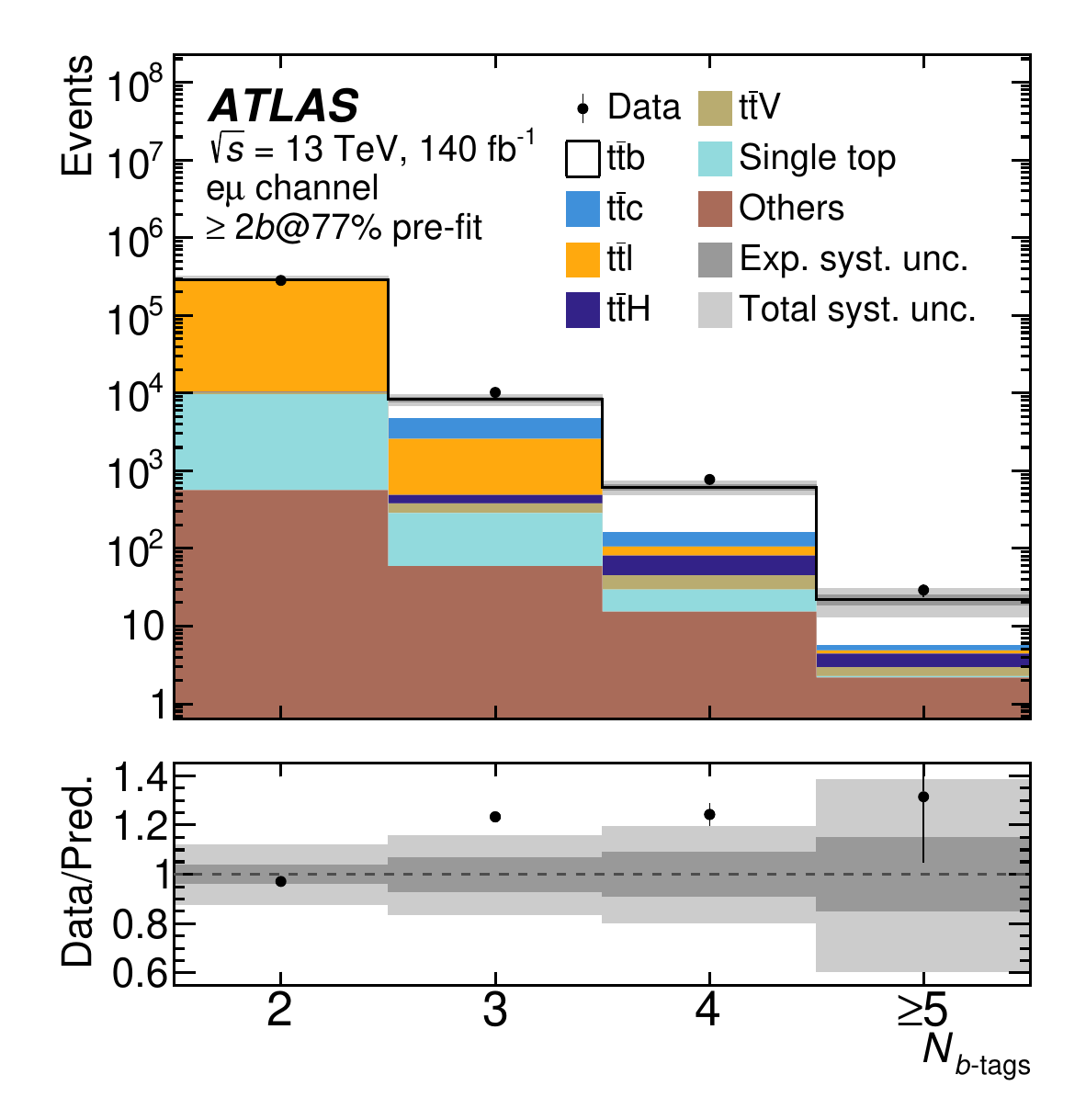}\label{fig:reco-dataMC-plots_nbjets_uncorr}}
\subfloat[]{\includegraphics[width=0.44\textwidth]{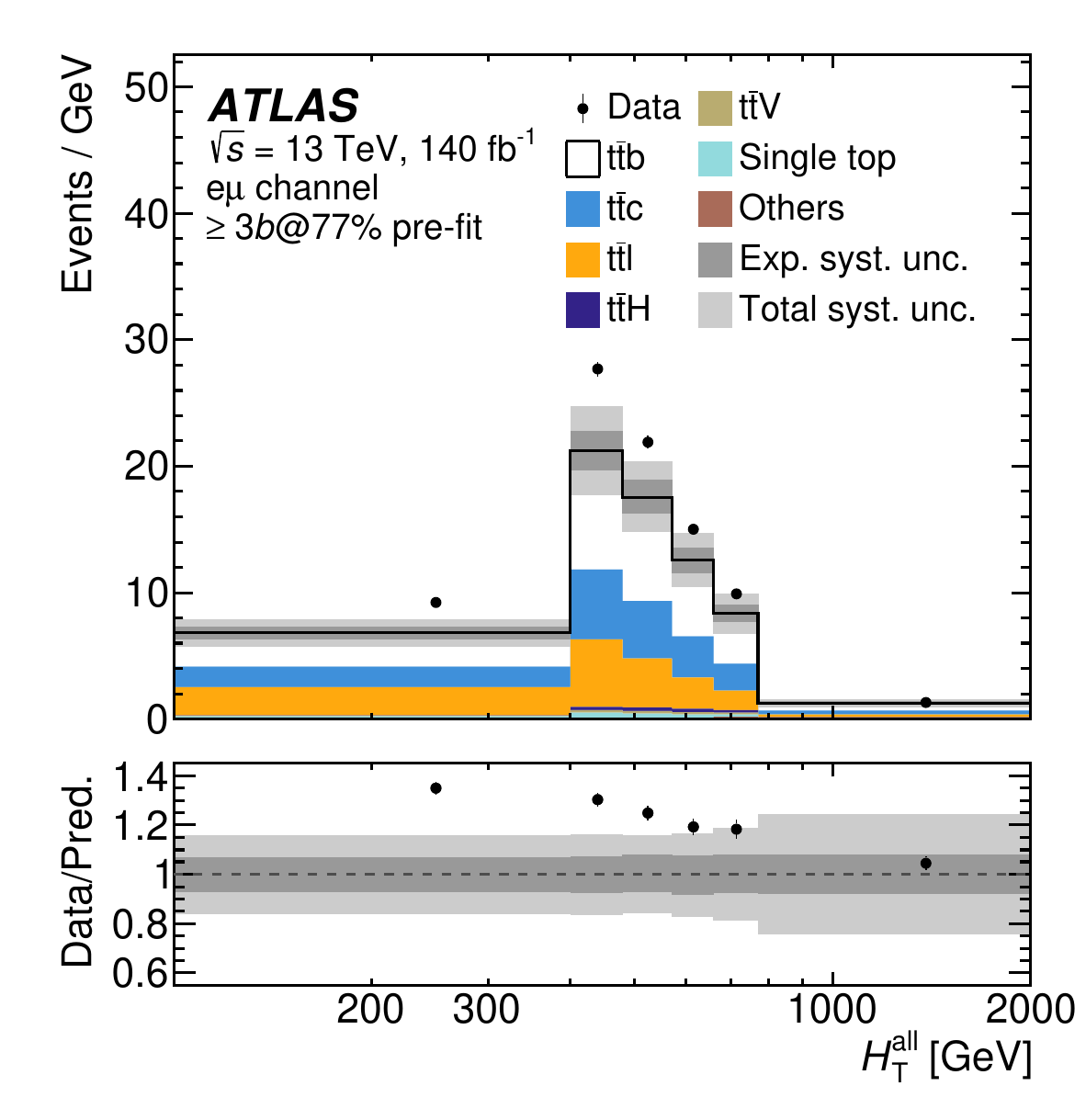}\label{fig:reco-dataMC-plots_uncorr_3j3b_HT}} \\
\subfloat[]{\includegraphics[width=0.44\textwidth]{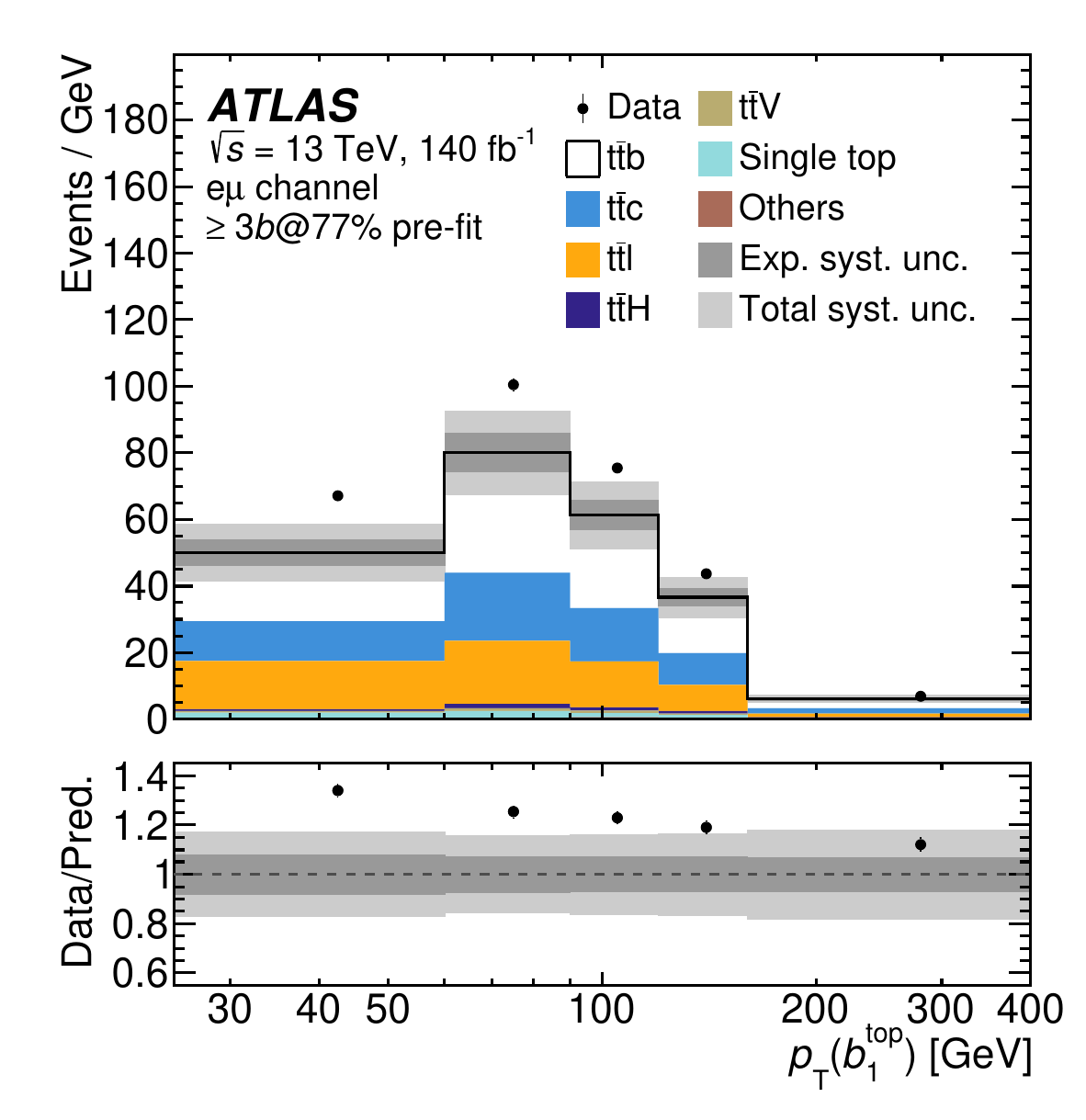}\label{fig:reco-dataMC-plots_uncorr_3j3b_TopbJetPt_1}}
\subfloat[]{\includegraphics[width=0.44\textwidth]{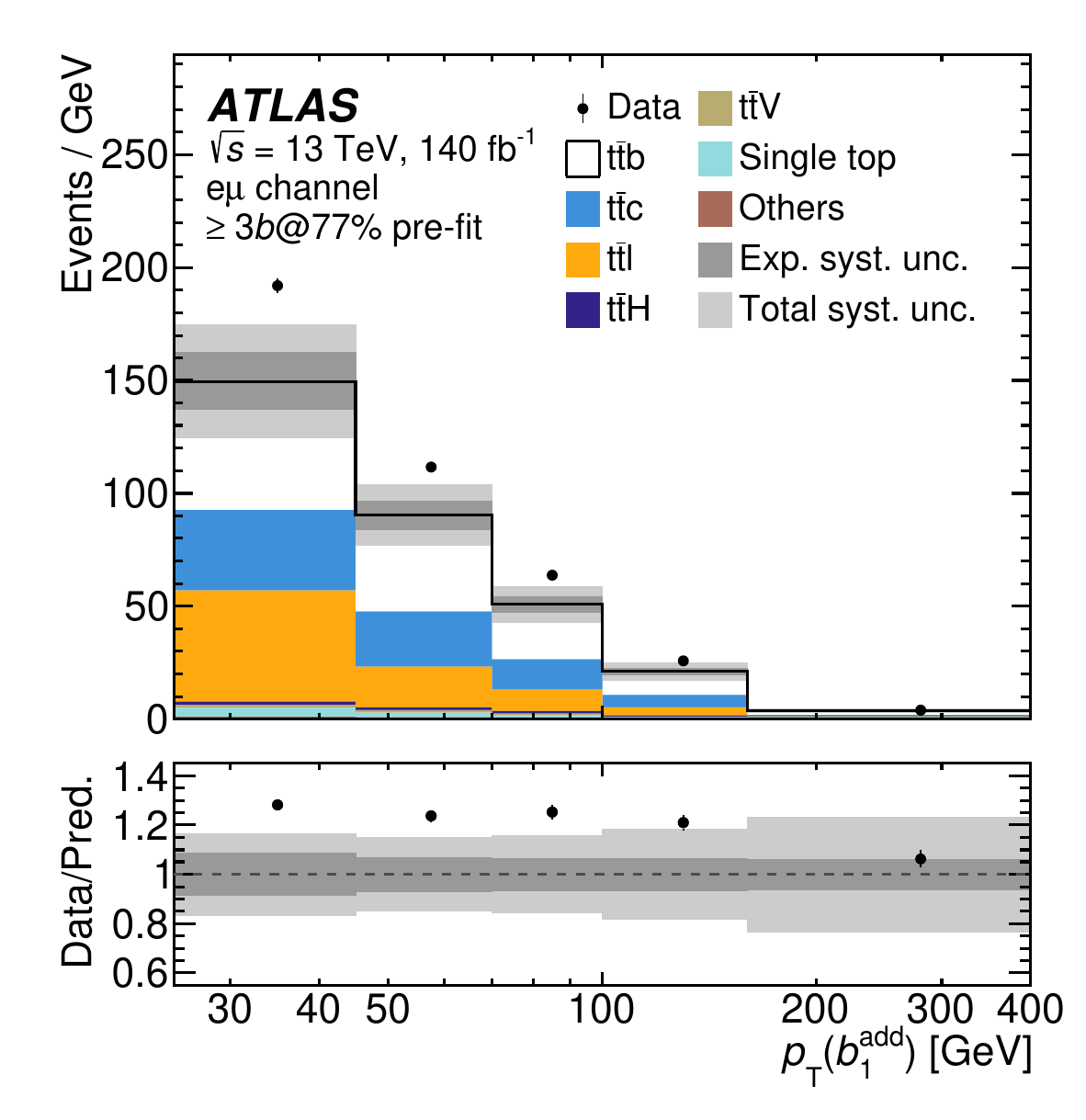}\label{fig:reco-dataMC-plots_uncorr_3j3b_AddbJetPt_1}} \\
\caption{ Comparison of data and predictions for the distribution of \protect \subref{fig:reco-dataMC-plots_nbjets_uncorr} the number of $b$-tagged jets for events with at least two $b$-tagged jets ($N_{b\mathrm{-tags}}$), \protect \subref{fig:reco-dataMC-plots_uncorr_3j3b_HT} $H_{\mathrm{T}}^{\mathrm{all}}$, \protect \subref{fig:reco-dataMC-plots_uncorr_3j3b_TopbJetPt_1} \pt$(b_1^{\text{top}})$, and \protect \subref{fig:reco-dataMC-plots_uncorr_3j3b_AddbJetPt_1} \pt$(b_1^{\text{add}})$ in reconstructed events with three or more $b$-tagged jets before the global scale factors are applied for different event categories as discussed in the text (see Section~\ref{subsec:flavCompCorrections}). The entries in each bin are divided by the bin width in \protect \subref{fig:reco-dataMC-plots_uncorr_3j3b_HT}, \protect \subref{fig:reco-dataMC-plots_uncorr_3j3b_TopbJetPt_1} and \protect \subref{fig:reco-dataMC-plots_uncorr_3j3b_AddbJetPt_1}. The lower panels show the ratios of the data to the predictions. The inner band includes the uncertainties due to limited MC statistics and due to various detector effects such as the jet energy scale and resolution, and the $b$-tagging, while the outer band also includes the theoretical uncertainties in the signal and background modelling. The last bin includes the overflow.}
\label{fig:reco-dataMC-plots_prefit}
\end{figure}

%
The particle-level distributions for different observables in the fiducial phase space defined in Section~\ref{subsec:fiddef} are extracted from the data distributions obtained after the event selections described in Section~\ref{subsec:evtsel}. The number of events meeting the baseline selection criteria with two or more $b$-tagged jets, primarily comprising \ttbar events, agrees well with the predictions, as detailed in Table~\ref{tab:yields}. However, there is about a $20\%$ underestimate of the number of events featuring three or more $b$-tagged jets.
This discrepancy, while noticeable in Figure~\ref{fig:reco-dataMC-plots_prefit}, is not significant given the large systematics uncertainties in the $b$-tagged jet multiplicity distribution in Figure~\ref{fig:reco-dataMC-plots_nbjets_uncorr}.
The data distributions of the scalar sum of the \pt of charged leptons and jets, the \pt of the $b$-tagged jets assigned to the top quark, and the \pt of the additional $b$-tagged jets show deviations from the predictions.  The measurement of \ttbj production is dependent on the estimate of background originating from other \ttbar processes, such as the mis-tagged jets in \ttc and \ttl events that give substantial background contributions.  To address this discrepancy, data-driven scale factors are derived to simultaneously adjust the predictions of additional \cjet or light jet backgrounds alongside the additional \bjets in the \ttbar MC simulation, as outlined in Section~\ref{subsec:flavCompCorrections}.
The number of background events produced by \ttH, \ttV, and non-\ttbar processes and the non-prompt lepton background described in Section~\ref{sec:bkgd} are subtracted from the data.
The data are then unfolded using the corrected MC simulation as described in Section~\ref{subsec:unfolding}.

\subsection{Data-driven correction factors for flavour composition of additional jets in \ttbar events}
\label{subsec:flavCompCorrections}
The mis-tagging of jets in \ttc and \ttl events contributes significantly to the background in the \ttbj process. For example, only about $50\%$ of simulated events selected at detector level with at least three $b$-tagged jets contain at least three true \bjets at the particle level in the fiducial phase space. The remaining events include at least one \cjet or light-flavour jet that is misidentified as a $b$-tagged jet. Moreover, the \ttc production cross-sections have not yet been precisely measured~\cite{CMS-TOP-20-003}.
To account for these significant background contributions, a template fit to data is performed to simultaneously extract the normalisation factors for \ttbj, \ttc and \ttl particle-level event categories and subsequently correct their compositions in \ttbar simulated samples.  This is done to reduce the impact of systematic uncertainties in \ttc and \ttl background estimates.

The template fit is performed using a binned maximum-likelihood approach, with the likelihood function modeled as a Poisson distribution:
\begin{equation}
\label{eq:poisson_likelihood}
\mathcal{L}(\vec{\alpha} | x_1, \ldots, x_n) =
\prod_{k}^{n}\frac{{\mathrm{e}}^{-\nu_k(\vec{\alpha})} \nu_k(\vec{\alpha})^{x_k}}{x_k!}~,
\end{equation}
where $x_k$ represents the number of events in bin $k$ from the data template, and
$\nu_k(\vec{\alpha})$ is the expected number of events, which depends on a set of free parameters,
$\vec{\alpha}$. Only the statistical uncertainty in the data is considered in the fit. Three fit parameters, $\alpha_b^s$, $\alpha_c^s$ and $\alpha_l^s$, are used in the maximum-likelihood fit, such that the expected number of events in bin $k$ of region $s$ is
\begin{equation}
\label{eq:partial_likelihood}
\nu_k(\alpha_b^s, \alpha_c^s, \alpha_l^s) =
\alpha_b^s N^{k,s}_{\ttb} + \alpha_c^s N^{k,s}_{\ttc} +  \alpha_l^s N^{k,s}_{\ttl}
+ N^{k,s}_{\textrm{non-}\ttbar}~,
\end{equation}

where $N^{k,s}_{\ttb}$, $N^{k,s}_{\ttc}$, $N^{k,s}_{\ttl}$ and
$N^{k,s}_{\textrm{non-}\ttbar}$ are the numbers of events in bin $k$ and in the region $s$, denoted $\geq 3j \geq 2b@77\%$, of the \ttb,
\ttc, \ttl and non-\ttbar background templates, respectively.

\begin{figure}[htbp]
\centering
\includegraphics[width=0.40\textwidth]{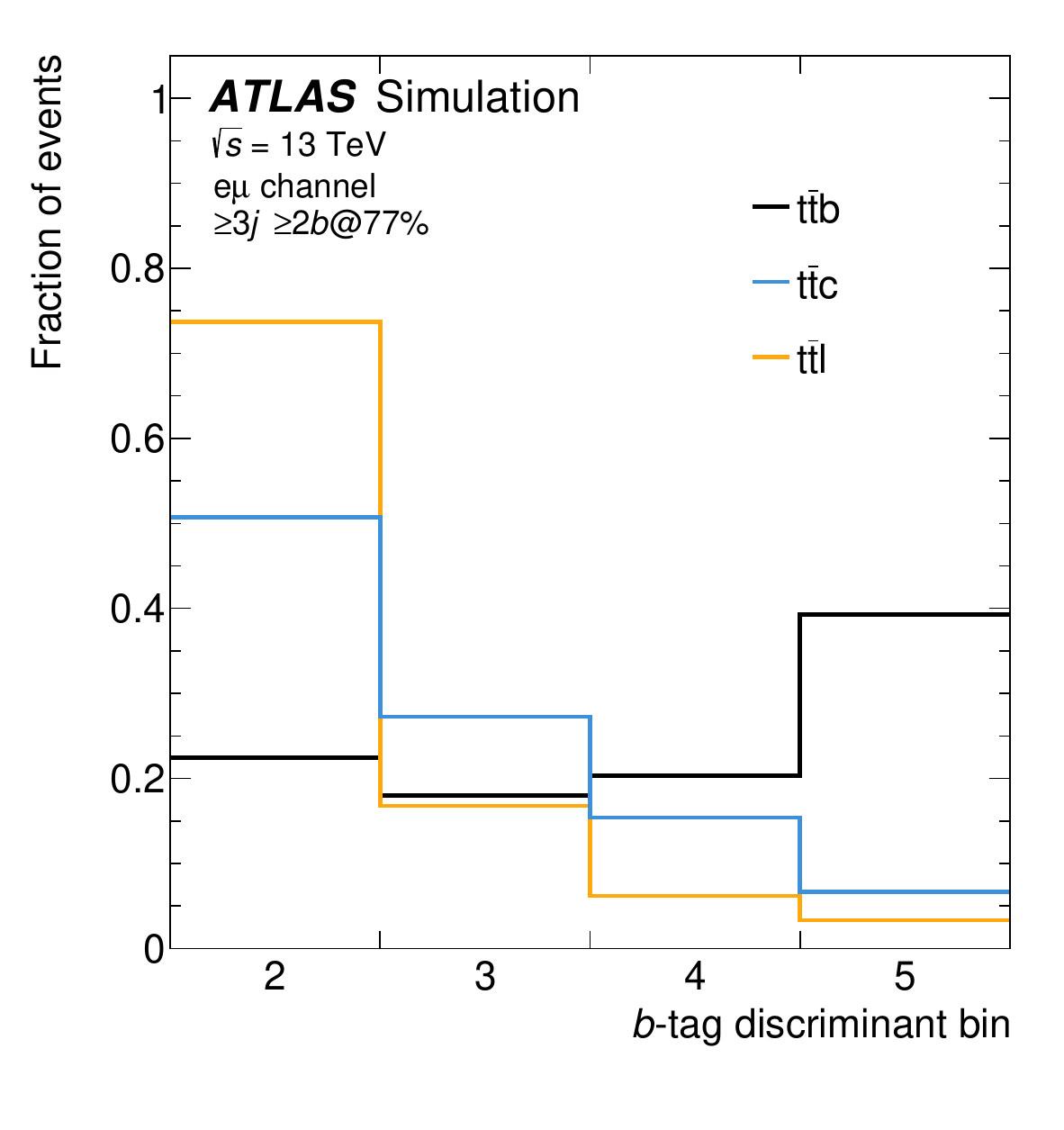}
\caption{
Templates of \ttbj, \ttc, and \ttl processes in the inclusive detector-level region with $\geq 3j \geq 2b@77\%$, as considered in the fit. For each category, they depict the distribution of the third-highest $b$-tagging discriminant score among selected jets in the event.  The score ranging from 2 to 5, correspond to a certain range of $b$-tagging efficiencies defined by the working points: 85\%--77\%, 77\%--70\%, 70\%--60\%, and $<60\%$, respectively.}
\label{fig:templates}
\end{figure}

Templates are created for each category using the third-highest $b$-tagging discriminant score in the event. This variable is a proxy for the flavour of an additional jet and helps discriminate between the \ttbj, \ttc and \ttl processes. Figure~\ref{fig:templates} shows the shape comparisons of these templates.  Two fitting approaches are considered:

\begin{enumerate}
\item \textit{Global fit}: Normalisation factors for the \ttbj, \ttc and \ttl templates are fitted in the inclusive region. This is the nominal approach and is used to correct the normalisation of these components.

\item \textit{Kinematic-dependent fit}: Normalisation factors are fitted in specific regions based on jet multiplicity and the \pt of the third-hardest jet in the reconstructed events. This approach is used only to assess the shape uncertainty of the \ttc and \ttl backgrounds.
\end{enumerate}

\begin{figure}[htbp]
\centering
\subfloat[]{\includegraphics[width=0.45\textwidth]{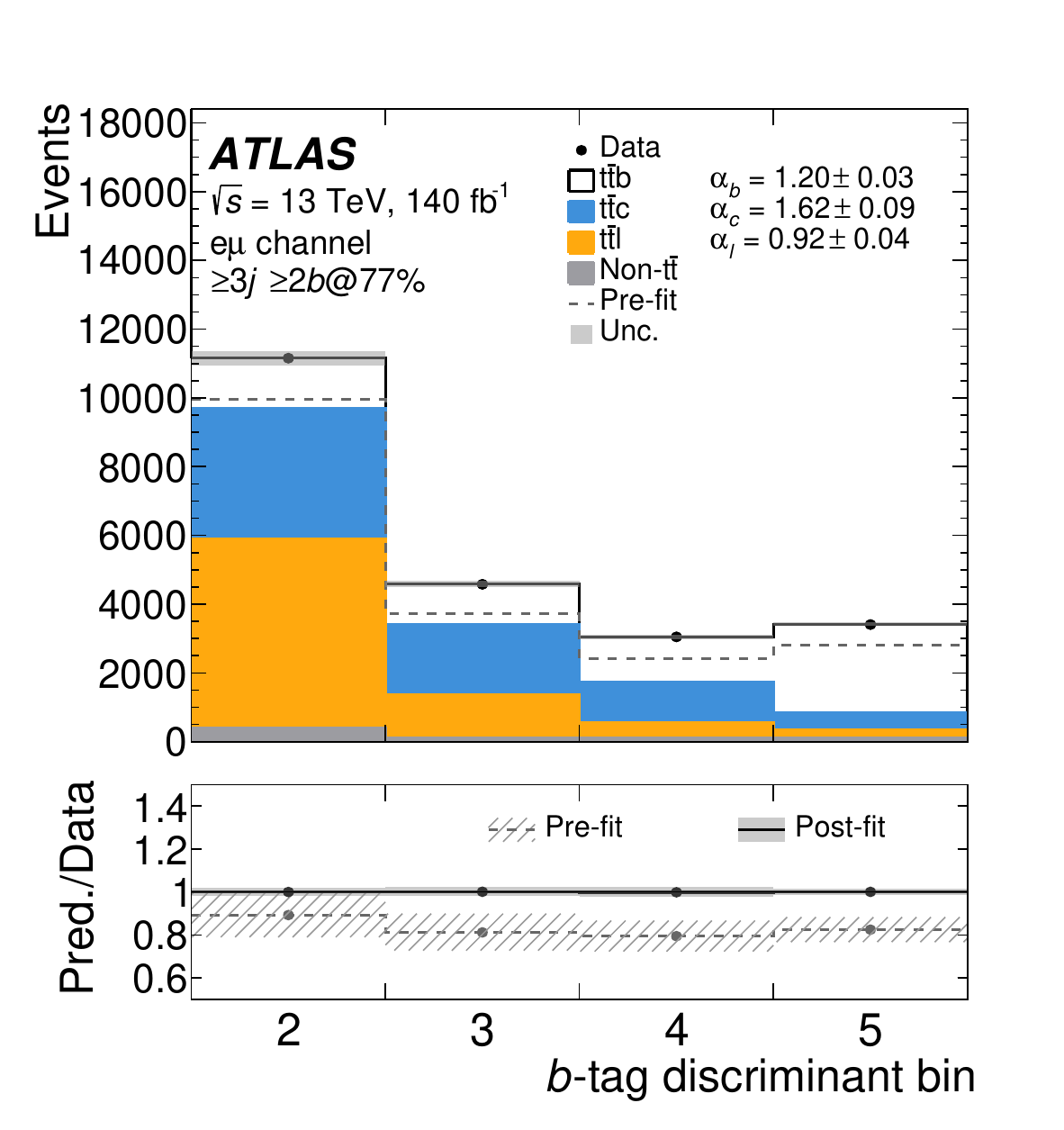}\label{fig:FCFResult_3j2b_a}}
\subfloat[]{\includegraphics[width=0.45\textwidth]{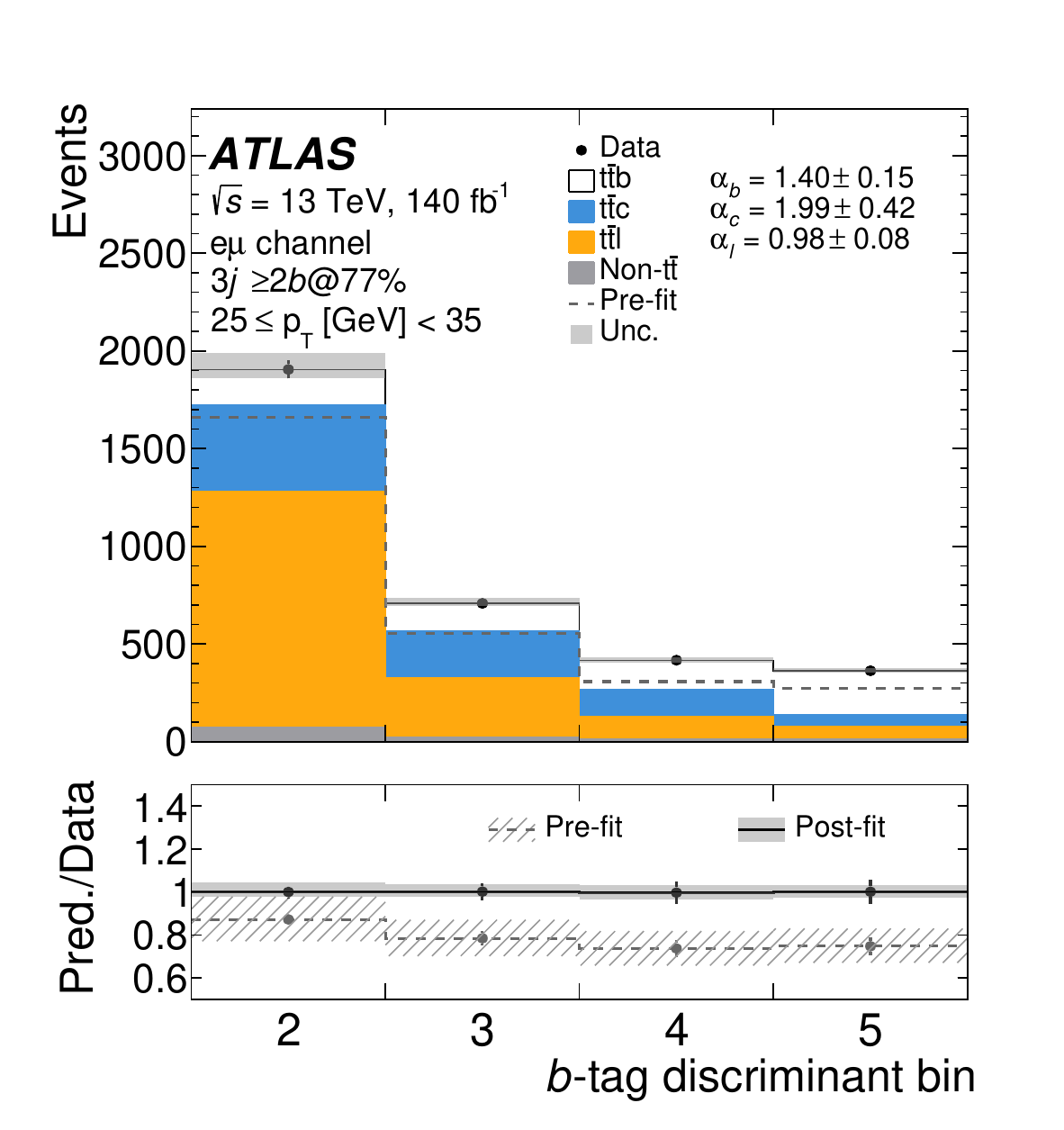}\label{fig:FCFResult_3j2b_b}} \\
\subfloat[]{\includegraphics[width=0.45\textwidth]{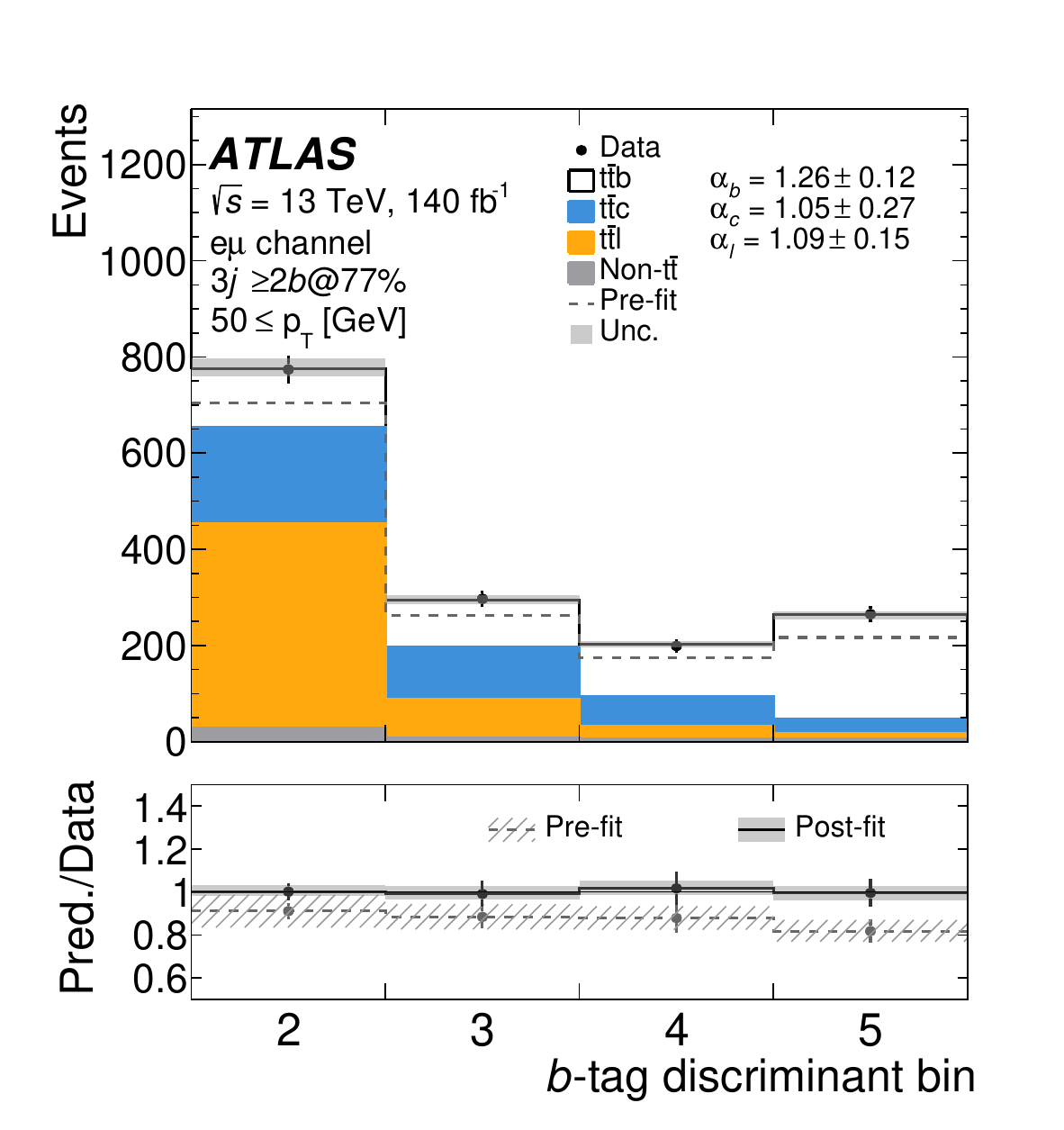}\label{fig:FCFResult_3j2b_c}}
\subfloat[]{\includegraphics[width=0.45\textwidth]{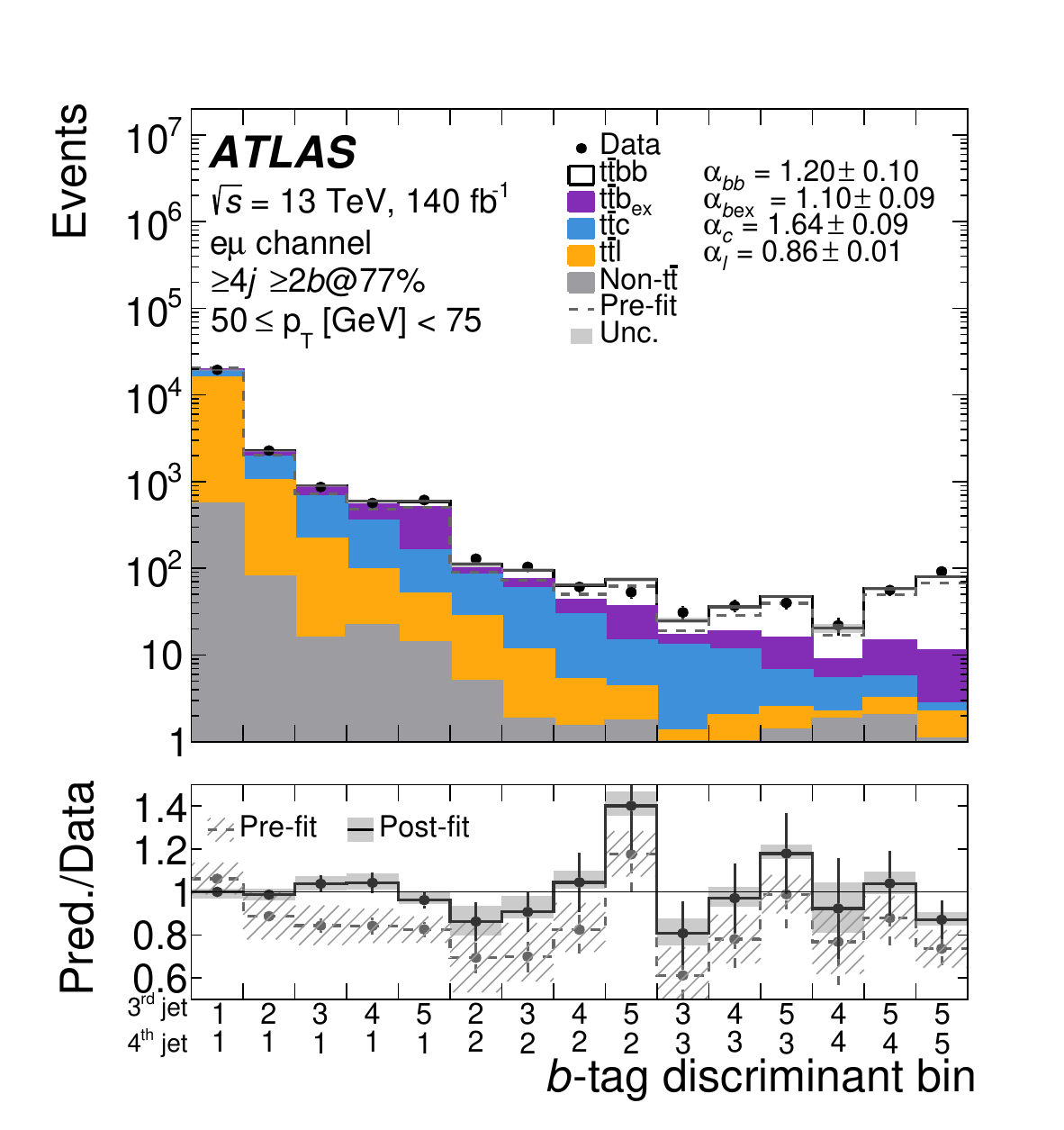}\label{fig:FCFResult_4j2b_d}} \\
\caption{Comparison of data and predictions for the $b$-tagging discriminant score distributions in \protect \subref{fig:FCFResult_3j2b_a} the inclusive detector-level $\geq 3j\ge2b@77\%$ region considered for the \textit{Global} fit. The score ranging from 1 to 5, correspond to a certain range of $b$-tagging efficiencies defined by the working points: 100\%--85\%, 85\%--77\%, 77\%--70\%, 70\%--60\%, and $<60\%$, respectively.  \protect \subref{fig:FCFResult_3j2b_b}, \protect \subref{fig:FCFResult_3j2b_c} and \protect \subref{fig:FCFResult_4j2b_d} show the distributions of discriminant score in the multiple binned sub-regions used for \textit{Kinematic-dependent} fit in the \pt ranges of the third highest-\pt jet in the event. \protect \subref{fig:FCFResult_3j2b_b} 25--35~\GeV\ and \protect \subref{fig:FCFResult_3j2b_c} $>$ 50~\GeV for the jet \pt sub-regions in the $3j\ge2b@77\%$ region, and \protect \subref{fig:FCFResult_4j2b_d} for the jet \pt between 50--75~\GeV\ in the region $\ge 4j\ge2b@77\%$. The two dimensional $\ge4j\ge2b@77\%$ templates representing the third and fourth $b$-tagging discriminant-ranked jet are flattened into one dimension. The dashed lines in the upper panel show the pre-fit total predictions, and the stacked histograms show the contributions scaled according to the results of the fit.  The ratio of total predictions before and after the fit to the data are shown in the lower panel. The vertical bars in each ratio represents only the statistical uncertainty, and the shaded bands represent the total error including the systematic uncertainties from experimental sources. The extracted scale factors $\alpha_{l}$, $\alpha_{c}$, $\alpha_{b}$ and $\alpha_{bb}$ are given considering only statistical uncertainties.}
\label{fig:FCFResult}
\end{figure}

The \textit{Global fit} is chosen as the nominal approach because it improves the modelling of the \ttc and \ttl backgrounds while maintaining correlations between the normalisation factors for the \ttbj and \ttc and \ttl processes. Figure~\ref{fig:FCFResult_3j2b_a} shows the template distributions both before and after applying the best-fit scale factors, with the results summarised in the first row of Table~\ref{tab:SFsFits}, where the quoted uncertainties are statistical only. This method ensures that the signal efficiency and acceptance corrections derived from simulation remain intact. On the other hand, the \textit{Kinematic-dependent fit} can improve both the normalisation and shape of the \ttc and \ttl backgrounds but may introduce biases into the unfolding corrections. Therefore, the \textit{Kinematic-dependent fit} is used only for evaluating shape uncertainties.

In the case of the \textit{Global fit} approach, events are selected with at least three reconstructed jets, at least two of which are $b$-tagged. This region labelled as $\geq 3j \geq 2b@77\%$ and listed in Table~\ref{tab:fitclass} is used to derive the nominal scale factors.  The templates are created for MC events in three different particle-level categories, labelled as \ttb, \ttc, and \ttl and described in Table~\ref{tab:fitclass} for the $\geq 3j \geq 2b@77\%$ region, using the $b$-tagging discriminant value of the jet with the third-highest $b$-tagging discriminant.  A small fraction ($\sim 5\%$) of events in the signal templates contain at least one additional \cjet and at least one additional \bjet.  Likewise, the signal templates contain a relatively small fraction ($\sim 8\%$) of events where at least one of the two \bjets originating from a top quark is outside of the fiducial volume.  The fiducial cross-section measurements remain largely unaffected by the choice of `truth'-heavy-flavour classifications of individual templates considered in the nominal fit, due to the common fiducial definition used in the scale factor evaluation and the cross-section measurement. %
One additional template is created from the sum of all non-\ttbar backgrounds described in Section~\ref{sec:bkgd}.

A selected jet in the analysis can be sorted into one of five possible bins (labelled 1 to 5 in the following) that correspond to a certain range of $b$-tagging efficiencies defined by the working points: 100\%--85\%, 85\%--77\%, 77\%--70\%, 70\%--60\%, and $<60\%$ respectively. The bin with 100\%--85\% efficiency contains only untagged jets mainly coming from the \ttl category, and is hence not used in the fit.  The remaining bins with substantial contributions from the \ttc and \ttb categories are considered in the \textit{Global fit} approach.

The \textit{Global fit} procedure for fitting the normalisation factors in the $\geq 3j \geq 2b@77\%$ region is repeated for each individual uncertainty considered in the analysis.  The corresponding scale factors for each uncertainty are then propagated to the unfolding step as discussed in Section~\ref{subsec:unfolding}.

\begin{figure}[htbp]
\centering
\subfloat[]{\includegraphics[width=0.44\textwidth]{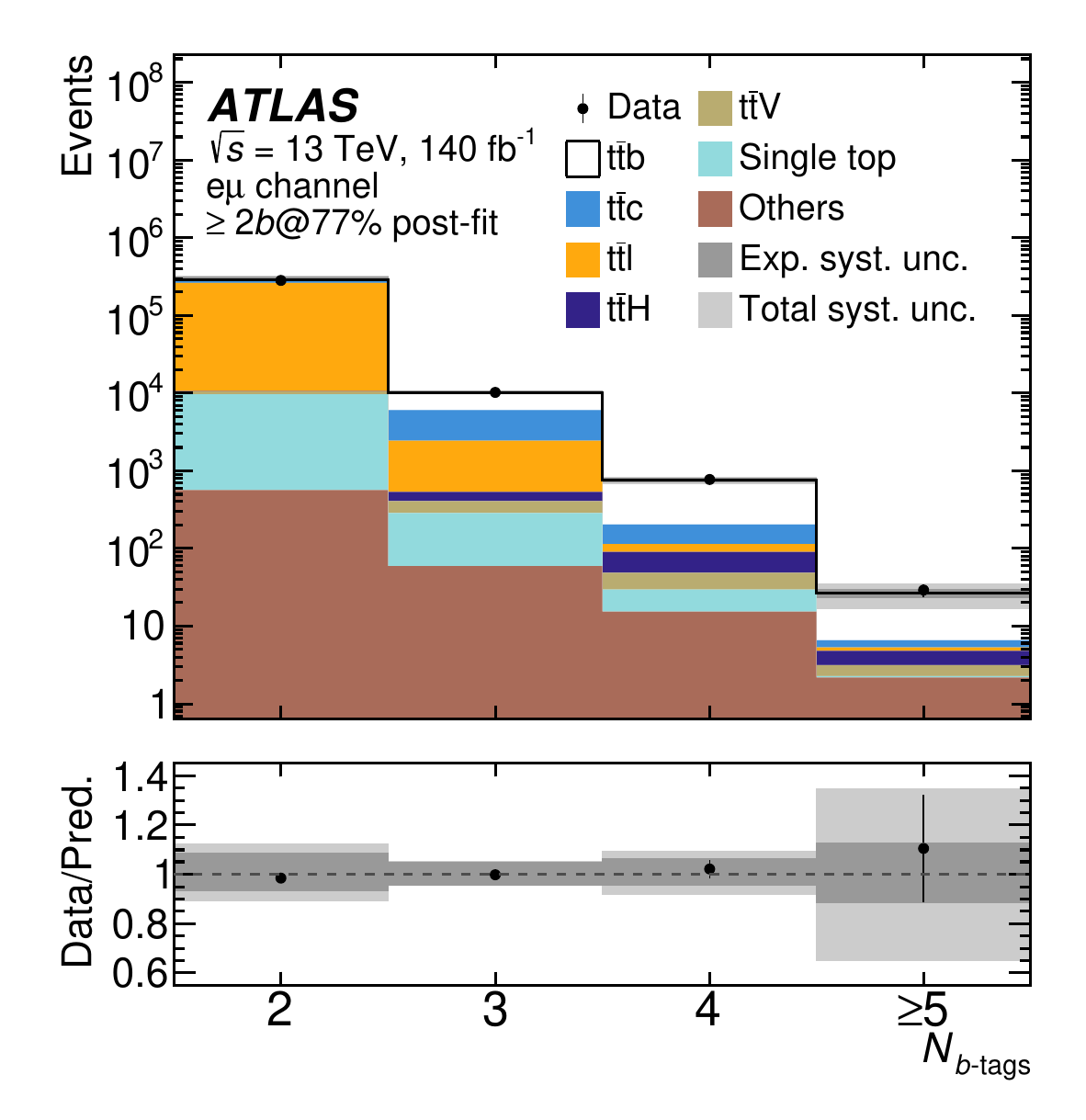}\label{fig:reco-dataMC-plots_nbjets_corr}}
\subfloat[]{\includegraphics[width=0.44\textwidth]{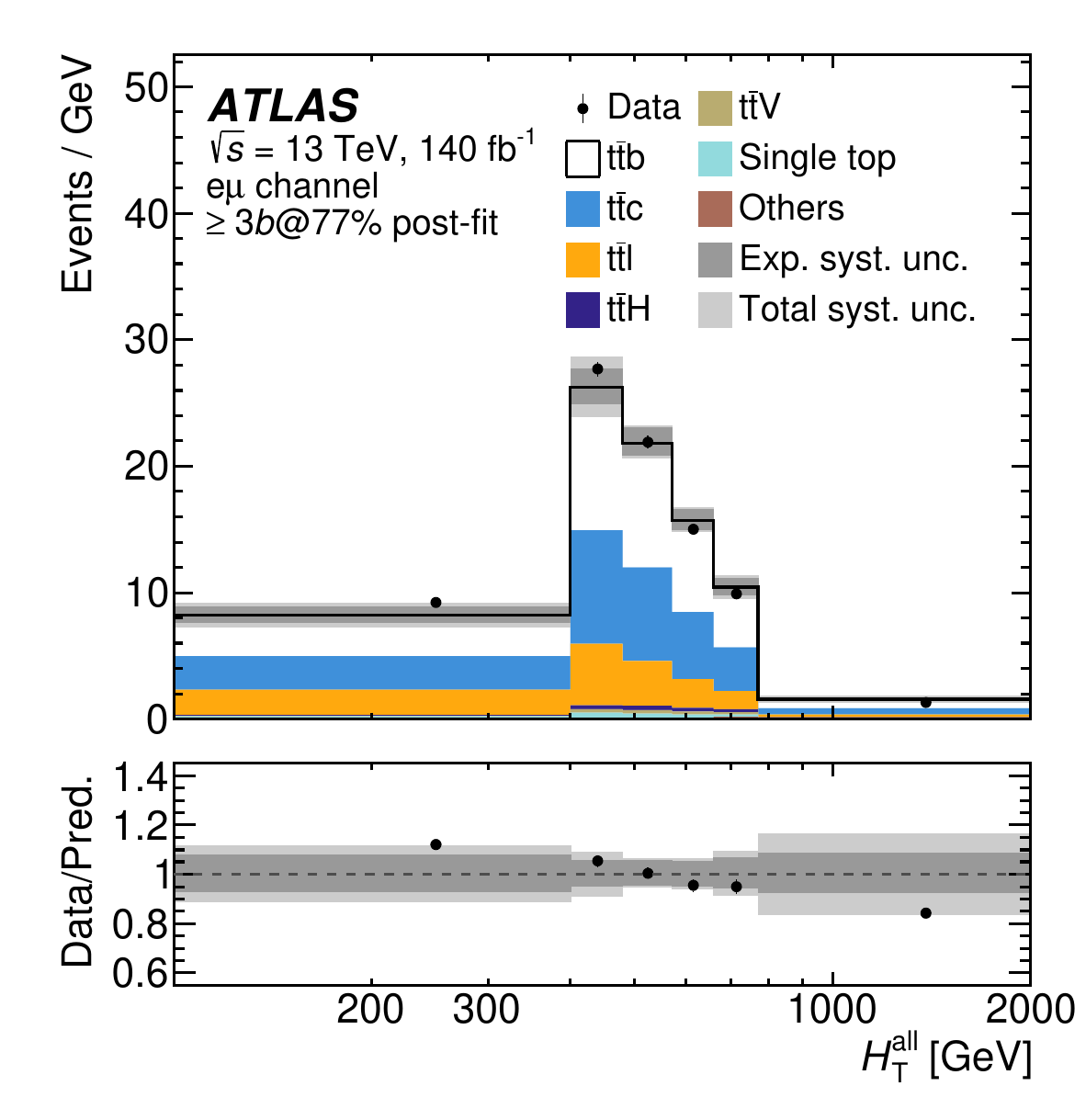}\label{fig:reco-dataMC-plots_corr_3j3b_HT}} \\
\subfloat[]{\includegraphics[width=0.44\textwidth]{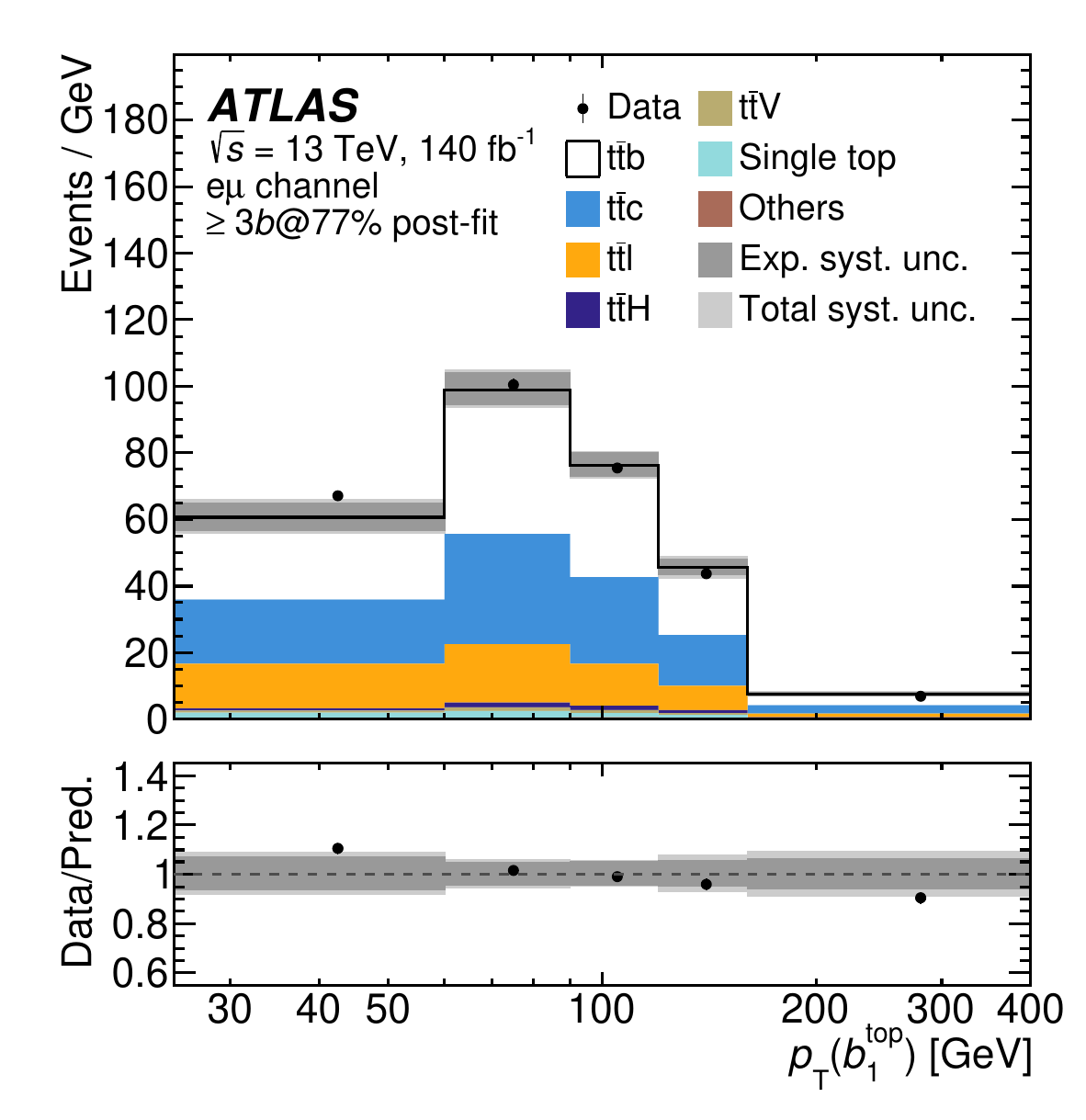}\label{fig:reco-dataMC-plots_corr_3j3b_TopbJetPt_1}}
\subfloat[]{\includegraphics[width=0.44\textwidth]{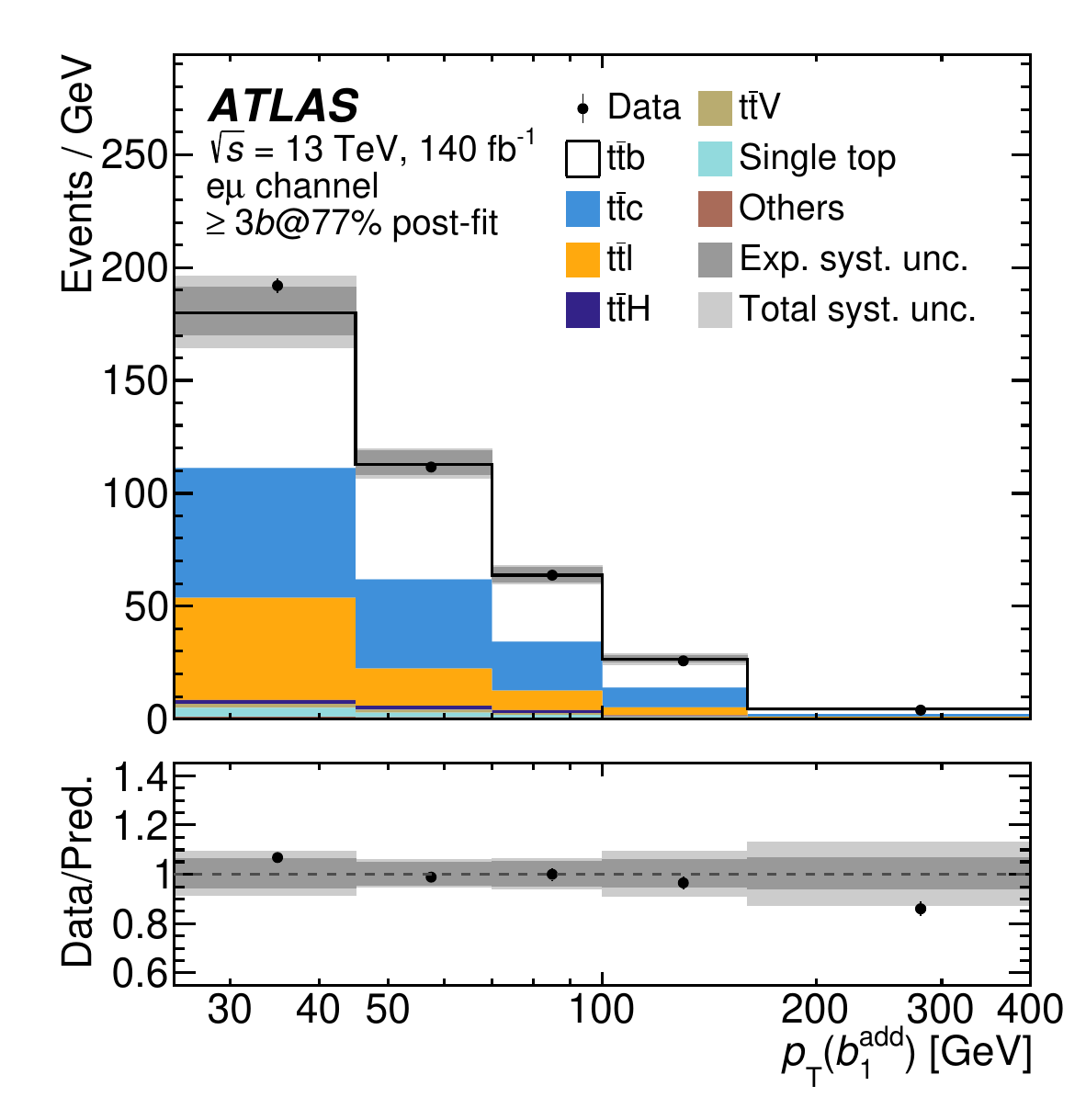}\label{fig:reco-dataMC-plots_corr_3j3b_AddbJetPt_1}} \\
\caption{Comparison of data and predictions for the distribution of \protect \subref{fig:reco-dataMC-plots_nbjets_corr} the number of $b$-tagged jets for events with at least two $b$-tagged jets ($N_{b\mathrm{-tags}}$), \protect \subref{fig:reco-dataMC-plots_corr_3j3b_HT} $H_{\mathrm{T}}^{\mathrm{all}}$, \protect \subref{fig:reco-dataMC-plots_corr_3j3b_TopbJetPt_1} \pt$(b_1^{\text{top}})$, and \protect \subref{fig:reco-dataMC-plots_corr_3j3b_AddbJetPt_1} \pt$(b_1^{\text{add}})$ in reconstructed events with three or more $b$-tagged jets after the \textit{Global} scale factors are applied for different event categories as discussed in text (see Section~\ref{subsec:flavCompCorrections}).  The entries in each bin are divided by the bin width in the \protect \subref{fig:reco-dataMC-plots_corr_3j3b_HT}, \protect \subref{fig:reco-dataMC-plots_corr_3j3b_TopbJetPt_1} and \protect \subref{fig:reco-dataMC-plots_corr_3j3b_AddbJetPt_1} distributions. The lower panels show the ratios of the data to the predictions. The inner band includes the uncertainties due to limited MC statistics and due to various detector effects such as the jet energy scale and resolution, the $b$-tagging, and the fit method, while the outer band also includes the theoretical uncertainties in the signal and background modelling. The last bin includes the overflow.}
\label{fig:reco-dataMC-plots_postfit}
\end{figure}

Figure~\ref{fig:reco-dataMC-plots_postfit} compares the data with predictions for $b$-tagged jet multiplicity in events with at least two $b$-tagged jets and the $H_{\mathrm{T}}^{\mathrm{all}}$, \pt$(b_1^{\text{top}})$, and \pt$(b_1^{\text{add}})$ spectra in events with three or more $b$-tagged jets. The \textit{Global fit} scale factors, derived from the $\geq 3j \geq 2b@77\%$ region, are applied to the reconstructed events for the corresponding \ttc, \ttl and \ttb event categories, depending on the observables. The uncertainty band includes the uncertainties from various experimental sources and the theoretical modelling uncertainties. Compared to the pre-fit distributions shown in Figure~\ref{fig:reco-dataMC-plots_prefit}, the post-fit predictions show significantly improved agreement with the data after corrections to the individual components.

\begin{table}
\caption{Truth categorisations, defined using particle level information, of the reconstructed events in the regions with selections $\geq 3j \geq 2b@77\%$, $3j \geq 2b@77\%$, and $\geq 4j \geq 2b@77\%$ of the \ttbar MC samples. %
The $3j \geq 2b@77\%$ and $\geq 4j \geq 2b@77\%$ regions are sub-divided in terms of the \pt of the third-leading \pt reconstructed jet.  Reconstructed events in each region and sub-region are categorised based on the particle-level selections of the number of $b$-jets, $c$-jets and light-flavour jets. All particle-level categories are listed in the first column and their definitions are given in the following columns under the respective reconstruction-level region.  The dashes (--) imply that the corresponding event category is not present for that region.}
\small
\begin{center}
\begin{scalebox}{0.90}{
\begin{tabular}{c||c|c|c|c|c|c|c}
\toprule
& \multicolumn{1}{c|}{Inclusive region} & \multicolumn{6}{c}{Regions in terms of jet multiplicity and third-highest-\pt jet-\pt} \\
& \textit{Global} approach  & \multicolumn{6}{c}{ \textit{Kinematic-dependent} approach} \\ \hline
& (nominal)  & \multicolumn{6}{c}{(systematic)} \\ \hline
Category & $\geq 3j \geq 2b@77\%$ & \multicolumn{3}{c|}{$3j \geq 2b@77\%$} & \multicolumn{3}{c}{$\geq 4j \geq 2b@77\%$}  \\
& $\geq 25$ \GeV & 25--35 \GeV & 35--50 \GeV & $\geq 50$ \GeV & 25--50 \GeV & 50--75 \GeV & $\geq 75$ \GeV  \\
\midrule
\ttb     & $\ge$3 \bjets & \multicolumn{3}{c|}{$\ge$3 \bjets}                     & \multicolumn{3}{c}{--}   \\
$\ttb_{\mathrm{ex}}$ & -- & \multicolumn{3}{c|}{--}                    & \multicolumn{3}{c}{exactly 3 \bjets}     \\
\ttbb    &  -- & \multicolumn{3}{c|}{--}                               & \multicolumn{3}{c}{$\ge$4 \bjets}     \\
\ttc     & $<$ 3 \bjets and $\ge$ 1 \cjet & \multicolumn{3}{c|}{$<$ 3 \bjets and $\ge$ 1 \cjet}    & \multicolumn{3}{c}{$<$ 3 \bjets and $\ge$ 1 \cjet} \\
\ttl     & \multicolumn{1}{c|}{events that do not meet}  & \multicolumn{3}{c|}{events that do not meet} & \multicolumn{3}{c}{events that do not meet}  \\
&  \multicolumn{1}{c|}{above criteria} & \multicolumn{3}{c|}{above criteria} & \multicolumn{3}{c}{above criteria}  \\
\bottomrule
\end{tabular}
}\end{scalebox}
\end{center}
\label{tab:fitclass}
\end{table}

For the evaluation of \textit{Kinematic-dependent fit} scale factors, the selected events are further split into multiple orthogonal regions depending on the number of reconstructed jets and the \pt of the third-highest-\pt jet in the event, as summarised in Table~\ref{tab:fitclass}.  Three \pt sub-regions are chosen with sufficient and similar numbers of events.
A total of 21 templates for the multiple binned scenario are created for the region with exactly three reconstructed jets ($3j \geq 2b@77\%$) using the third-highest $b$-tagging discriminant score, and for the region with four or more reconstructed jets ($\geq 4j \geq 2b@77\%$) using the $b$-tagging discriminant values of the third- and fourth-highest score in the event. The $3j \geq 2b@77\%$ detector-level region has three particle-level categories: \ttb, \ttc, and \ttl, while the $\geq 4j \geq 2b@77\%$ region has four particle-level categories: $\ttb_{\mathrm{ex}}$, \ttbb, \ttc, and \ttl with the same definitions for the \ttc and \ttl background categories in all regions, as described in Table~\ref{tab:fitclass}.  The $\ttb_{\mathrm{ex}}$ category contains exactly three particle-level \bjets, while the \ttb (\ttbb) category includes three (four) or more particle-level \bjets.  One-dimensional templates with four bins are formed starting from bin~2 for the jet with the third highest $b$-tagging discriminant value in $3j \geq 2b@77\%$ detector-level regions, while in the $\geq 4j \geq 2b@77\%$ detector-level regions two-dimensional templates are created using the third- and fourth-highest $b$-tagging discriminant values in the full range for the two jets.

In the $3j \geq 2b@77\%$ detector-level region, three fit parameters, $\alpha_b^s$, $\alpha_c^s$ and $\alpha_l^s$, are used in the maximum-likelihood fit following Eq.~(\ref{eq:poisson_likelihood}) and Eq.~(\ref{eq:partial_likelihood}) for each of sub-region $s$ given in Table~\ref{tab:fitclass}. In the $\geq 4j \geq 2b@77\%$ regions, four fit parameters $\alpha_{b_{\mathrm{ex}}}^s$, $\alpha_{bb}^s$, $\alpha_c^s$ and $\alpha_l^s$ are used, such that the expected number of events in bin $k$ of a given sub-region $s$ defined in Table~\ref{tab:fitclass} is
\begin{equation*}
\nu_k(\alpha_{b_{\mathrm{ex}}}^s, \alpha_{bb}^s, \alpha_c^s, \alpha_l^s) =
\alpha_{b\mathrm{ex}}^s N^{k,s}_{\ttb_{\mathrm{ex}}} + \alpha_{bb}^s N^{k,s}_{\ttbb} + \alpha_{c}^s N^{k,s}_{\ttc} +  \alpha_l^s N^{k,s}_{\ttl}
+ N^{k,s}_{\textrm{non-}\ttbar}~,
\end{equation*}
where $N^{k,s}_{\ttb_{\mathrm{ex}}}$ and $N^{k,s}_{\ttbb}$ are the numbers of events in bin $k$ and each sub-region $s$ of the $\ttb_{\mathrm{ex}}$ and \ttbb templates, respectively. The fit is performed in each sub-region listed in Table~\ref{tab:fitclass}.
%

\begin{table}
\caption{Best-fit values of the \ttb, $\ttb_{\mathrm{ex}}$, \ttbb, \ttc, and \ttl scale factors determined from dedicated fit regions. The quoted uncertainties are statistical only.}
\small
\begin{center}
\begin{tabular}{c|ccccc|c}
\toprule
& \multicolumn{5}{c|}{Fitted values of scale factors} & \multicolumn{1}{c}{Type} \\
Regions  & $\alpha_{b}^s$ & $\alpha_{b\mathrm{ex}}^s$ & $\alpha_{bb}^s$  & $\alpha_{c}^s$ & $\alpha_{l}^s$ &  \\
\midrule
$\geq 3j \geq 2b$; $\geq 25$ \GeV & 1.20 $\pm$ 0.03 & -- & -- & 1.62 $\pm$ 0.09 & 0.92 $\pm$ 0.04 & \textit{Global} \\ \hline
$3j \geq 2b$; (25--35) \GeV  & 1.40 $\pm$ 0.15 & -- & -- & 1.99 $\pm$ 0.42 & 0.98 $\pm$ 0.08 & \\
$3j \geq 2b$; (35--50) \GeV  &  1.30 $\pm$ 0.11 & -- & -- & 1.74 $\pm$ 0.27 & 0.77 $\pm$ 0.11 & \multirow{1}{*}{} \\
$3j \geq 2b$;  $\geq 50$ \GeV  & 1.26 $\pm$ 0.12 & -- & -- & 1.05 $\pm$ 0.27 & 1.09 $\pm$ 0.15 & \multirow{1}{*}{\textit{Kinematic-}} \\
$\geq 4j \geq 2b$; (25--50) \GeV  & -- & 1.31 $\pm$ 0.10 & 1.15 $\pm$ 0.14 & 1.93 $\pm$ 0.11 & 0.92 $\pm$ 0.01 &  \multirow{1}{*}{\textit{dependent}}\\
$\geq 4j \geq 2b$; (50--75) \GeV  & -- & 1.10 $\pm$ 0.09 & 1.20 $\pm$ 0.10 & 1.64 $\pm$ 0.09 & 0.86 $\pm$ 0.01 & \\
$\geq 4j \geq 2b$; $\geq 75$ \GeV & -- & 1.10 $\pm$ 0.10 &  1.09 $\pm$ 0.10 &  1.25 $\pm$ 0.10 & 0.83 $\pm$ 0.02 & \\
\bottomrule
\end{tabular}
\end{center}
\label{tab:SFsFits}
\end{table}


The result of the \textit{Kinematic-dependent} scale factor fits in some regions are presented in Figure~\ref{fig:FCFResult}, and the fit values are summarised in Table~\ref{tab:SFsFits} for each fit set-up, where the quoted uncertainties are statistical only.  The values of $\alpha_b^s$, $\alpha_{b\mathrm{ex}}^s$, and $\alpha_{bb}^s$ extracted from various regions suggest 10\%--30\% level corrections to the \powpy predictions. On the other hand the \ttc background components have scale factors of up to two extracted from the fit and they tend to decrease with the \pt of third-highest-\pt jet in the event.

The \ttc+\ttl background estimation using the \textit{Kinematic-dependent fit} gives up to $5\%$ shape variations relative to that obtained from the \textit{Global fit} estimation method, and this is accounted for as the shape uncertainty of this background.

\subsection{Unfolding}
\label{subsec:unfolding}

The measured distributions at the detector level are unfolded to the stable particle level.  The unfolding procedure corrects for resolution effects and for detector efficiencies and acceptances. An iterative Bayesian unfolding technique~\cite{DAgostini:1994fjx}, as implemented in the \textsc{RooUnfold} software package~\cite{Adye:2011gm}, is used.

First, for each of the observable distributions, the number of non-\ttbar, \ttV, \ttH and other rare background events in bin $k$ ($N_{\mathrm{bkg}}^{k}$), as described in Section~\ref{sec:bkgd}, are subtracted from the data distribution at the detector level ($N_{\mathrm{data}}^{k}$). This retains a mixture of signal and mis-tagged \ttc and \ttl background events.  A series of corrections are then applied, with all corrections derived from nominal simulated \ttbar events after the \ttc, \ttl and $\ttb$ event categories are scaled using the \textit{Global} scale factors as described in Section~\ref{subsec:flavCompCorrections}. %
These correction factors are determined separately for the \ttb and \ttbb categories of the fiducial definitions.
The background-subtracted data are first corrected for the mis-tagged \ttc and \ttl background events by applying the correction
($f_{\ttb}^k$) for the mis-tagged \ttc and \ttl background events defined as

\begin{equation*}
f_{\ttb}^k = \frac{\mathcal{S}^k_{\ttb,\text{reco}}}{\mathcal{S}^k_{\ttb,\text{reco}} + \mathcal{B}^k_{\ttb,\text{reco}}},
\end{equation*}

where $\mathcal{S}^k_{\ttb,\text{reco}}$ is the number of detector-level events belonging to the \ttb category, and $\mathcal{B}^k_{\ttb,\text{reco}}$ is the number of detector-level \ttc and \ttl events in bin $k$ as predicted by the MC simulation after the global scale factors are applied. Next, an acceptance correction, $f_{\mathrm{accept}}^k$, is applied, which corrects for the fiducial acceptance and is defined as the probability of a \ttb event satisfying the detector-level selection in a given bin $k$ ($\mathcal{S}^k_{\ttb, \text{reco}}$) to also fall within the fiducial particle-level phase space ($\mathcal{S}^k_{\ttb, \mathrm{reco}\wedge \mathrm{part}}$). It is estimated as

\begin{equation*}
f_{\mathrm{accept}}^{k} = \frac{\mathcal{S}^k_{\ttb, \mathrm{reco}\wedge \mathrm{part}}}{\mathcal{S}^k_{\ttb, \text{reco}}}
\end{equation*}

with values ranging from $0.97$ to $1$.

The remaining part of the unfolding procedure consists of inverting the migration matrix $\mathcal{M}$ to correct for the resolution effects and subsequently correcting for detector inefficiencies.  The matrix, $\mathcal{M}$, represents the probability for a particle-level event in bin $i$ to be reconstructed in bin $k$.  The chosen binning is optimised for each distribution to have a migration matrix with a large fraction ($\sim 60\%$) of events on the diagonal and a sufficient number of events in each bin. The Bayesian unfolding technique performs the effective matrix inversion, $\mathcal{M}^{-1}_{ik}$, iteratively.  Four iterations are used for all measured distributions.

Finally, the factor $f_{\mathrm{eff}}^{i}$ corrects for the reconstruction efficiency and is defined as

\begin{equation*}
f_{\mathrm{eff}}^{i} = \frac{\mathcal{S}^i_{\ttb, \mathrm{part}\wedge \mathrm{reco}}}{\mathcal{S}^i_{\ttb, \mathrm{part}}},
\end{equation*}

where $\mathcal{S}^i_{\ttb, \mathrm{part}}$ is the number of events from \ttb category passing the particle-level selection in bin $i$ and $\mathcal{S}^i_{\ttb, \mathrm{part}\wedge \mathrm{reco}}$ is the number of events from \ttb category in bin $i$ that also satisfy the detector-level selection.  This efficiency factor ranges from $20\%$ to $25\%$ ($10\%$ to $20\%$) for the observables defined in phase space with at least three (four) \bjets.  %

The unfolding procedure for an observable $X$ at particle level can be summarised by the following expression

\begin{equation}
N_{\text{unfold}}^i = \frac{1}{f_{\text{eff}}^i} \displaystyle\sum_{k} \mathcal{M}_{ik}^{-1}\,f_{\text{accept}}^k\,f_{\ttb}^k\,(N_{\text{data}}^{k} - N_{\text{bkg}}^{k}),
\label{eq:Unfolding_ttb}
\end{equation}

where $N_{\text{unfold}}^i$ is the number of events in bin $i$ of the unfolded distribution, and the summation runs over the bin $k$ of the reconstructed distribution. The total fiducial cross-sections ($\sigma^{\mathrm{fid}}$) are obtained from the expression for $N_{\text{unfold}}^i$ with the bin $i$ taken as the entire fiducial region and dividing this by intergrated luminosity $\mathcal{L}$. Results from the unfolded distributions are presented in terms of a normalised differential cross-section given by

\begin{equation*}
\frac{1}{\sigma^{\mathrm{fid}}}.\frac{\text{d} \sigma^{\text{fid}}}{\text{d}X^i} = \frac{N_{\text{unfold}}^i}{\Delta X^i \displaystyle\sum_{i} N_{\text{unfold}}^i},
\end{equation*}

where $\Delta X^{i}$ is the bin width.

The terms $f_{\text{eff}}^i$, $\mathcal{M}_{ik}^{-1}$, $f_{\text{accept}}^k$, $f_{\ttb}^k$ and $N_{\text{bkg}}^{k}$ are different  for the different systematic variations discussed in Section~\ref{sec:systs}. The nominal distributions of unfolding corrections, namely $f_{\text{eff}}^i$, $\mathcal{M}_{ik}^{-1}$, $f_{\text{accept}}^k$ and $f_{\ttb}^k$, are obtained using the nominal \ttbar MC samples after the \textit{Global} scale factors as given in Table~\ref{tab:SFsFits} are applied, which do not bias the efficiency and acceptance corrections or the migration matrix.  The \textit{Kinematic-dependent} scale factors given in Table~\ref{tab:SFsFits} are applied only as a systematic variation on the \ttl and \ttc background estimates due to the fit method.  The \textit{Global} method is used to re-derive the scale factors for each systematic uncertainty component considered, and the uncertainties in the unfolded results are evaluated as discussed in Section~\ref{sec:systs}.

Various tests are performed to check for any biases due to the choice of the nominal unfolding matrix.  The pseudo-data distributions are generated after reweighting the distributions from the nominal \ttbar sample to the alternative distributions.  Three different reweighting scenarios are considered, and the pseudo-data distributions obtained after the reweighting are unfolded using the nominal response matrix. In the first case, the weight functions are derived from the ratio of the detector-level distributions in the data to the nominal \ttbar sample, and the weights are applied to the nominal \ttbar events depending on the particle-level variables.  In the second case, the weights are obtained after taking the ratio of \sherpa \ttbar to the \powpy \ttbar predictions as a function of the observable in question and applied to the nominal \ttbar events as a function of particle-level variable.  In the third case, an alternative kinematic variable, unrelated to the distribution being unfolded, is selected, and the weights are derived from the ratio of data and MC simulated distribution of this variable. The events in the nominal sample are then reweighted using these weights to produce reweighted distributions.  The differences in each bin of the resulting unfolded distribution are found to be compatible with the corresponding expectations within systematic uncertainties, hence validating that the unfolding model does not introduce any biases in the measurements.

%

%
%
%


%

%
\section{Statistical and systematic uncertainties}
\label{sec:systs}

%
Various sources of uncertainties affecting the measurements are described in this section. The effects of finite numbers of data and MC events are evaluated after generating pseudo-experiments as described in Section~\ref{subsec:StatUnc}.  Experimental sources of uncertainties are described in Section~\ref{subsec:ExpUnc}, sources of uncertainties in \ttbar modelling are described in Section~\ref{subsec:SigUnc}, and the uncertainties in the background estimates are described in Section~\ref{subsec:BkgUnc}.

\subsection{Statistical uncertainties}
\label{subsec:StatUnc}
The impact of the statistical uncertainty in the data is evaluated after generating 1000 pseudo-experiments by fluctuating the data in each bin $i$ of the reconstructed distribution based on a Poisson distribution $P(N_{\mathrm{obs}}^i$) with a mean value of $N_{\mathrm{obs}}^i$, which represents the observed number of events in bin $i$.  The reconstructed distribution obtained from each pseudo-experiment is unfolded using the nominal unfolding matrix, and the standard deviation in each bin of the resulting particle level distributions is taken as the statistical uncertainty in that bin. A similar procedure is used to estimate the effect of the statistical uncertainties in the MC simulation-based quantities in Eq.~(\ref{eq:Unfolding_ttb}), where the predictions in each bin are fluctuated according to a Gaussian distribution to generate the pseudo-experiments. %

\subsection{Experimental uncertainties}
\label{subsec:ExpUnc}

As discussed in Section~\ref{subsec:detobjs}, the efficiencies of physics object reconstruction and identification may differ between data and the MC simulation. The uncertainties in the scale factors to correct for the efficiency differences between data and simulation in lepton trigger~\cite{TRIG-2018-05,TRIG-2018-01}, reconstruction, identification and isolation~\cite{EGAM-2018-01,MUON-2022-01} %
are taken into account by varying the scale factors within their uncertainties. These efficiencies are estimated in data using a tag-and-probe technique in $Z\rightarrow e^+ e^-$ and $Z\rightarrow \mu^+ \mu^-$ events. The electron (muon) momentum scale and resolution are determined using the measurement of the position and width of the $Z$ boson peak in $Z\rightarrow e^+ e^- (\mu^+\mu^-)$ events. The lepton-related uncertainties are considerably smaller than the jet energy scale and flavour-tagging efficiency uncertainties discussed below.

Jets are calibrated using a series of simulation-based corrections and in situ techniques~\cite{JETM-2018-05}. %
The uncertainties due to the jet energy scale (JES) are estimated by using a combination of simulation, test-beam data and in situ measurements.  Contributions from the jet-flavour composition, $\eta$-intercalibration, leakage of the hadron showers beyond the extent of the hadronic calorimeters (punch-through), single-particle response, calorimeter response to different jet flavours, and pile-up are taken into account, resulting in 30 independent uncertainty components.  The total uncertainty due to the JES is one of the dominant uncertainties in this analysis.

The jet energy resolution (JER) is measured using both data and simulation~\cite{JETM-2018-05}.  First, the resolution in the simulation is determined by comparing the particle-level and reconstructed jet \pt in simulation as a function of the jet \pt and $\eta$.  Second, an in situ measurement of the JER is made using the \textit{dijet balance} method in dijet data events. The resolution in data and simulation are compared and the energies of jets in the simulation are smeared to match the resolution observed in data. In total eight independent uncertainty components are used. The uncertainties in the JER stem from uncertainties in both the MC modelling of the JER and the data-driven method.

The JVT is calibrated using $Z(\rightarrow \mu\mu)+$jet events where the jet balances the \pt of the $Z$ boson.  Scale factors binned in jet \pt are applied to each event to correct for small differences in the JVT efficiency between the data and simulation. The uncertainty in the efficiency to satisfy the JVT requirement is evaluated by varying the scale factors within their uncertainties~\cite{PERF-2014-03}.%

Differences in the $b$-tagging efficiencies between the data and simulation are corrected using scale factors derived from events containing two leading jets in the dilepton \ttbar enriched sample~\cite{FTAG-2018-01} as a function of jet \pt. The $c$-jet mistagging efficiency calibration is derived from lepton+jets \ttbar ~\cite{FTAG-2020-08}
as a function of jet \pt with uncertainties in the range of 12\%--19\%.  A negative tag method is used to calibrate mis-tagged light-flavour jets with a precision of 18\%--31\%~\cite{FTAG-2019-02}.
The scale factors are measured for different efficiency working points.  The associated flavour-tagging uncertainties, including the high-\pt extrapolations, are computed by varying the scale factors within their uncertainties. In total, there are 49 components related to the $b$-tagging efficiencies and 22 (24) components related to the mis-tag rates of $c$-jets (light-flavour jets).

The uncertainty in the pile-up reweighting of the reconstructed events in the simulation is estimated by comparing the distribution of the number of primary vertices in the simulation with the one in the data as a function of the instantaneous luminosity.  Differences between these distributions are adjusted by scaling the mean number of $pp$ interactions per bunch crossing in the simulation and the $\pm 1$ standard deviation (s.d.) uncertainties are assigned to these scaling factors. The pile-up weights are recalculated after varying the scale factors within their uncertainties.

The uncertainty in the combined 2015-2018 integrated luminosity is $0.83\%$~\cite{DAPR-2021-01}, obtained using the LUCID-2 detector~\cite{LUCID2} for the primary luminosity measurements, complemented by measurements using the inner detector and calorimeters.

All experimental uncertainties affecting the backgrounds ($N_{\text{bkg}}^{k}$), the corresponding fitted values of \ttb, %
\ttc and \ttl scale factors, and the unfolding correction factors ($f_{\text{eff}}^i$, $\mathcal{M}_{ik}^{-1}$, $f_{\text{accept}}^k$, $f_{\ttb}^k$) are propagated to the unfolded results after repeating the fitting and reweighting procedure for each systematic variation.  Various methods are explored for the evaluation of the impact of dominant uncertainty components, such as the JES, to the unfolded results, and the outcomes are compared with the well established approach, where the data distribution is unfolded using Eq.~(\ref{eq:Unfolding_ttb}) with varied simulation-based quantities and where the uncertainty is taken as the difference between the result and the nominal unfolded distribution.  In the end, a technically convenient approach is taken for each experimental uncertainty component without affecting the estimated size of its impact on the unfolded results.  The detector-level pseudo-data distribution constructed from the predicted signal and estimated background with a varied uncertainty component is unfolded using the nominal unfolding distributions, and the result is compared to the corresponding particle-level distribution ($\mathcal{S}^i_{\ttb, \mathrm{part}}$) of the reweighted signal sample, and the relative difference is taken as an uncertainty.  %

The statistical uncertainties in the \ttb, %
\ttc and \ttl global scale factors fits are taken into account as $\pm 1$ s.d. variations in those scale factors as given in Table~\ref{tab:SFsFits}. Individual components of reconstruction- and particle-level distributions of the nominal MC sample are reweighted with varied scale factors.  The detector-level distribution from that reweighted sample is unfolded with the nominal unfolding corrections as described in Section~\ref{subsec:unfolding}, and the result is compared with the particle-level distribution from the corresponding reweighted sample. The relative difference in each bin is taken as the systematic uncertainty.  Additional uncertainties are attributed to the shape differences of the \ttc and \ttl background estimates where the alternative fit set-up depending on the jet \pt and multiplicity, as presented in Table~\ref{tab:SFsFits}, is used.  These are evaluated as the relative difference between the nominal unfolded data distribution and the unfolded distribution obtained when the \ttc and \ttl background estimates are taken from the alternative fit set-up.

\subsection{Signal modelling uncertainties}
\label{subsec:SigUnc}

Uncertainties due to the choice of the \ttbar MC generator settings affecting the \ttc and \ttl background predictions and the signal modelling are evaluated as follows.  Individual components in each alternative \ttbar sample are first fit to data using the procedure described in Section~\ref{subsec:flavCompCorrections}, the corresponding samples are then reweighted according to their corresponding fitted scale factors following the same procedure as applied to the nominal \powpy sample.  Unless stated otherwise, the uncertainties in the results are evaluated after unfolding the detector-level pseudo-data distributions from alternative \ttbar samples, described in Section~\ref{sec:mcsim}, using the nominal unfolding set-up as described in Section~\ref{subsec:unfolding}. %
The unfolded distributions are then compared with the particle-level distributions from the alternative sample and the relative difference in each bin is taken as the systematic uncertainty.

Uncertainties related to the QCD scale variations in the matrix element predictions are evaluated by varying the renomalisation ($\mu_{\mathrm{R}}$) and factorisation ($\mu_{\mathrm{F}}$) scale settings in the \powpy sample by a factor of 0.5 or 2 independently. The momentum scale, $\mu_{\mathrm{R}}^{\mathrm{ISR}}$, for the initial-state radiation produced by parton shower is varied using the \texttt{var3cUp} / \texttt{var3cDown} parameter settings in the A14 \PYTHIA[8] tune. The uncertainties in the tuning of final-state radiation in the parton shower are obtained by changing the momentum scale settings of $\mu_{\mathrm{R}}^{\mathrm{FSR}}$ associated with this radiation in \PYTHIA[8] by a factor of 0.625 or 2 relative to the nominal scale.  A higher scaling factor than 0.5 is used for the downward variation to avoid spurious generator weights and to keep the upward and downward variations symmetrical. %
All of these variations are obtained using internal MC generator weights available in the nominal \powpy sample, leading to eight alternative models of \powpy with independently varied hard-scatter QCD scales in the matrix element and the parton shower scales of the initial- and final-state radiation.  The maximum uncertainty of the up or down variations is taken for each model and then added in quadrature.

The uncertainty in the modelling of first hard emission at the matrix element is evaluated using the \powpy sample with the $h_{\mathrm{damp}}$ parameter set to twice its nominal value.
Uncertainties due to the choice of parton shower, hadronisation and underlying-event model are evaluated using the \powhwseven sample. %
The uncertainty in the choice of matching between matrix element and parton shower is evaluated using the alternative \ppyeight $p_{\mathrm{T}}^{\mathrm{hard}}$ sample. %
The uncertainty in the modelling of second and subsequent gluon radiation from $b$-quarks in $t\rightarrow bW$ decay is evaluated by using the \ppyeight \texttt{recoilToTop} sample that predicts slightly different energy clustering around the \bjets.  The uncertainty due to the choice of PDF is evaluated following the PDF4LHC prescription~\cite{Butterworth:2015oua}. %

An additional uncertainty is assigned due to discrepancies in the simulated top and anti-top quark \pt and $m_{t\bar{t}}$ spectra of \ppyeight relative to the predictions from the NNLO QCD + NLO EW theory calculations~\cite{Czakon_2017}. The \ppyeight events are first reweighted iteratively to match the parton-level distributions of the \pt of top and anti-top quark and the $m_{t\bar{t}}$ distribution, and then the flavour composition scale factors are extracted using the nominal fit set-up as discussed in Section~\ref{subsec:flavCompCorrections}.  The data distributions are unfolded using these reweighted and corrected distributions, the results are then compared to the nominal unfolded data distributions and the relative difference is taken as the uncertainty.

\subsection{Background modelling uncertainties}
\label{subsec:BkgUnc}

The uncertainties associated with the background estimations are propagated to the unfolded distributions. These are evaluated by independently varying the estimates upward or downward by their uncertainties before subtracting them from the data distributions. The unfolded distribution obtained from the varied background-subtracted detector-level data distribution is compared with the nominal unfolded data distribution and the relative difference in each bin of the distribution is taken as the systematic uncertainty.

The uncertainty in the number of single-top-quark events due to the interference between \ttbar and $tW$ amplitudes is evaluated by comparing the nominal $tW$ background prediction (diagram-removal scheme) with an alternative sample generated with the diagram-subtraction scheme~\cite{Frixione:2008yi}. The uncertainties due to QCD scales $\mu_{\mathrm{R}}$ and $\mu_{\mathrm{F}}$ dependence in the matrix elements and due to the scale dependence of the initial- and final-state radiation predicted by the parton shower model is taken into account. The uncertainties in parton shower and hadronisation models are evaluated by comparing the nominal estimate to that obtained using the \powhwseven sample. The matching uncertainty in the matrix element prediction is evaluated by comparing the nominal estimate to that obtained from \mcnlopy. %

The uncertainties in \ttH predictions are evaluated by independently varying the $\mu_{\mathrm{R}}$ and $\mu_{\mathrm{F}}$ scales in the matrix element by factors of 0.5 and 2.  The $\mu_{\mathrm{R}}^{\mathrm{ISR}}$ scale for the initial-state radiation produced by the parton shower is varied using the \texttt{Var3c} variations.  The $\mu_{\mathrm{R}}^{\mathrm{FSR}}$ scale uncertainty for the final-state radiation in the parton shower is evaluated by varying this scale by 0.625 or 2.  The parton shower, hadronisation and underlying-event uncertainties are taken from comparisons with the \powhwseven sample and the NLO matrix element matching uncertainty is evaluated using \mcnlopy. Additionally, the $10\%$ theoretical uncertainty in the inclusive cross-section~\cite{deFlorian:2016spz}, is taken into account.

A normalisation uncertainty of $12\%$ due to theoretical uncertainty in the inclusive cross-section calculations of \ttV processes is considered similar to that reported in Ref.~\cite{TOPQ-2018-08,TOPQ-2020-20,CMS-TOP-18-009}. The $\mu_{\mathrm{R}}$ and $\mu_{\mathrm{F}}$ QCD scales in matrix element are varied by factors of 0.5 and 2 to evaluate the shape uncertainty in kinematic distributions. An additional conservative uncertainty of $30\%$ is applied due to the modelling of the matrix element matching with the parton shower algorithm based on the comparisons of different MC generator predictions in various kinematics distributions as given in Ref.~\cite{TOPQ-2018-08,TOPQ-2019-30,TOPQ-2020-20}.  %

The uncertainty in the non-prompt lepton background is obtained by varying the transfer factor, as discussed in Section~\ref{subsec:fakelepbkg}, by 25\%--30\% depending on the leading lepton \pt and the multiplicity of \bjets to account for the effects of showering, hadronisation model and matrix-element matching.  Furthermore, the modelling uncertainty in the same-sign prompt event yields are evaluated by comparing the nominal MC prediction with those from the alternative MC samples.  An additional $6\%$ uncertainty is attributed to the same-sign prompt lepton estimate from the simulation, which is significantly affected by the electron charge misidentification. This uncertainty is estimated from the overall discrepancy in the data distribution of the same-sign events as compared to the MC estimate.

A $35\%$ uncertainty is attributed to the very small background contribution from $Z$+jets.  This uncertainty is based on the estimates of the normalisation factor of this background in various phase space regions involving heavy-flavour jets.  The normalisation factors differ by $15\%$ to $35\%$ from unity, and a conservative estimate of $35\%$ is taken as an uncertainty in the MC prediction of $Z$+jets background with negligible impact on the measurements~\cite{EXOT-2021-15,HDBS-2020-19}. The uncertainty due to the backgrounds from diboson and other rare processes is evaluated by independently varying their estimates by conservative uncertainties of $50\%$. These uncertainties have no impact on the results.


%

%
\section{Results}
\label{sec:result}

%
The unfolded results in this section are presented as fiducial cross-sections and as normalised fiducial differential cross-sections at the particle level as a function of the \bjet multiplicity, additional $l/c$-jet multiplicity, global event properties, and kinematic variables. Table~\ref{tab:res_xsec} lists the measured fiducial cross-sections for \ttbj production in the phase space with three \bjets ($\geq 3b$), at least three \bjets and one or more $l/c$-jet ($\geq 3b$ $\geq 1l/c$), at least four \bjets ($\geq 4b$) and least four \bjets with one or more $l/c$-jet ($\geq 4b$ $\geq 1l/c$).  Table~\ref{tab:xsec_unc} lists the contributions to the uncertainty in fiducial cross-sections, among which the most precise measurement in the $\geq 3b$ fiducial phase space has an uncertainty of $8.5\%$.
The uncertainties are predominantly systematic, stemming mainly from $b$-tagging, JES, and \ttbar modelling. The precision of the results presented here surpasses that of previous fiducial measurements of \ttbj production using partial 13~\TeV\ ATLAS data~\cite{TOPQ-2017-12}. This improvement is due to increased number of data events and a significant reduction in uncertainties related to parton shower, hadronisation, and underlying event models as implemented in \PYTHIA[8] and \HERWIG[7]. Additionally, the updated matrix element uncertainty recommendations, as detailed in Refs.~\cite{ATL-PHYS-PUB-2023-029,ATL-PHYS-PUB-2024-002,ATL-PHYS-PUB-2020-023}, and improved luminosity calibrations~\cite{DAPR-2021-01} have contributed to this enhancement.
The largest gain in precision results from the upgrade in the MC modelling leading to a reduction in the parton shower, hadronisation and underlying event modelling uncertainty by a factor of two.  The relative size of the uncertainties due to experimental sources such as the JES and $b$-tagging are comparable to those in the previous ATLAS results.  The measured values of fiducial cross-sections are not fully comparable due to slight differences between the particle-level lepton \pt thresholds.

\begin{table}[!htb]
\centering
\caption{Measured and predicted fiducial cross-section results for additional
\bjet production in four phase-space regions. The dashes (--) indicate that the predictions are not available. The differences between the various MC generator predictions are smaller than the size of theoretical uncertainties (20\%--50\%, not presented here) in the predictions.} %

\sisetup{round-mode=figures, round-precision=2}
\begin{tabular}{l
S[table-format=4.1, table-number-alignment=right]
S[table-format=4.1, table-number-alignment=right]
S[table-format=4.1, table-number-alignment=right]
S[table-format=4.1, table-number-alignment=right]
S[table-format=4.1, table-number-alignment=right]
S[table-format=4.1, table-number-alignment=right]
S[table-format=4.1, table-number-alignment=right]
S[table-format=4.1, table-number-alignment=right]
}
\toprule
& \multicolumn{8}{c}{Fiducial cross-sections [fb]} \\
Fiducial phase space  & \multicolumn{2}{c}{$\geq 3b$} & \multicolumn{2}{c}{$\geq 3b$ $\geq 1l/c$} & \multicolumn{2}{c}{$\geq 4b$} & \multicolumn{2}{c}{$\geq 4b$ $\geq 1l/c$} \\
\cmidrule(r){1-1}\cmidrule(lr){2-3} \cmidrule(lr){4-5} \cmidrule(lr){6-7} \cmidrule(lr){8-9}
\multirow{3}{*}{Measured} & \multicolumn{2}{r}{143} & \multicolumn{2}{r}{87}  & \multicolumn{2}{r}{22} & \multicolumn{2}{r}{14} \\
& \multicolumn{2}{r}{$\pm1 \mathrm{~(stat)}$} & \multicolumn{2}{r}{$\pm 1 \mathrm{~(stat)}$} & \multicolumn{2}{r}{$\pm1 \mathrm{~(stat)}$} & \multicolumn{2}{r}{$\pm 1 \mathrm{~(stat)}$} \\
& \multicolumn{2}{r}{$\pm 12 \mathrm{~(syst)}$} & \multicolumn{2}{r}{$\pm8 \mathrm{~(syst)}$} & \multicolumn{2}{r}{$\pm3 \mathrm{~(syst)}$} & \multicolumn{2}{r}{$\pm2 \mathrm{~(syst)}$} \\
\midrule
\powpy \ttbb (4FS) & \multicolumn{2}{r}{$132$} & \multicolumn{2}{r}{$78$} & \multicolumn{2}{r}{$23$} & \multicolumn{2}{r}{$14$} \\
\powpy \ttbb $h_{bzd}$ (4FS) & \multicolumn{2}{r}{$129$} & \multicolumn{2}{r}{$74$} & \multicolumn{2}{r}{$21$} & \multicolumn{2}{r}{$13$} \\
\powpy \ttbb dipole (4FS) & \multicolumn{2}{r}{$128$} & \multicolumn{2}{r}{$71$} & \multicolumn{2}{r}{$22$} & \multicolumn{2}{r}{$13$} \\
\powpy \ttbb $p_{\mathrm{T}}^{\mathrm{hard}}$ (4FS) & \multicolumn{2}{r}{$129$} & \multicolumn{2}{r}{$68$} & \multicolumn{2}{r}{$21$} & \multicolumn{2}{r}{$12$} \\
\powhwseven \ttbb (4FS) & \multicolumn{2}{r}{$130$} & \multicolumn{2}{r}{$77$} & \multicolumn{2}{r}{$22$} & \multicolumn{2}{r}{$14$} \\
\sherpaOL (4FS) &  \multicolumn{2}{r}{$135$} & \multicolumn{2}{r}{$90$} & \multicolumn{2}{r}{$21$} & \multicolumn{2}{r}{$15$} \\
\hline
\textsc{Helac-NLO} (off-shell) $e\mu+4b$ &  \multicolumn{2}{r}{--} & \multicolumn{2}{r}{--} & \multicolumn{2}{r}{$20$} & \multicolumn{2}{r}{--} \\
\hline
\powpy \ttbar (5FS) & \multicolumn{2}{r}{$120$} & \multicolumn{2}{r}{$74$} & \multicolumn{2}{r}{$18$} & \multicolumn{2}{r}{$11$} \\
\powhwseven \ttbar (5FS) & \multicolumn{2}{r}{$128$} & \multicolumn{2}{r}{$75$} & \multicolumn{2}{r}{$18$} & \multicolumn{2}{r}{$11$} \\
\mcnlopy \ttbar (5FS) & \multicolumn{2}{r}{$122$} & \multicolumn{2}{r}{$72$} & \multicolumn{2}{r}{$18$} & \multicolumn{2}{r}{$11$} \\
\mcnlohwseven \ttbar (5FS) & \multicolumn{2}{r}{$110$} & \multicolumn{2}{r}{$66$} & \multicolumn{2}{r}{$13$} & \multicolumn{2}{r}{$~8$} \\
\SHERPA[2.2.12] \ttbar (5FS) & \multicolumn{2}{r}{$124$} & \multicolumn{2}{r}{$73$} & \multicolumn{2}{r}{$16$} & \multicolumn{2}{r}{$10$} \\

\bottomrule
\end{tabular}


\label{tab:res_xsec}
\end{table}

%
\begin{table}[ht!]
\caption{Main systematic uncertainties in percentage for particle-level
measurement of fiducial cross-sections in the $\ge 3b$, $\ge 3b$ $\ge 1l/c$, $\ge 4b$, and $\ge 4b$ $\ge 1l/c$ phase
space.}
\centering
\sisetup{table-number-alignment=right}
\begin{tabular}{l
S[table-format=2.1, table-number-alignment=right]
S[table-format=2.1, table-number-alignment=right]
S[table-format=2.1, table-number-alignment=right]
S[table-format=2.1, table-number-alignment=right]
rrrr}
\toprule
Source  & \multicolumn{4}{c}{Fiducial cross-section phase space} \\
\cmidrule(r){2-3}\cmidrule(l){4-5}
& {$\ge 3 b$}   & {$\ge 3b$ $\ge 1l/c$}    & {$\ge 4b$}   & {$\ge 4b$ $\ge 1l/c$}   \\
& {Unc. [$\%$]} & {Unc. [$\%$]} & {Unc. [$\%$]} & {Unc. [$\%$]} \\
\midrule %
Data statistical uncertainty           & 1.0         & 1.2         & 3.9         & 4.8         \\
\midrule %
Luminosity                & 0.8        & 0.8        & 0.8        & 0.8         \\
Jet                       & 3.4         & 5.2         & 6.6         & 8.5         \\
$b$-tagging               & 5.1         & 4.9         & 6.5         & 6.4         \\
Lepton and trigger        & 1.4         & 1.4         & 1.2         & 1.2         \\
Pile-up                   & 0.9         & 0.7         & 0.6         & 0.3         \\
$\ttc / \ttl$ fit variation  & 1.7      & 1.7         & 0.8         & 0.8  \\
$\ttc / \ttl$ shape variation  & 0.2    & 0.5         & 0.3         & 1.6  \\
$\ttH / \ttV$ and non-\ttbar background  & 1.1        & 1.1         & 2.2         & 2.4         \\
\midrule %
Detector+background total syst. & 6.7   & 7.6         & 9.7         & 11.2          \\
\midrule %
Parton shower and hadronisation             & 2.9         & 3.5         & 1.5         & 3.6         \\
$\mu_{\mathrm{R}}$ and $\mu_{\mathrm{F}}$ scale variations & 0.7    & 0.6         & 0.2         & 0.3         \\
Matrix element matching ($p_{\mathrm{T}}^{\mathrm{hard}}$)  & 1.3         & 1.1         & 4.8         & 7.0        \\
$h_{\mathrm{damp}}$       & 1.8         & 1.5         & 2.9         & 3.2         \\
ISR                       & 0.1         & 0.4         & 0.2         & 0.3         \\
FSR                       & 3.1         & 3.6         & 3.3         & 3.1         \\
RecoilToTop               & 1.8         & 1.9         & 2.4         & 3.4         \\
PDF                       & 0.2         & 0.2         & 0.1         & 0.1         \\
NNLO reweighting             & 0.6         & 0.5         & 0.5         & 0.5         \\
MC statistical uncertainty      & 0.2         & 0.2         & 0.5         & 0.6         \\
\midrule %
\ttbar modelling total syst. & 5.2      & 5.7         & 7.2         & 9.7         \\
\bottomrule %
\toprule %
Total syst.               & 8.5          & 9.6        & 12.1          & 14.8       \\
Total                     & 8.5          & 9.6        & 12.7          & 15.5       \\
\bottomrule %
\end{tabular}
\label{tab:xsec_unc}
\end{table}

\begin{figure}[htb]
\centering
\includegraphics[width=1.0\textwidth]{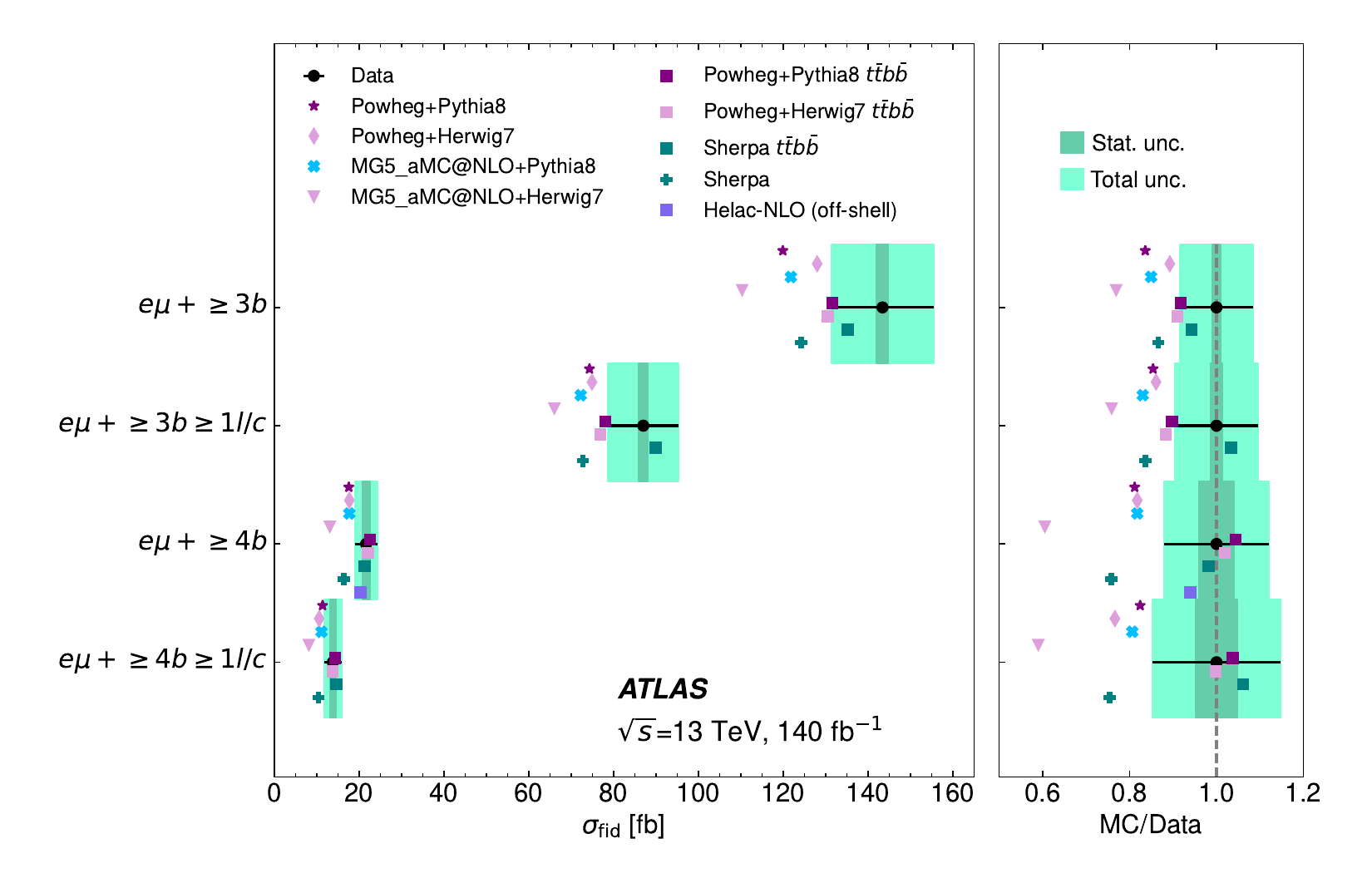}
\caption{Measured fiducial cross-sections for \emuthreeb, \emuThreebOnelight, \emufourb, and \emuFourbOnelight compared with the central values of \ttbb NLO predictions from \powpy \ttbb, \powhwseven \ttbb, \sherpaOL, and the \textsc{Helac-NLO} (off-shell) calculations.  %
Comparisons with \ttbar NLO MC predictions from \powpy, \powhwseven, \amcnlopyeight, \mcnlohwseven, and \sherpat are also made.
The right hand panel shows the ratios of the MC predictions to the data.
The inner uncertainty band (dark) is the statistical uncertainty in the data and the outer band (light) includes all uncertainties due to the instrumental and theoretical sources.  The differences between the various MC generator predictions are smaller than the size of theoretical uncertainties (20\%--50\%, not shown) in the predictions.}
\label{fig:xsec_summary}
\end{figure}

The achieved precision in the fiducial measurements is better than the current uncertainties in the theoretical predictions of fiducial cross-sections, which typically range from 20\% to 50\%.  The results are summarised in Figure~\ref{fig:xsec_summary}, and compared with \ttbb NLO MC predictions from \powpy \ttbb, \powhwseven \ttbb, \sherpaOL, and the \textsc{Helac-NLO} $e\mu+4b$ calculations including the offshell effects, as well as with the \ttbar NLO MC predictions from \powpy, \powhwseven, \amcnlopyeight, \mcnlohwseven, and \sherpat.
Predictions that calculate the \ttbar production matrix element at NLO, but rely on the parton shower for the high jet multiplicities (\powpy, \powhwseven, \amcnlopyeight, \mcnlohwseven) show slightly lower rates of events with three or four \bjets as compared to those where the additional massive $b$-quarks are included in the matrix element calculations and all the additional radiations are produced at LO accuracy in the parton shower model (\powpy \ttbb, \powhwseven \ttbb, \sherpaOL).
The \sherpat (5FS), which models the \ttbar pair with one parton at NLO accuracy and up to four additional partons at LO accuracy, also gives a relatively low rate of events with three or four \bjets. %
The differences between the MC generator predictions are smaller than the theoretical uncertainties in the predictions. %

The \ttbb (4FS) predictions (\powpy \ttbb, \powhwseven \ttbb, \sherpaOL) describe the data very well in the $\geq 4b$ phase spaces.  Slightly smaller cross-sections are predicted by \powpy \ttbb and \powhwseven \ttbb models in the $\geq 3b$ fiducial phase spaces as shown in Figure~\ref{fig:xsec_summary}, however, they are still consistent within the measurement uncertainties.  %
The \textsc{Helac-NLO} $e\mu+4b$ fixed order parton-level cross-section calculations that includes off-shell effects are compared after correcting for the non-perturbative effects such as multiple parton interactions, beam remnants and hadronisation.  These corrections are evaluated by comparing the generator level distribution obtained from the \powpy \ttbb MC between set-ups with or without turning off the non-perturbative effects.  The size of these corrections reach the level of $10\%$ depending on the observable values.  %

The central values of the measured fiducial cross-sections remain unaffected if the $\geq 3b$ particle-level events in the \powpy (5FS) \ttbar sample are replaced by those in the \powpy \ttbb (4FS) sample for the unfolding corrections.  The precision of the measurements depends significantly on the modelling of misidentified background from \ttj %
events. With the improvement in the background MC modelling, various uncertainties arising from the relative flavour composition corrections in the MC samples are reduced.  Theoretical developments in the modelling of \ttj for all additional-jet flavours at the higher orders in QCD will benefit future measurements.  %

\begin{figure}[htbp]
\centering
\includegraphics[width=0.7\textwidth]{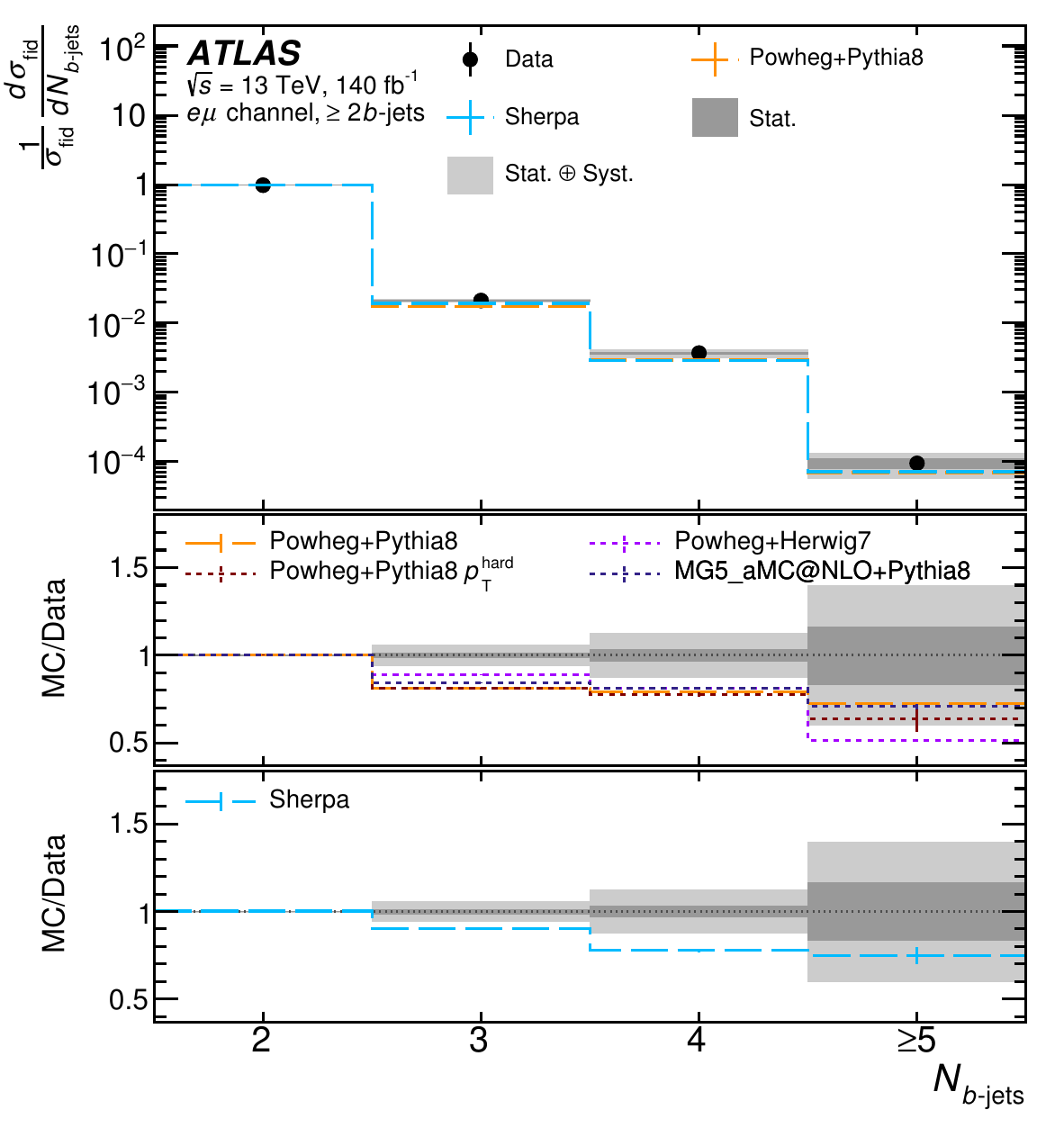}
\caption{Measured normalised differential cross-section in the phase space with at least two \bjets as a function of the number of \bjets compared with predictions. The lower panels show the ratios of various predictions to the data. The shaded regions show the statistical (dark) and total uncertainties (light) in the measurement. The vertical line on the MC predictions represents the statistical uncertainty.  The last bin contains the overflow.}
\label{fig:unfolded_2j2b_global}
\end{figure}

\begin{figure}[htbp]
\centering
\subfloat[]{\includegraphics[width=0.45\textwidth]{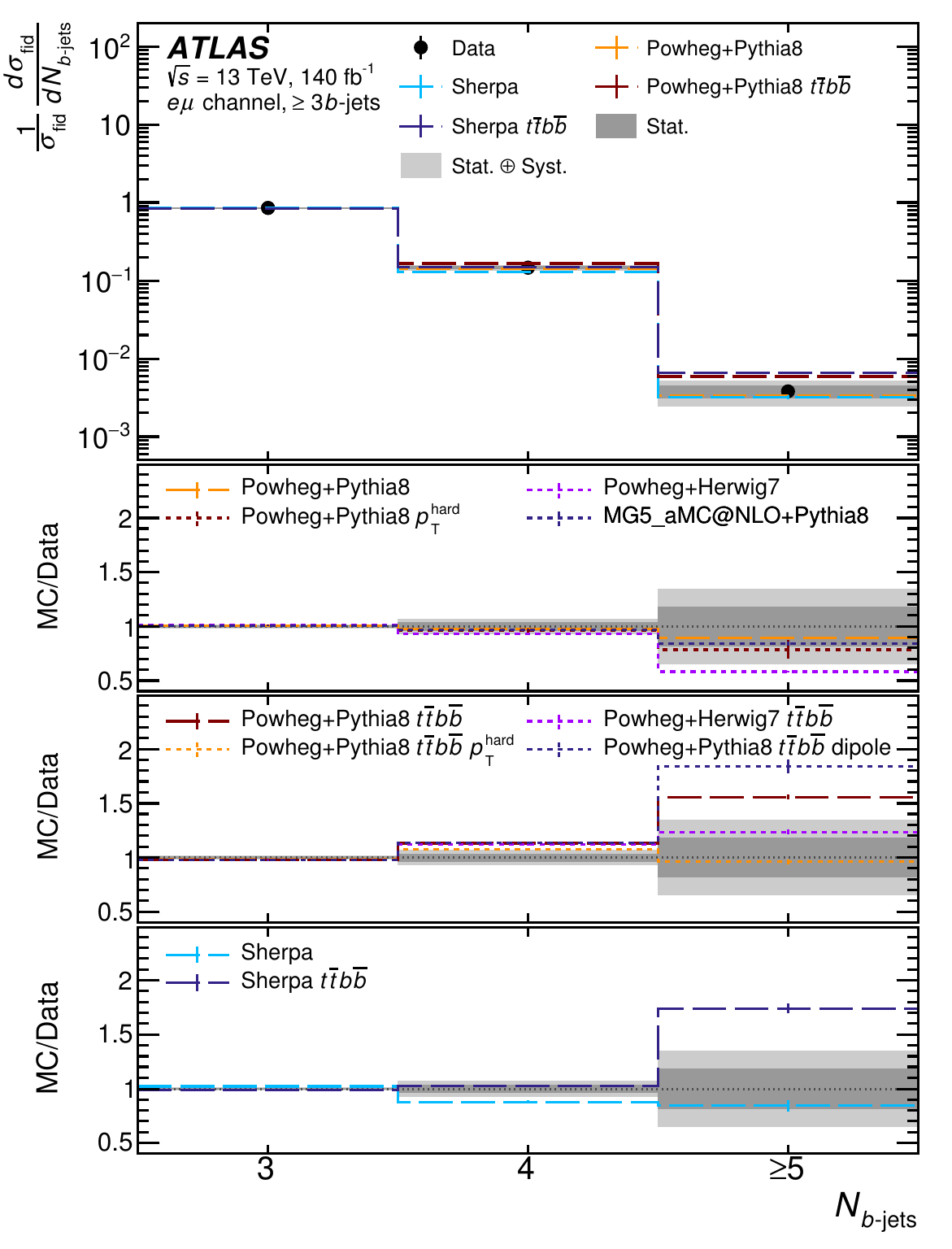}\label{fig:3j3b_nbjets}}
\subfloat[]{\includegraphics[width=0.45\textwidth]{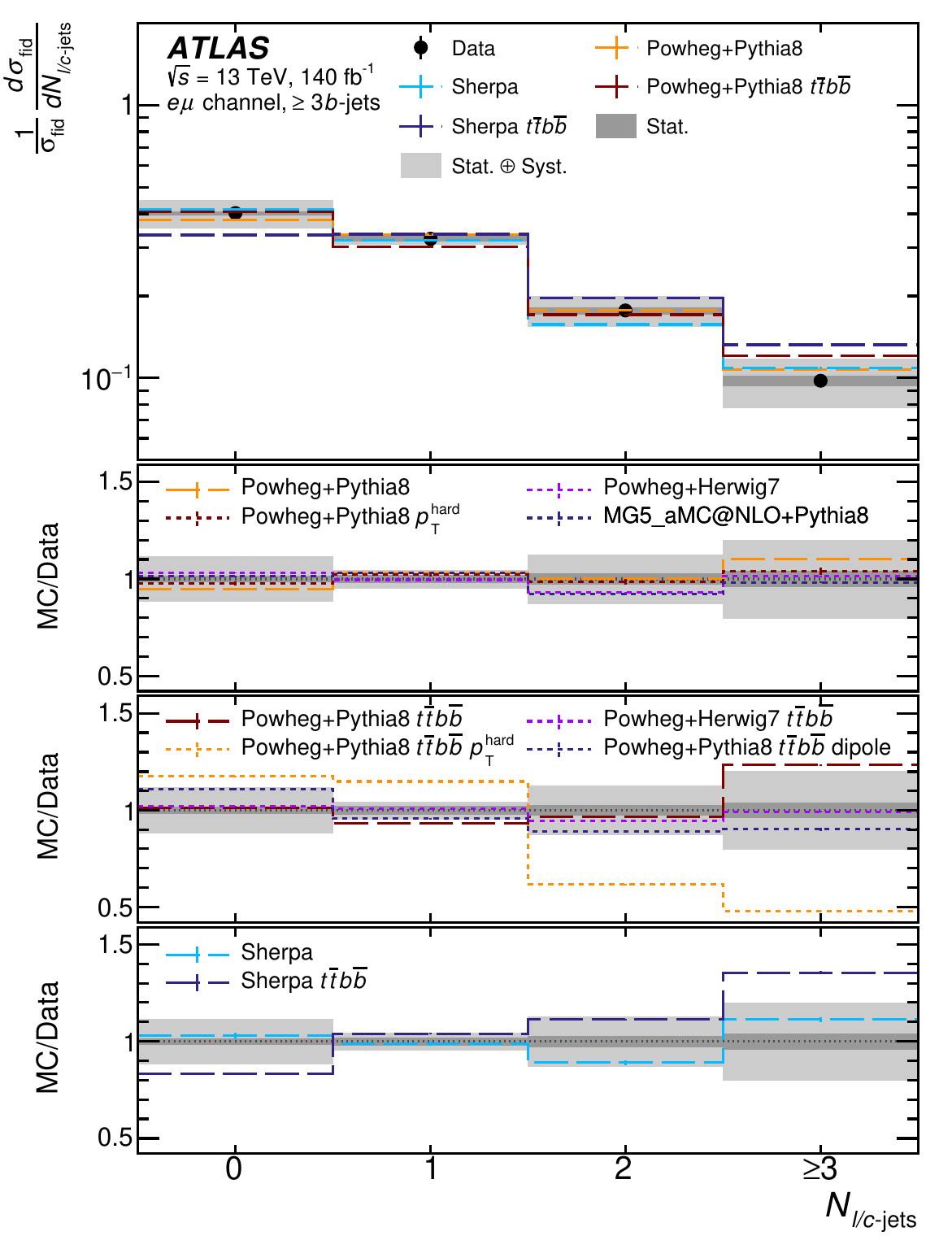}\label{fig:3j3b_nlight}}\\
\subfloat[]{\includegraphics[width=0.45\textwidth]{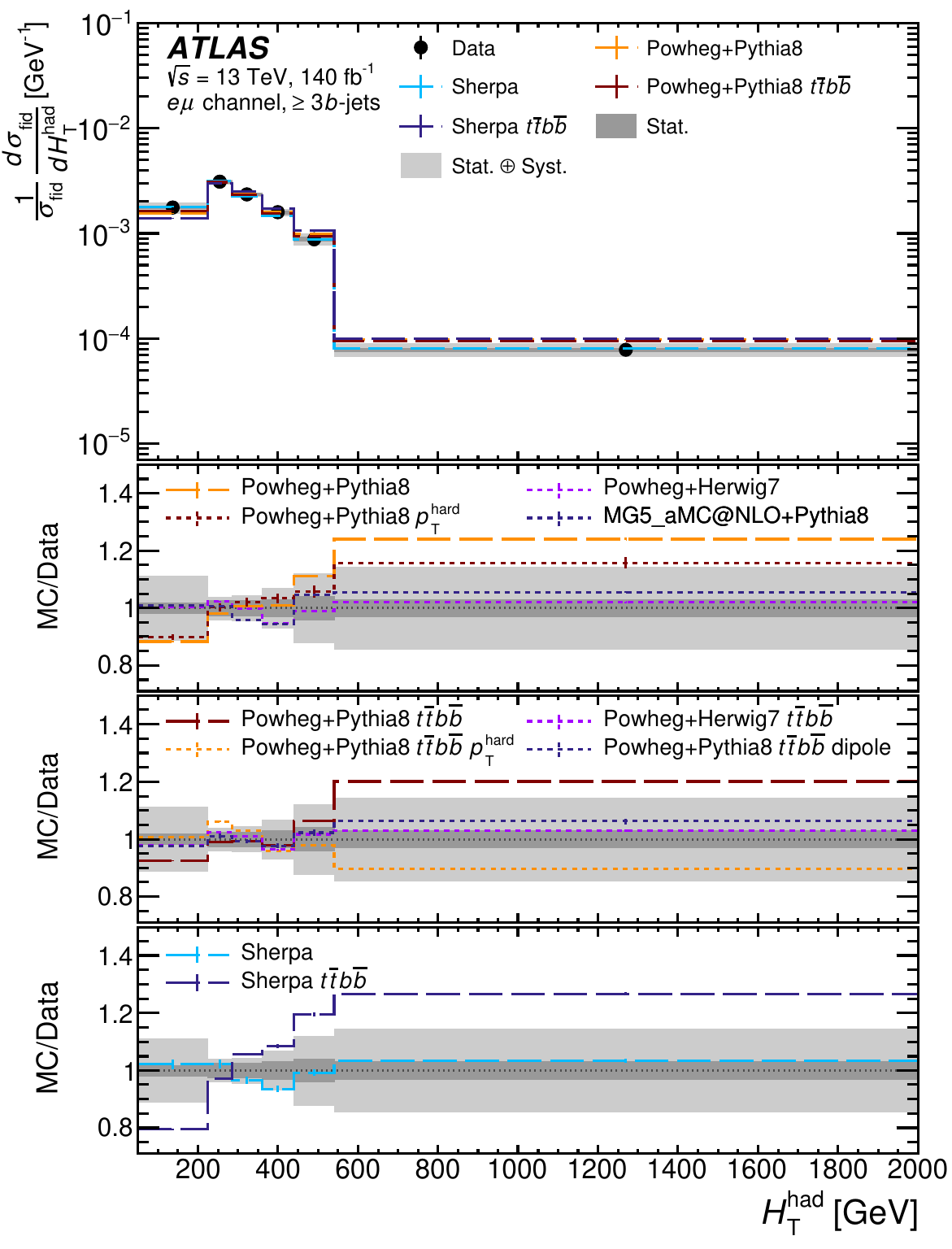}\label{fig:3j3b_HThad}}
\subfloat[]{\includegraphics[width=0.45\textwidth]{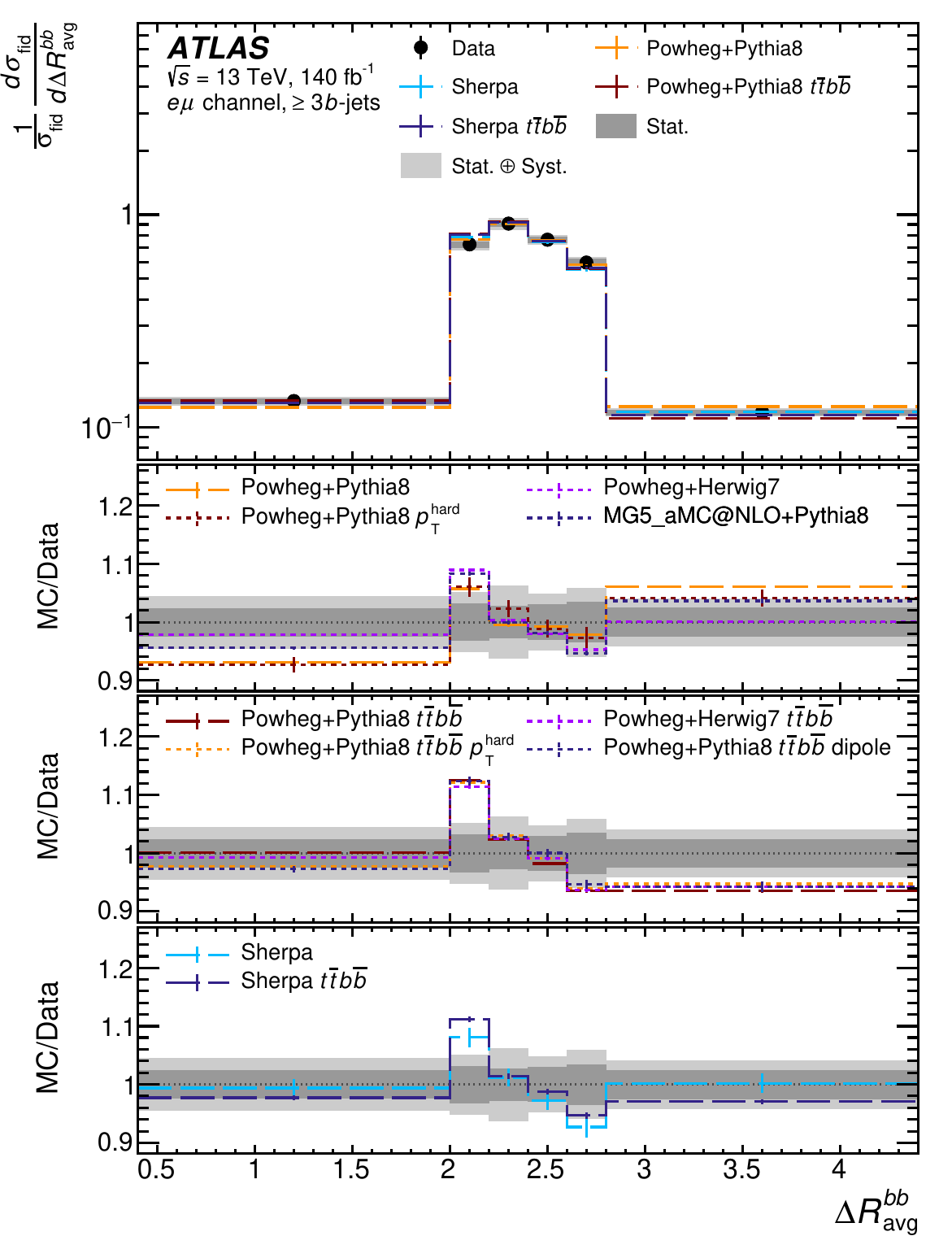}\label{fig:3j3b_dRbb}}\\
\caption{Measured normalised differential cross-section in the phase space with at least three \bjets as a function of \protect \subref{fig:3j3b_nbjets} \bjet multiplicity, \protect \subref{fig:3j3b_nlight} $l/c$-jet multiplicty, \protect \subref{fig:3j3b_HThad} $H_{\mathrm{T}}^{\mathrm{had}}$, and \protect \subref{fig:3j3b_dRbb} $\Delta R_{\text{avg}}^{bb}$ compared with predictions.  The lower panels show the ratios of various predictions to the data. The shaded regions show the statistical (dark) and total uncertainties (light) in the measurement.  The vertical line on the MC predictions represents the statistical uncertainty.  The last bin contains the overflow.}
\label{fig:unfolded_3j3b_global}
\end{figure}

Figure~\ref{fig:unfolded_2j2b_global} shows the measured fiducial normalised distribution of the \bjet multiplicity in the phase space with at least two \bjets and the comparisons with various \ttbar (5FS) predictions. %
This observable indicates the relative fractions of the \ttbj rate with zero, one, two or three additional \bjets relative to the total \ttj rate in the fiducial phase space. All models are consistent with the data for events with exactly two \bjets, but they face challenges in accurately describing the additional \bjets production in events with three or more \bjets. %
Figure~\ref{fig:unfolded_3j3b_global} shows the measured normalised cross-sections as a function of the \bjet and $l/c$-jet multiplicities, $H_{\mathrm{T}}^{\mathrm{had}}$, and $\Delta R_{\text{avg}}^{bb}$ variables, the distributions are normalised to the fiducial cross-section in the phase space with at least three \bjets.
The distributions are compared with various MC predictions obtained at the particle level, and their ratios relative to the measurements are displayed in the lower panels.  The large uncertainties due to scale variations in the matrix elements are expected to largely cancel in the normalised predictions in fiducial phase space with three or more \bjets.  The \ttbb predictions from \powhwseven \ttbb describes the $H_{\mathrm{T}}^{\mathrm{had}}$ and jet multiplicities reasonably well.  The \sherpaOL yields a slightly higher number of events with additional $l/c$-jets.
Distributions such as $\Delta R_{\text{avg}}^{bb}$ show different levels of agreement with data between \ttbb simulations using the 4FS and the 5FS. Although the jets generated by the matrix element generally have higher \pt compared to those produced by the parton shower, the matching treatment between the matrix element and the parton shower results in a significant difference.

\begin{figure}[htbp]
\centering
\subfloat[]{\includegraphics[width=0.45\textwidth]{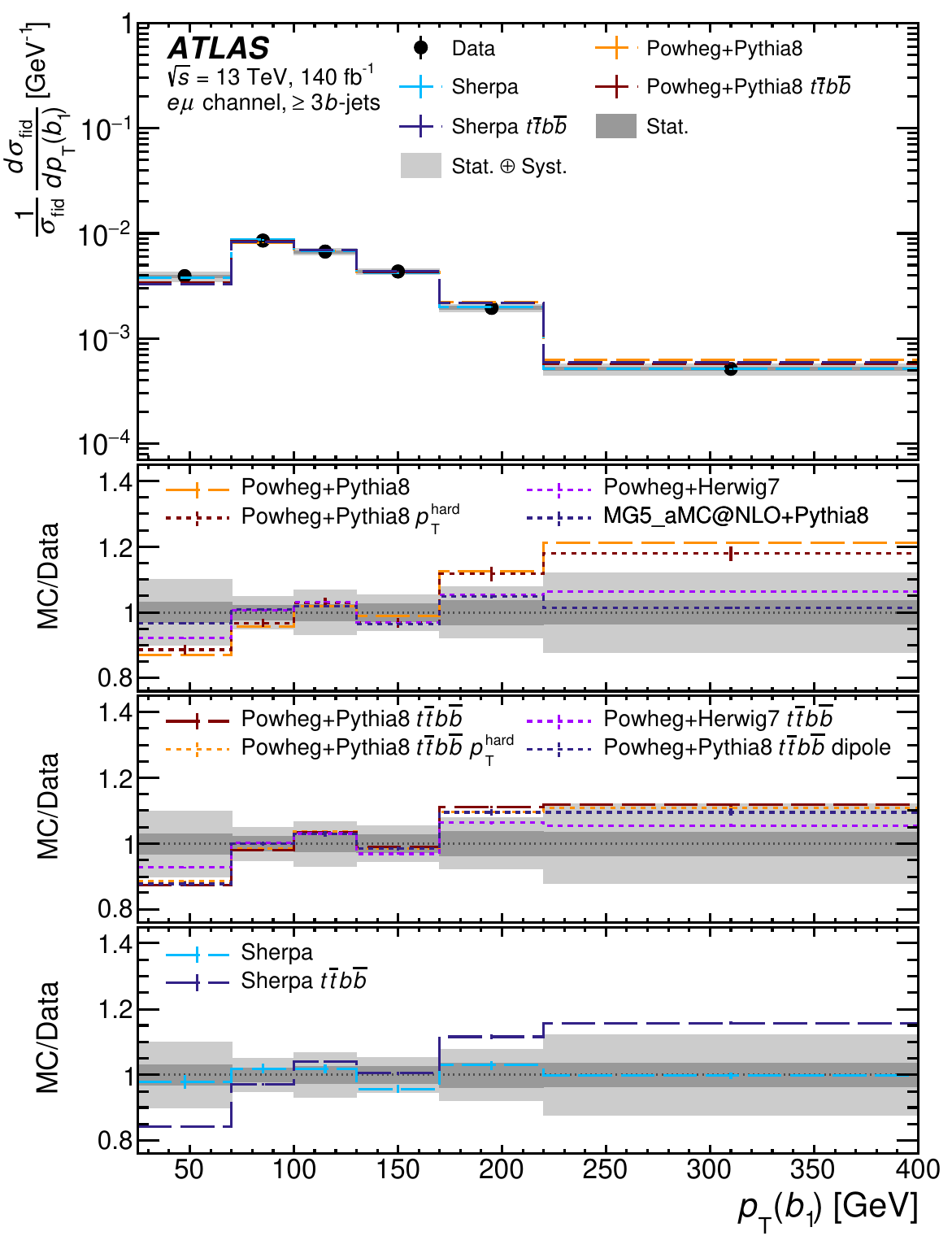}\label{fig:3j3b_b1pt}}
\subfloat[]{\includegraphics[width=0.45\textwidth]{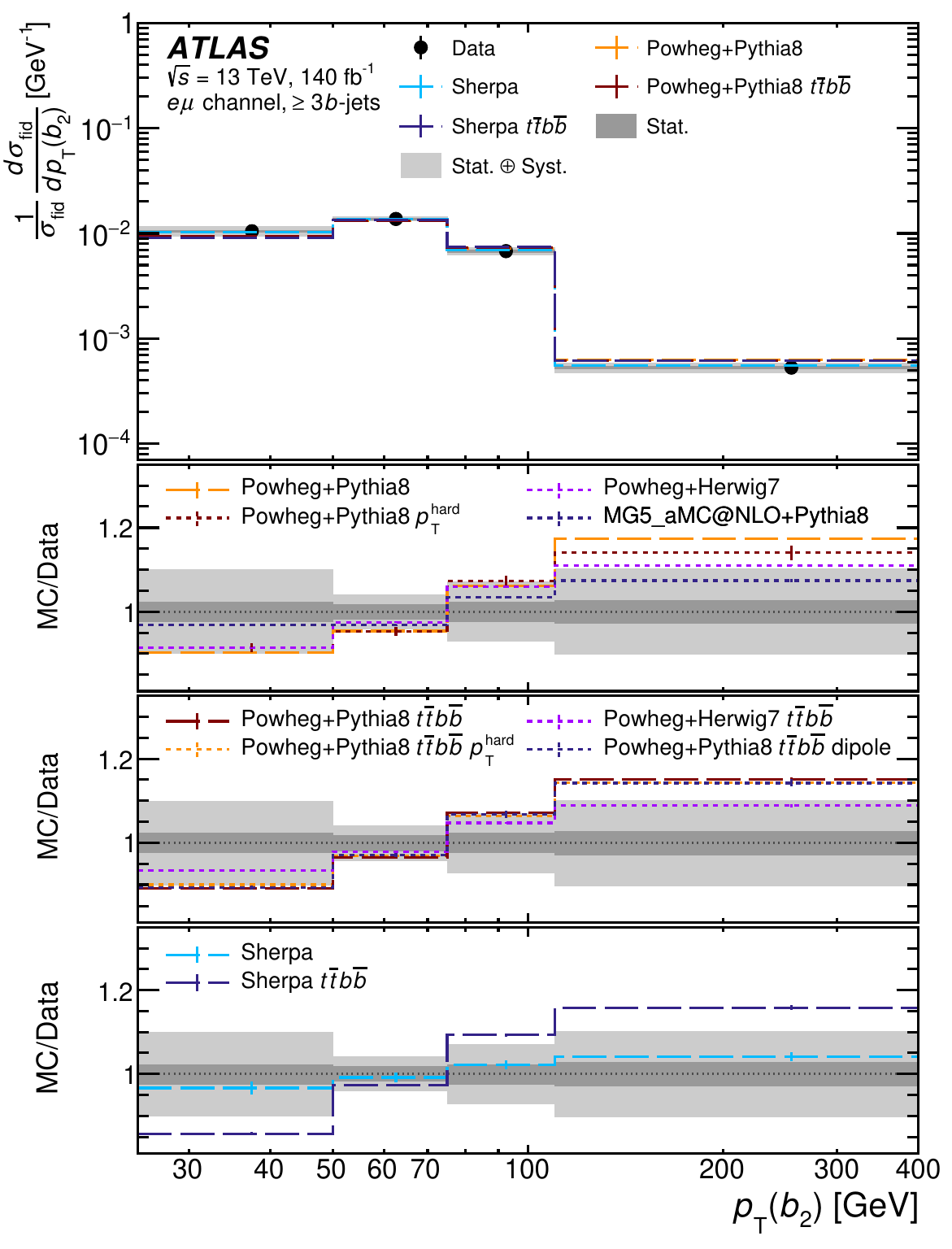}\label{fig:3j3b_b2pt}}\\
\subfloat[]{\includegraphics[width=0.45\textwidth]{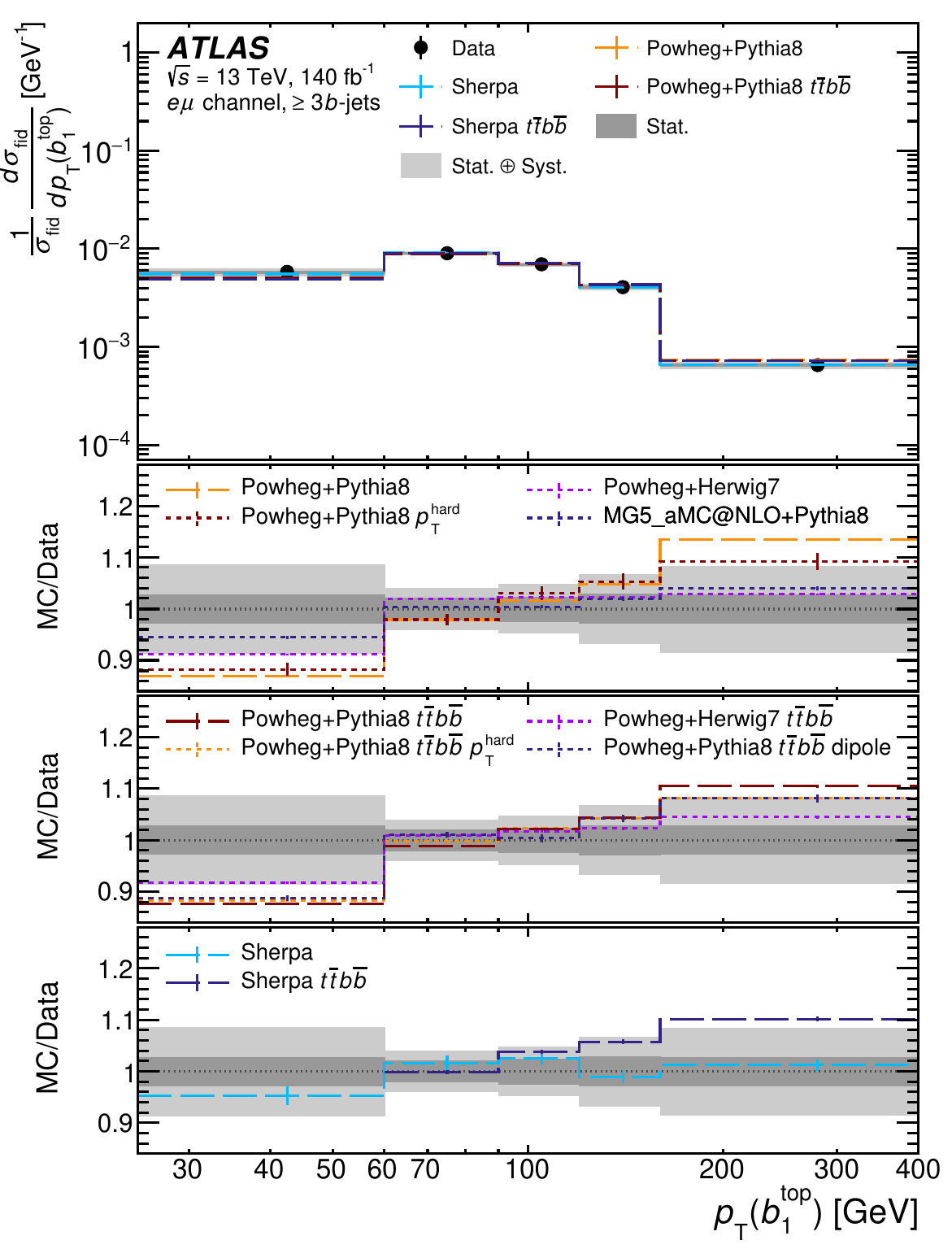}\label{fig:3j3b_topb1pt}}
\subfloat[]{\includegraphics[width=0.45\textwidth]{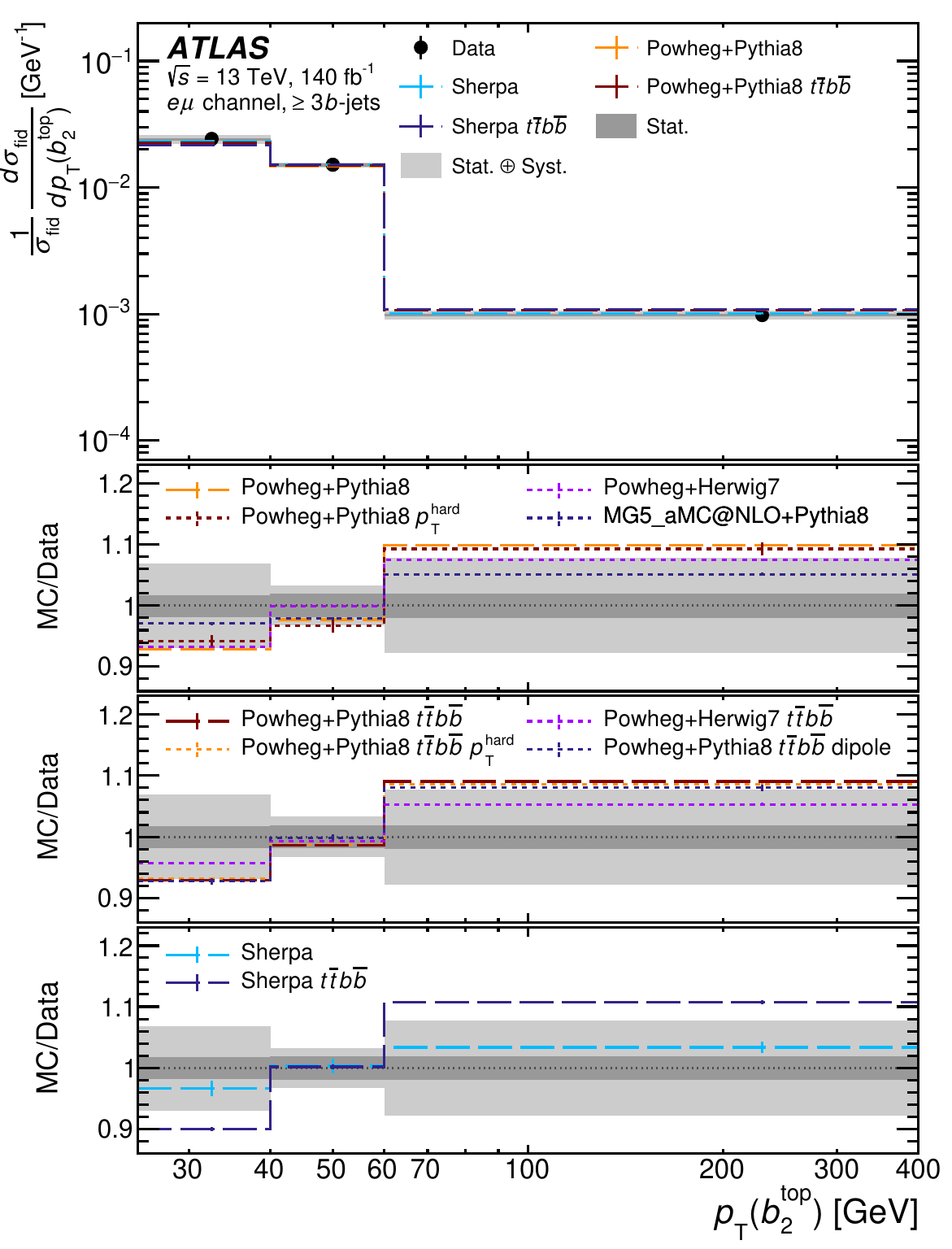}\label{fig:3j3b_topb2pt}}\\
\caption{Measured normalised differential cross-section in the phase space with at least three \bjets as a function of \protect \subref{fig:3j3b_b1pt} $\pt(b_1)$, \protect \subref{fig:3j3b_b2pt} $\pt(b_2)$, \protect \subref{fig:3j3b_topb1pt} $\pt(b_1^{\text{top}})$, and \protect \subref{fig:3j3b_topb2pt} $\pt(b_2^{\text{top}})$ compared with predictions. The lower panels show the ratios of various predictions to the data. The shaded regions show the statistical (dark) and total uncertainties (light) in the measurement. The vertical line on the MC predictions represents the statistical uncertainty.  The last bin contains the overflow.}
\label{fig:unfolded_3j3b_b1Andb2pt}
\end{figure}

\begin{figure}[htbp]
\centering
\subfloat[]{\includegraphics[width=0.45\textwidth]{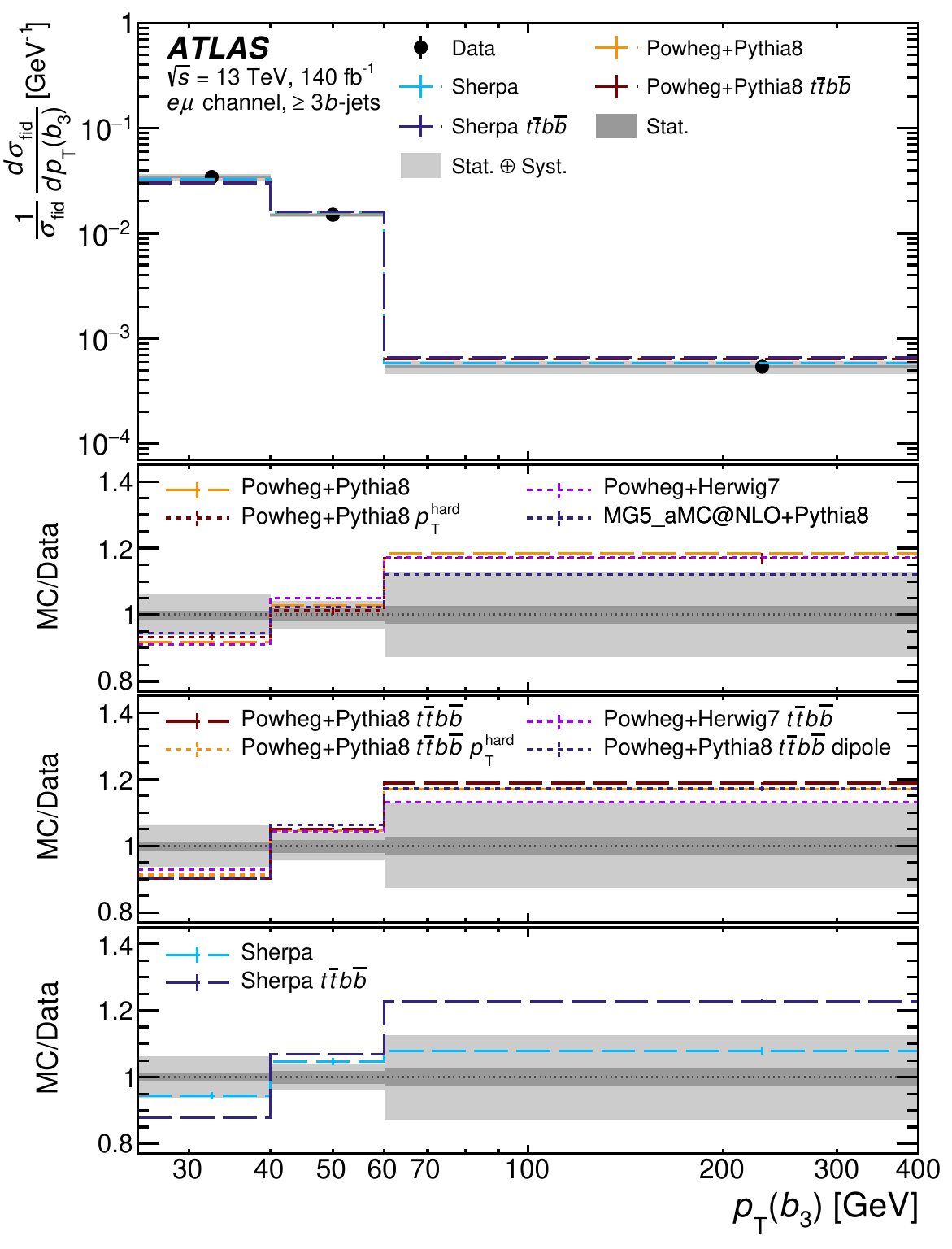}\label{fig:3j3b_b3pt}}
\subfloat[]{\includegraphics[width=0.45\textwidth]{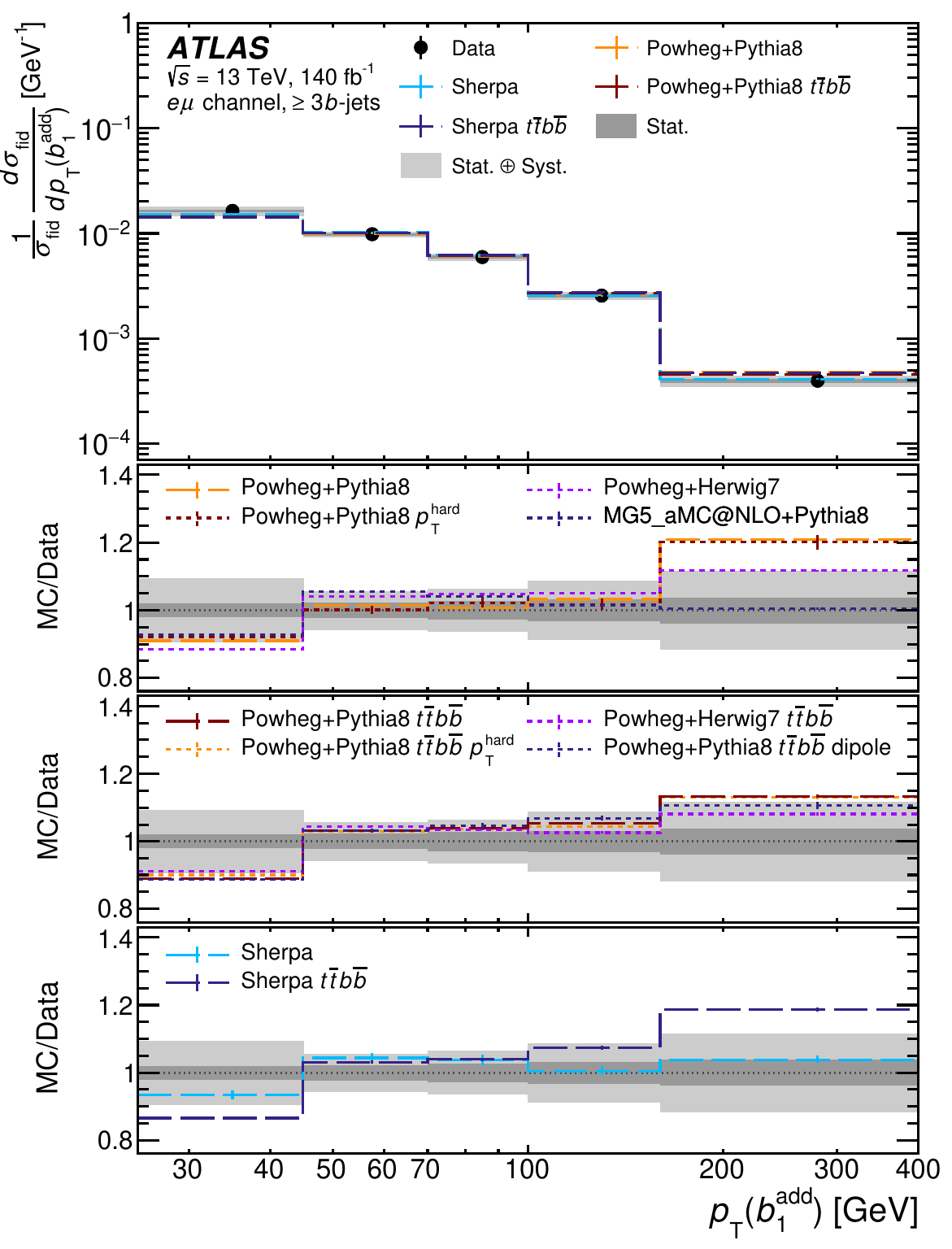}\label{fig:3j3b_b1addpt}}\\
\caption{Measured normalised differential cross-section in the phase space with at least three \bjets as a function of \protect \subref{fig:3j3b_b3pt} $\pt(b_3)$, and \protect \subref{fig:3j3b_b1addpt} $\pt(b_1^{\text{add}})$ compared with predictions.  The lower panels show the ratios of various predictions to the data. The shaded regions show the statistical (dark) and total uncertainties (light) in the measurement. The vertical line on the MC predictions represents the statistical uncertainty.  The last bin contains the overflow.}
\label{fig:unfolded_3j3b_b3Andbadd}
\end{figure}

\begin{figure}[htbp]
\centering
\subfloat[]{\includegraphics[width=0.45\textwidth]{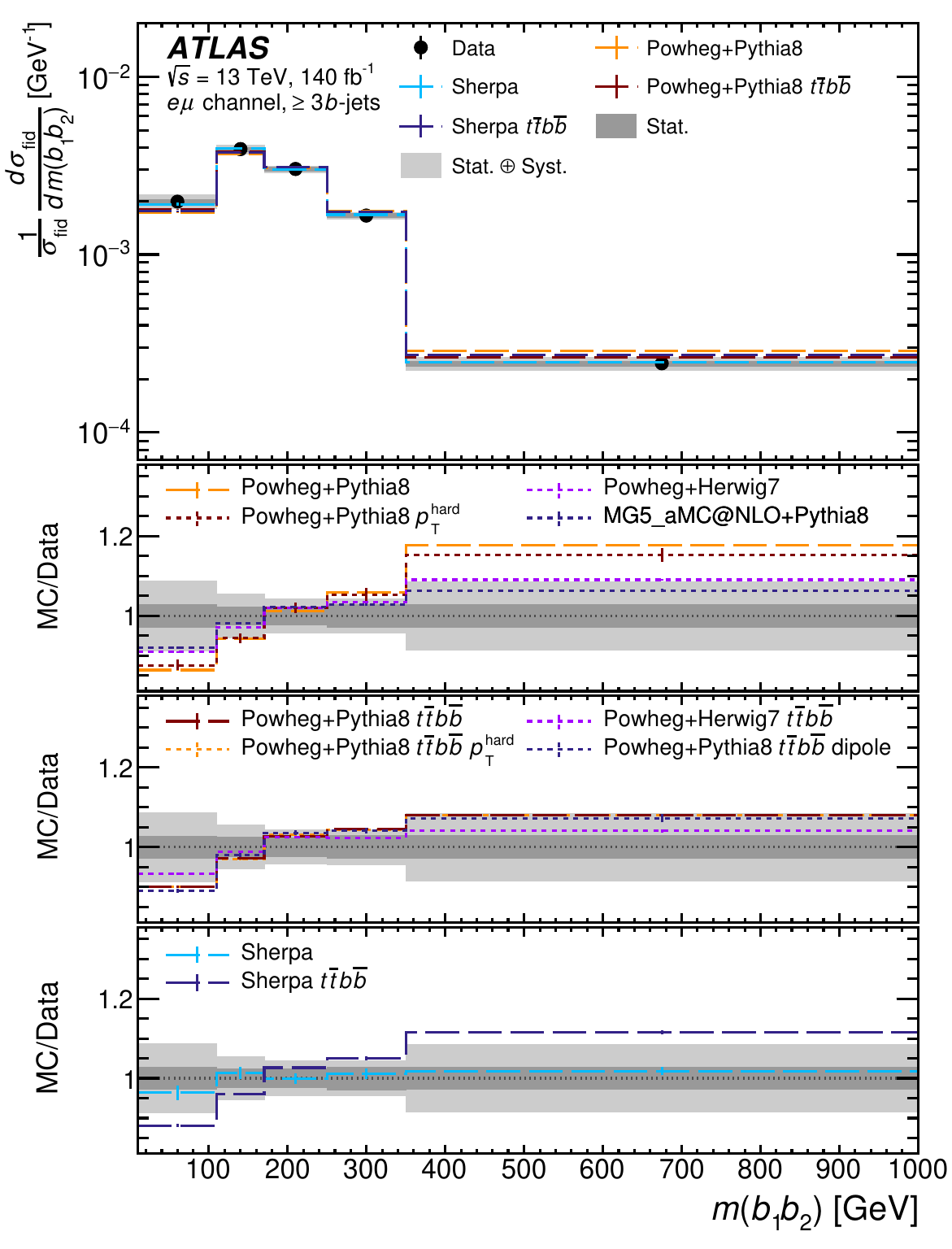}\label{fig:3j3b_mb1b2}}
\subfloat[]{\includegraphics[width=0.45\textwidth]{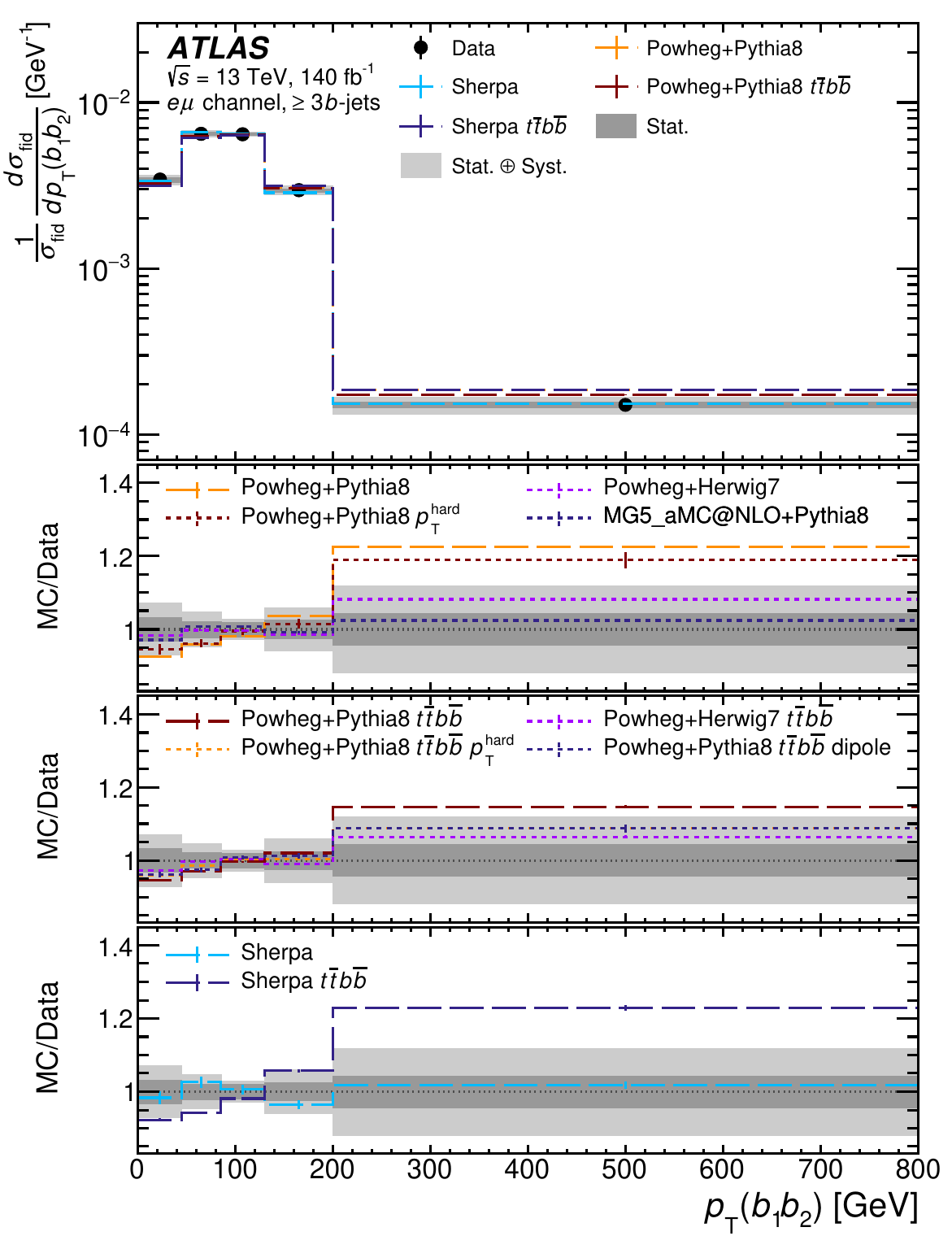}\label{fig:3j3b_pTb1b2}}\\
\subfloat[]{\includegraphics[width=0.45\textwidth]{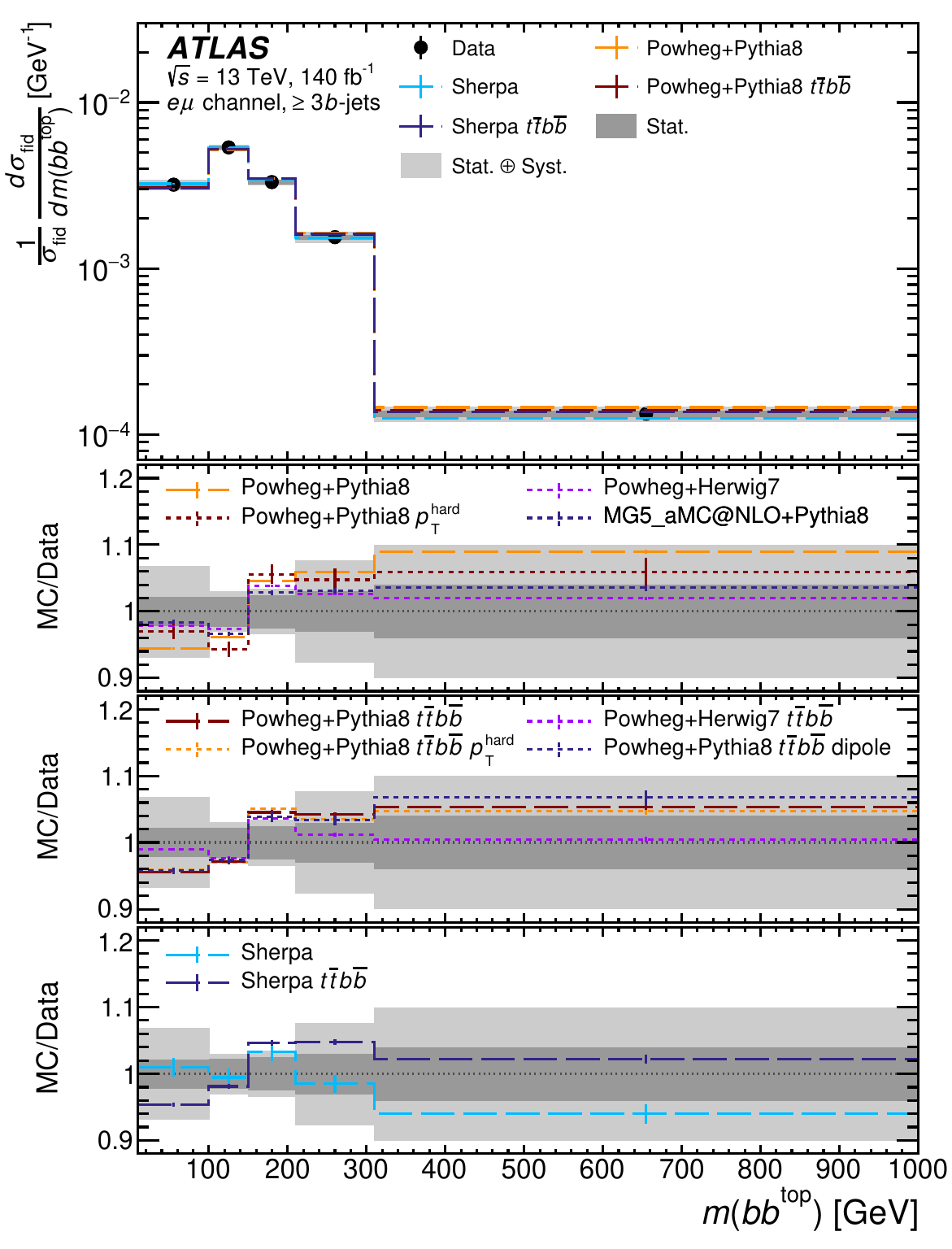}\label{fig:3j3b_Mbb_top}}
\subfloat[]{\includegraphics[width=0.45\textwidth]{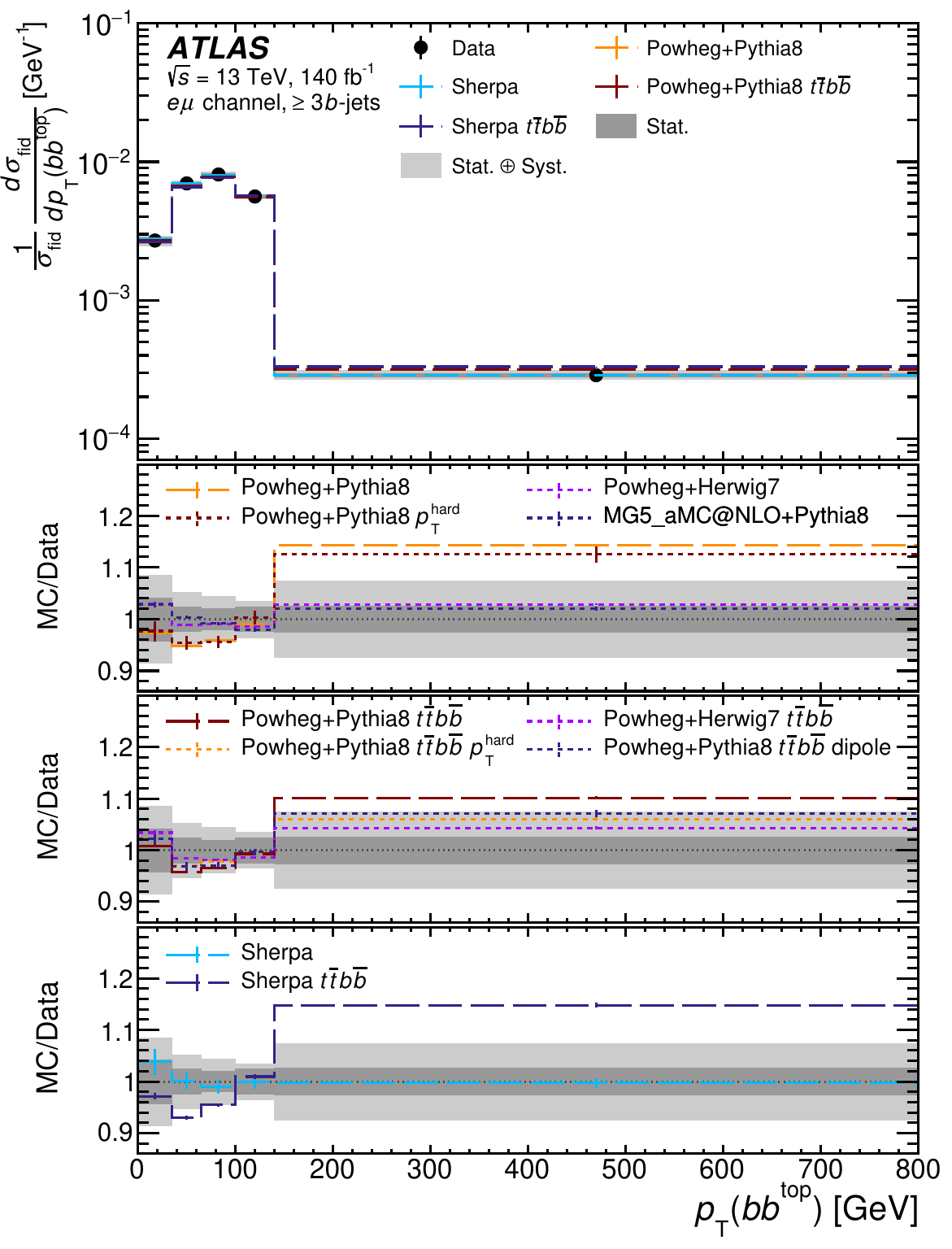}\label{fig:3j3b_pTbb_top}}\\
\caption{Measured normalised differential cross-section in the phase space with at least three \bjets as a function of \protect \subref{fig:3j3b_mb1b2}  $m(b_1b_2)$, \protect \subref{fig:3j3b_pTb1b2} $\pt(b_1b_2)$, \protect \subref{fig:3j3b_Mbb_top} $m(bb^{\text{top}})$, and \protect \subref{fig:3j3b_pTbb_top} $\pt(bb^{\text{top}})$ compared with predictions. The lower panels show the ratios of various predictions to the data. The shaded regions show the statistical (dark) and total uncertainties (light) in the measurement. The vertical line on the MC predictions represents the statistical uncertainty.  The last bin contains the overflow.}
\label{fig:unfolded_3j3b_bb_top}
\end{figure}

\begin{figure}[htbp]
\centering
\subfloat[]{\includegraphics[width=0.45\textwidth]{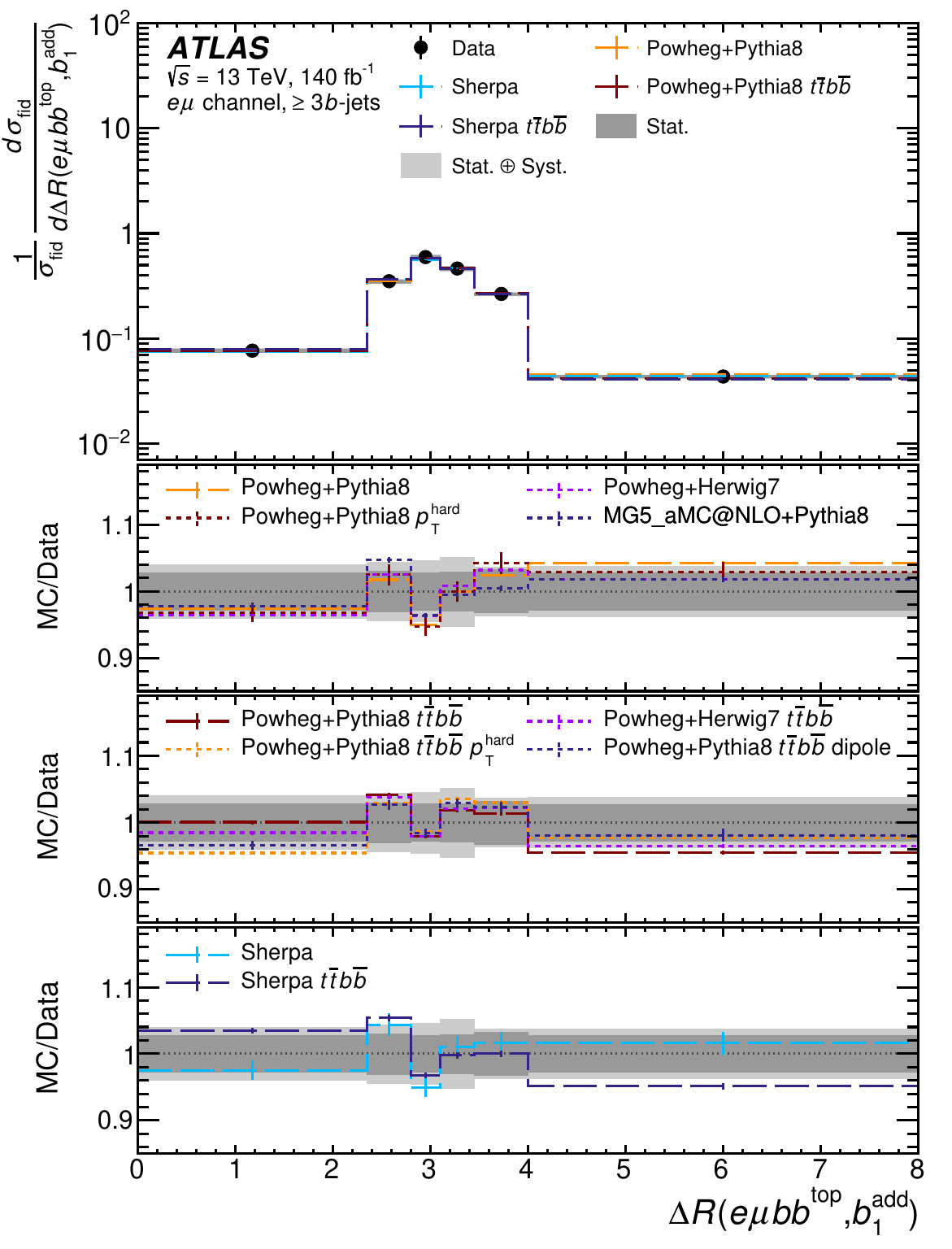}\label{fig:3j3b_dRemubb_ab1}}
\subfloat[]{\includegraphics[width=0.45\textwidth]{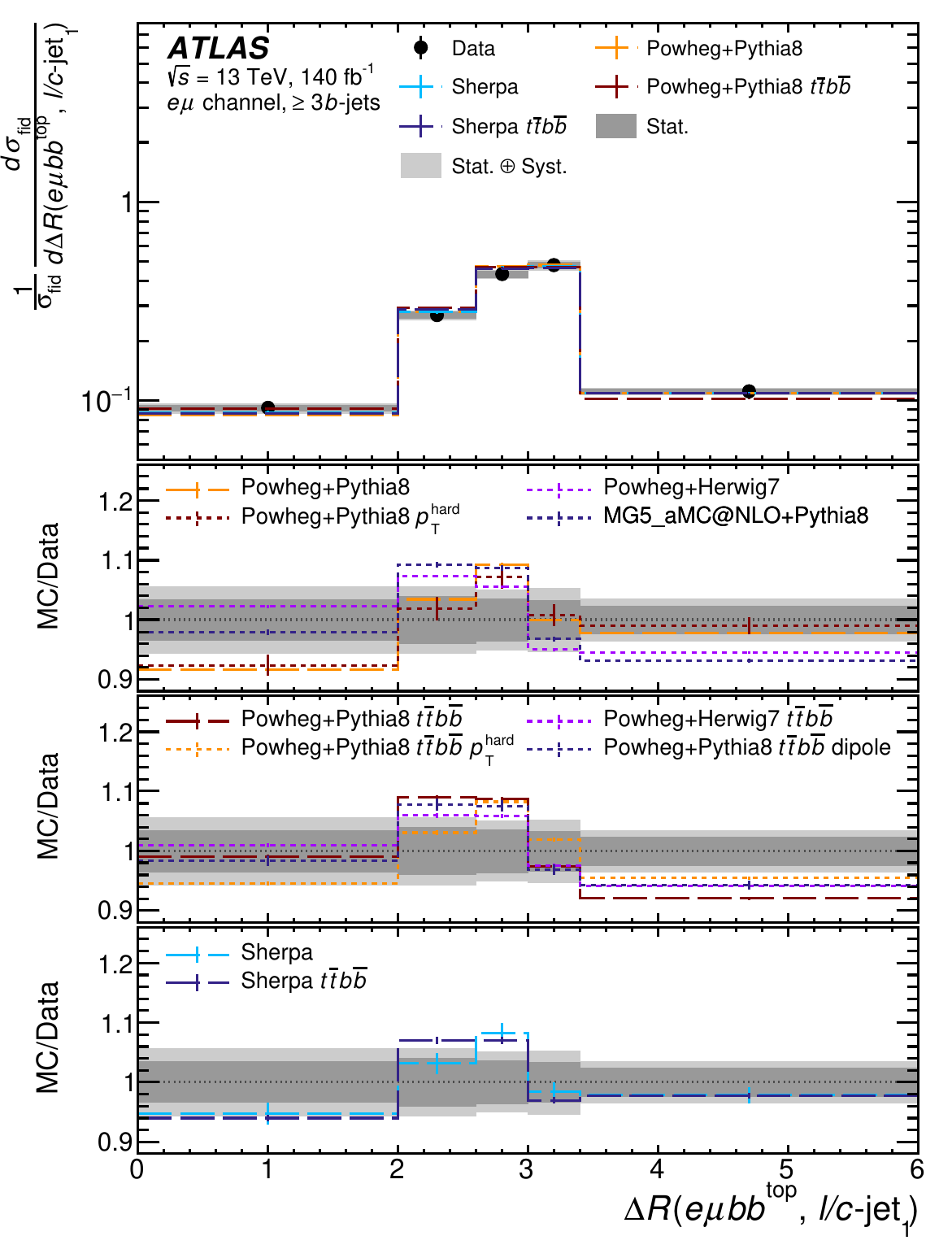}\label{fig:3j3b_dRemubb_alj1}}\\
\subfloat[]{\includegraphics[width=0.45\textwidth]{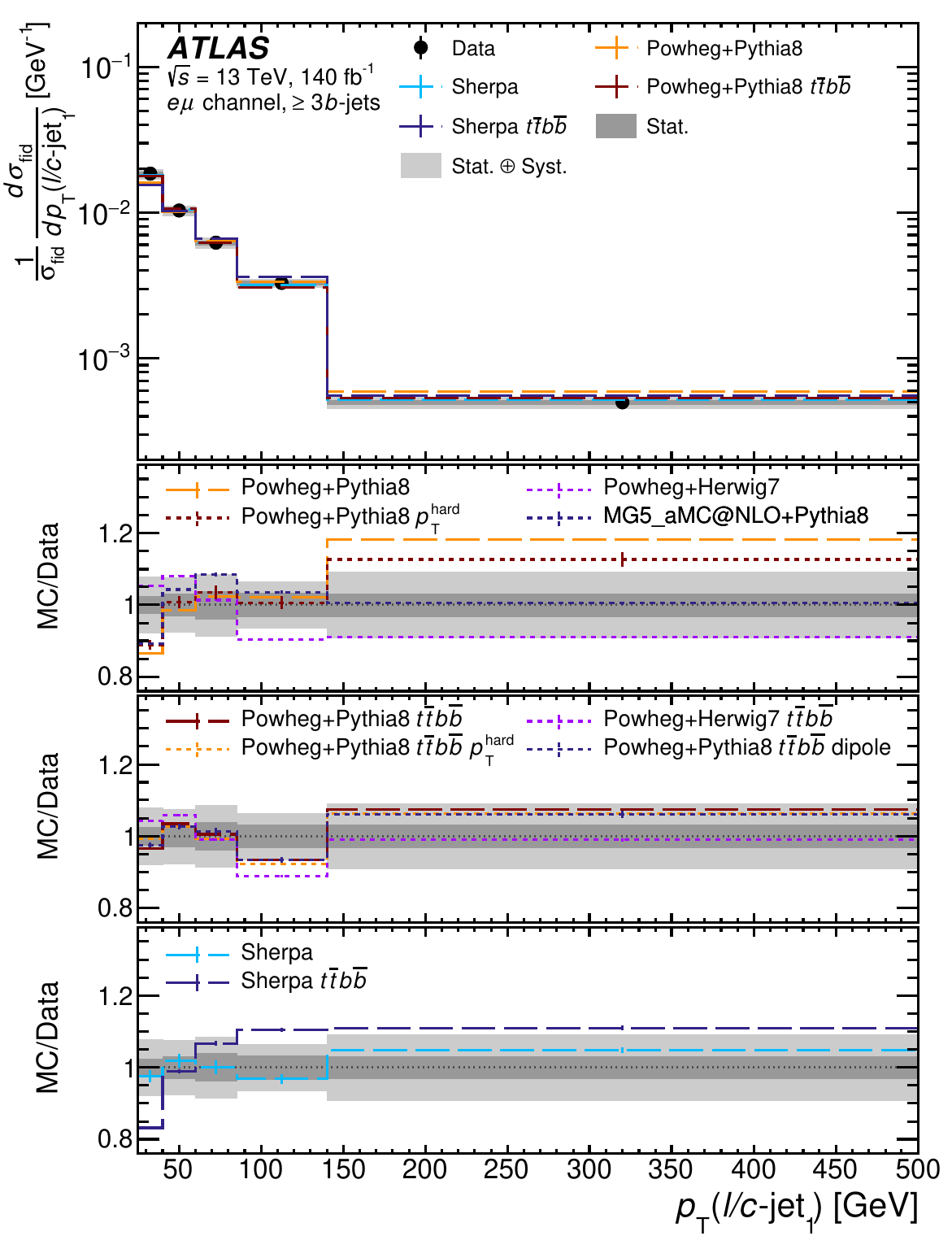}\label{fig:3j3b_lj1Pt}}
\subfloat[]{\includegraphics[width=0.45\textwidth]{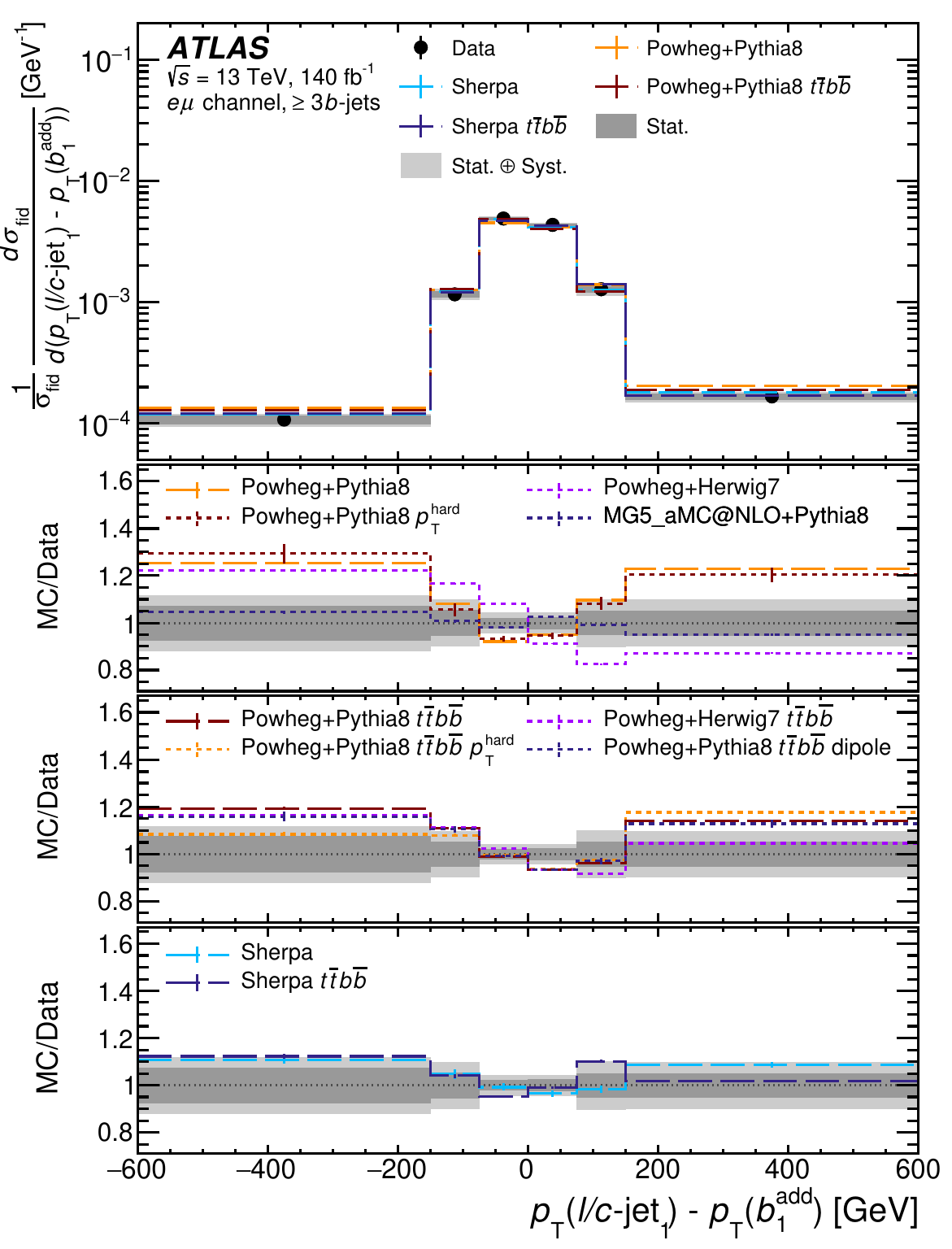}\label{fig:3j3b_ptdiffLjaddjet}}\\
\caption{Measured normalised differential cross-section in the phase space with at least three \bjets as a function of \protect \subref{fig:3j3b_dRemubb_ab1} $\Delta R(e\mu bb^{\mathrm{top}}, b_1^{\mathrm{add}})$, \protect \subref{fig:3j3b_dRemubb_alj1} $\Delta R (e\mu bb^{\text{top}}, l/c\text{-jet}_1)$, \protect \subref{fig:3j3b_lj1Pt} $\pt(l/c\text{-jet}_1)$, and \protect \subref{fig:3j3b_ptdiffLjaddjet} $\pt(l/c\text{-jet}_1) - \pt(b_1^{\text{add}})$ compared with predictions.  The lower panels show the ratios of various predictions to the data. The shaded regions show the statistical (dark) and total uncertainties (light) in the measurement. The vertical line on the MC predictions represents the statistical uncertainty.  The last bin contains the overflow.}
\end{figure}

Figures~\ref{fig:unfolded_3j3b_b1Andb2pt}--\ref{fig:unfolded_3j3b_b3Andbadd} show the kinematics of three leading-\pt \bjets and of the \bjets either assigned to the top quarks or to the additional gluon, which are well predicted by \amcnlopyeight, \sherpat and \powhwseven \ttbb.  Other predictions give varying degrees of consistency with the data distributions within the measurement uncertainties.
Figure~\ref{fig:unfolded_3j3b_bb_top} shows the invariant mass and \pt of the system of two leading-\pt \bjets or of the \bjet pair assigned to top quark, where the predictions from \powhwseven \ttbb matrix element and \amcnlopyeight and \sherpat \ttbar predictions are closer to the data.
Figures~\ref{fig:3j3b_dRemubb_ab1} and~\ref{fig:3j3b_dRemubb_alj1} show the distributions of angular distance in $\Delta R$ of the additional \bjet and additional $l/c$-jet, respectively, relative to the momentum direction of the $e\mu bb$ system. Most predictions
agree reasonably with the data in the $\Delta R(e\mu bb^{\mathrm{top}}, b_1^{\mathrm{add}})$ distribution.  The $\Delta R (e\mu bb^{\text{top}}, l/c\text{-jet}_1)$ variable probing the dynamics of additional partons in the parton shower, is not so well described by the various MC set-ups.  Further compatibility of parton shower models predicting the additional $l/c$-jets \pt and $\eta$ distributions can be seen in Figures~\ref{fig:3j3b_lj1Pt} and~\ref{fig:3j3b_ptdiffLjaddjet}, where various parton shower models have either slightly softer or harder \pt spectra than the data. The best agreement is found with the \sherpat \ttbar model. A comparison of the relative \pt of the leading additional \bjet and the leading additional $l/c$-jet is made between data and predictions, as shown in Figure~\ref{fig:3j3b_ptdiffLjaddjet}, indicating that the additional $l/c$-jets are typically softer than the leading-\pt additional \bjet in the \pow\ \ttbb predictions.  On the other hand the \amcnlopyeight \ttbar, \sherpat \ttbar, and \sherpaOL give good descriptions of the data for this variable.

\begin{figure}[htbp]
\centering
\subfloat[]{\includegraphics[width=0.45\textwidth]{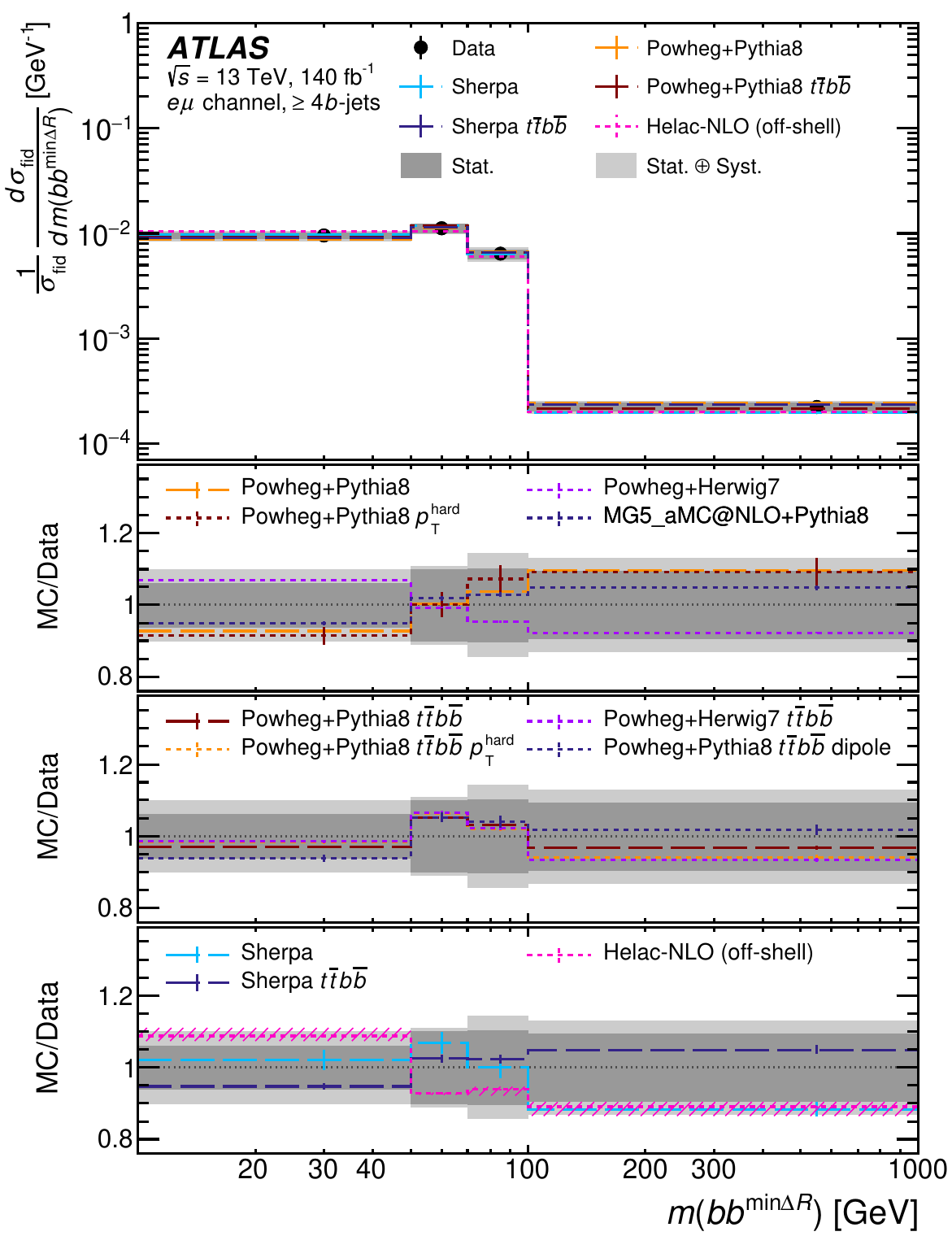}\label{fig:4j4b_mbbmindR}}
\subfloat[]{\includegraphics[width=0.45\textwidth]{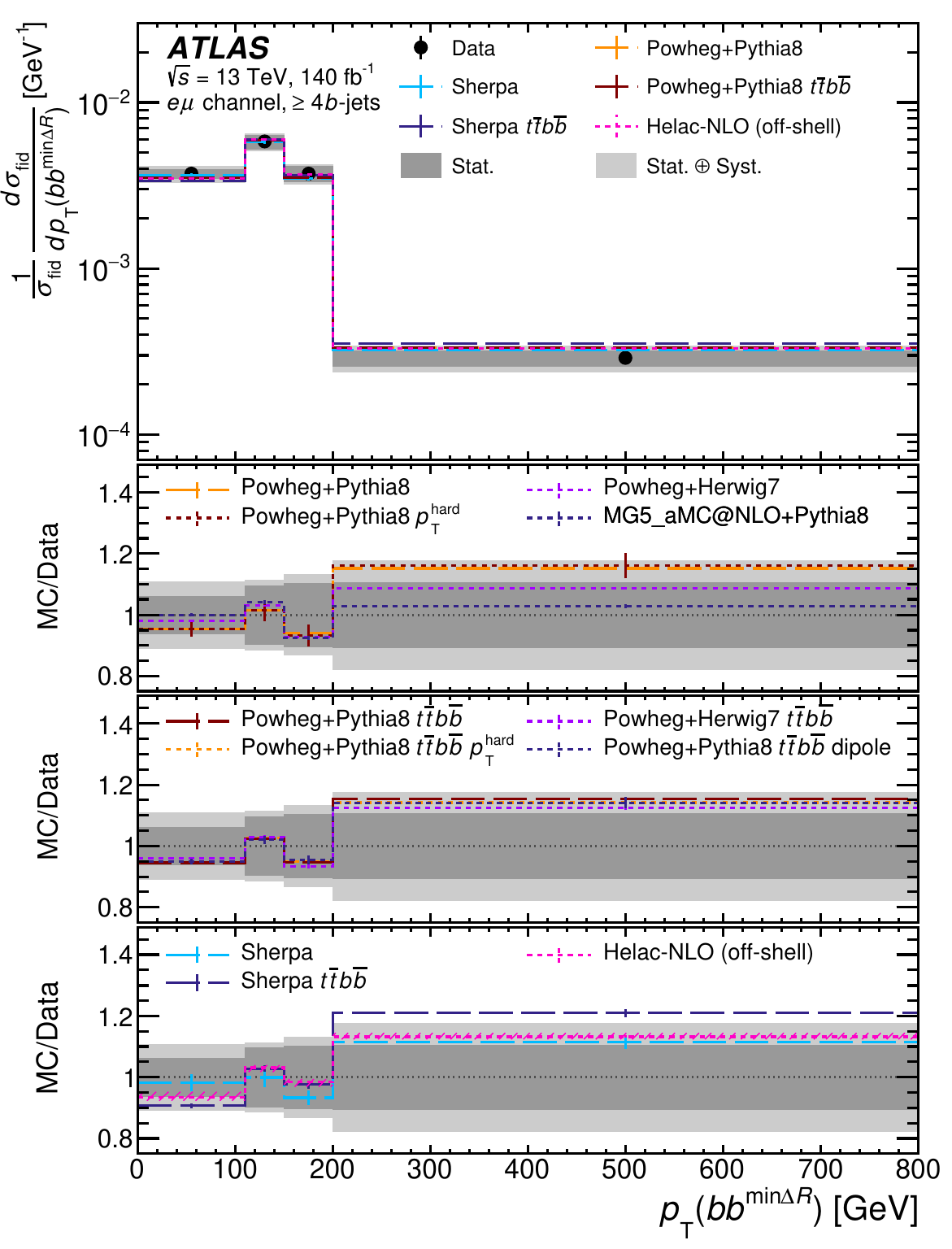}\label{fig:4j4b_pTbbmindR}}\\
\subfloat[]{\includegraphics[width=0.45\textwidth]{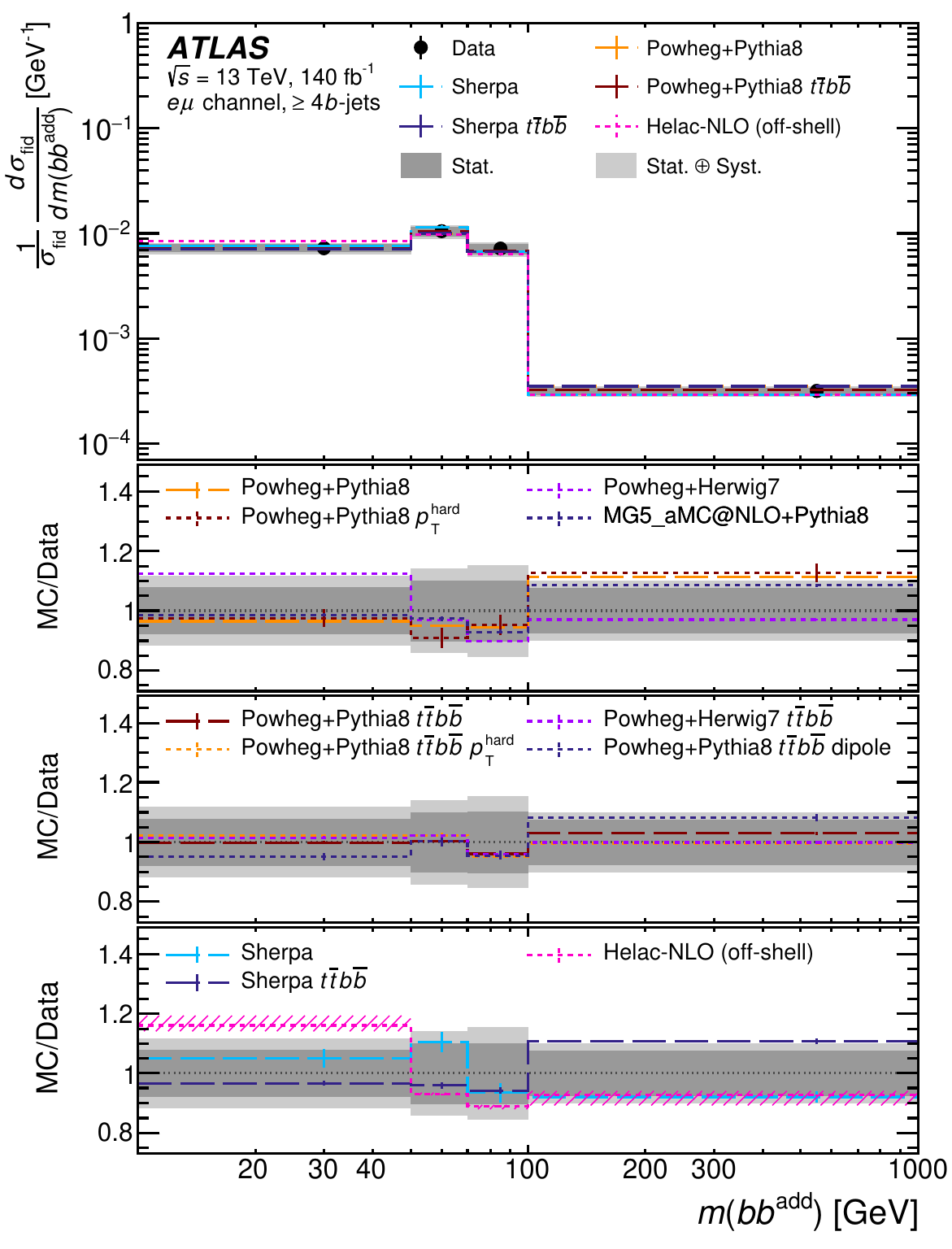}\label{fig:4j4b_mbbadd}}
\subfloat[]{\includegraphics[width=0.45\textwidth]{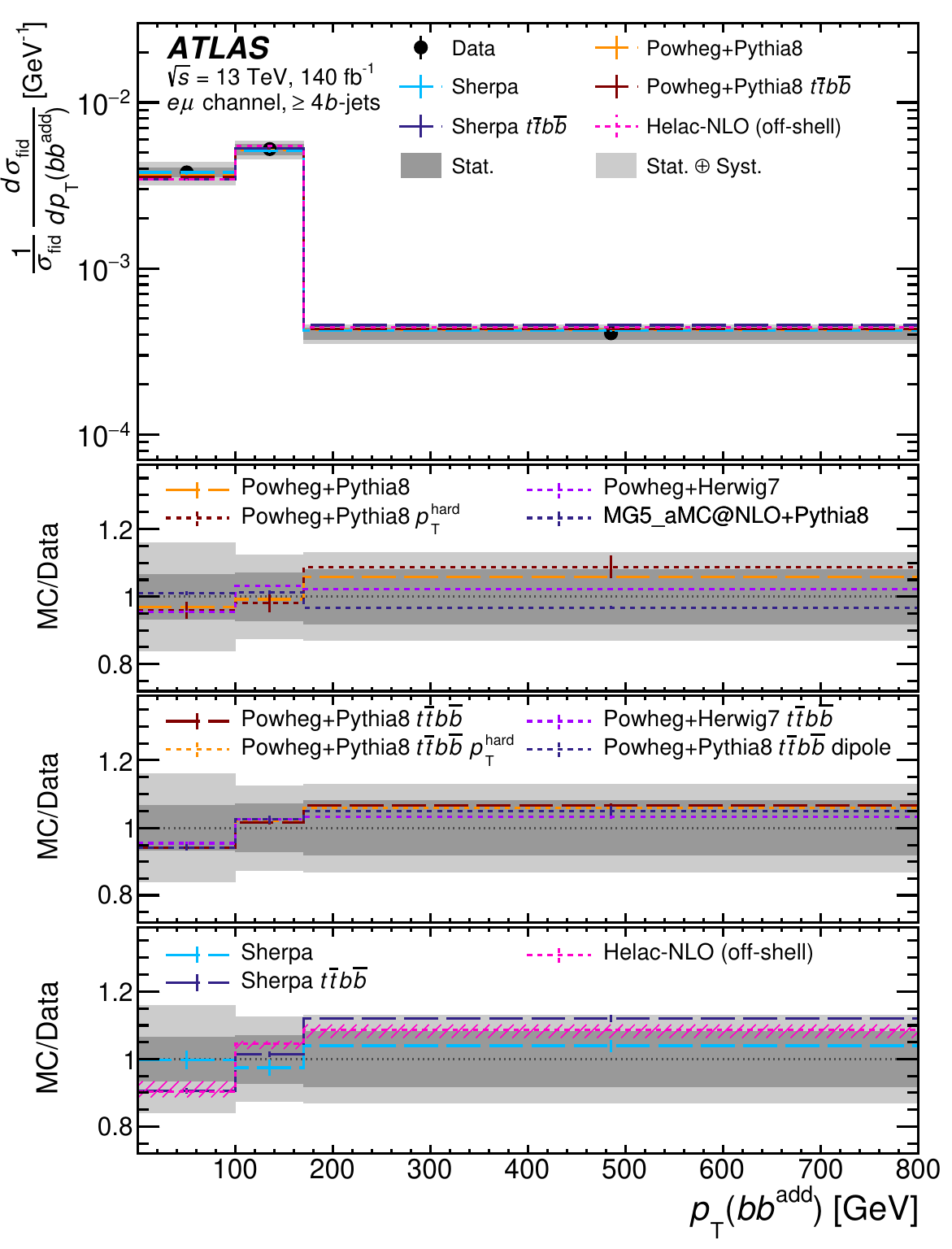}\label{fig:4j4b_pTbbadd}}\\
\caption{Measured normalised differential cross-section in the phase space with at least four \bjets as a function of \protect \subref{fig:4j4b_mbbmindR} $m (bb^{\text{min}\Delta R})$, \protect \subref{fig:4j4b_pTbbmindR} $p_{\text{T}}(bb^{\text{min}\Delta R})$, \protect \subref{fig:4j4b_mbbadd} $m (bb^{\text{add}})$, and \protect \subref{fig:4j4b_pTbbadd} $\pt (bb^{\text{add}})$ compared with predictions. The lower panels show the ratios of various predictions to the data. The hashed band around the \textsc{Helac-NLO} $e\mu+4b$ prediction represents the uncertainties obtained from the envelope of seven sets of $\mu_R$, $\mu_F$ QCD scale variations.  The shaded regions show the statistical (dark) and total uncertainties (light) in the measurement. The vertical line on the MC predictions represents the statistical uncertainty.  The last bin contains the overflow, except in the case of \textsc{Helac-NLO} (off-shell) $e\mu+4b$ prediction.}
\end{figure}

The measurements of normalised distributions of the invariant mass and the \pt of the two closest \bjets system in the phase space with at least four \bjets are compatible with various predictions as shown in Figures~\ref{fig:4j4b_mbbmindR}and~\ref{fig:4j4b_pTbbmindR}, respectively.  The invariant mass and the \pt distributions of the two additional \bjets pair as presented in Figures~\ref{fig:4j4b_mbbadd} and~\ref{fig:4j4b_pTbbadd} give similar pictures as expected, because the \bjets with smallest angular separation make significant contributions to the \bjet assignment scheme.

A quantitative assessment of the level of agreement between
data and the various predictions is performed by calculating a $\chi^2$ value for each
prediction. The $\chi^2$ value is defined as
\begin{equation*}
\chi^2 = S_{b}^T~V^{-1}~S_{b} ~,
\end{equation*}
where $V^{-1}$ is the inverse of the covariance matrix $V$, calculated for each
variable including all statistical and systematic uncertainties that affect the measurements and $S_{b}$ is
a vector of the differences between the measured and predicted cross-sections
being tested. The resulting value of the $\chi^2$ calculation is converted into a $p$-value using the number of degrees of freedom for each variable,
which is the number of bins minus one in the
case of the normalised differential cross-sections to reflect the
normalisation constraint.  The uncertainties associated with the MC predictions being tested are not taken into account while computing the $p$-values.

\begin{figure}[htbp]
\centering
\includegraphics[width=1.0\textwidth]{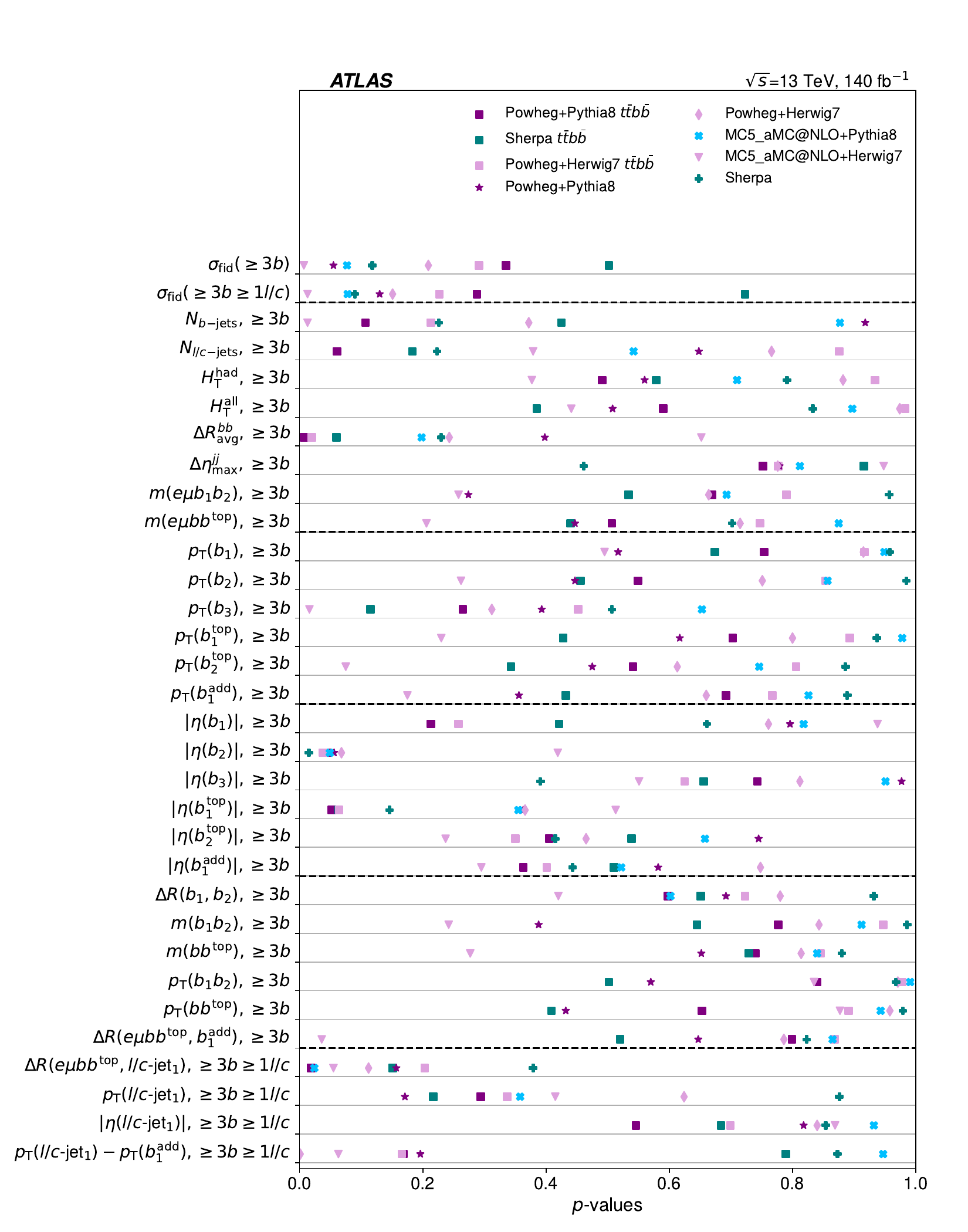}
\caption{Summary of quantitative comparisons for each observable measurement to predictions in form of $p$-values.  This summary refers to observables measured in the phase space with at least three \bjets. The theoretical uncertainties in the MC predictions are not taken into account when computing the $p$-values.}%
\label{fig:ttb_pvalue_summary}
\end{figure}

\begin{figure}[htbp]
\centering
\includegraphics[width=1.0\textwidth]{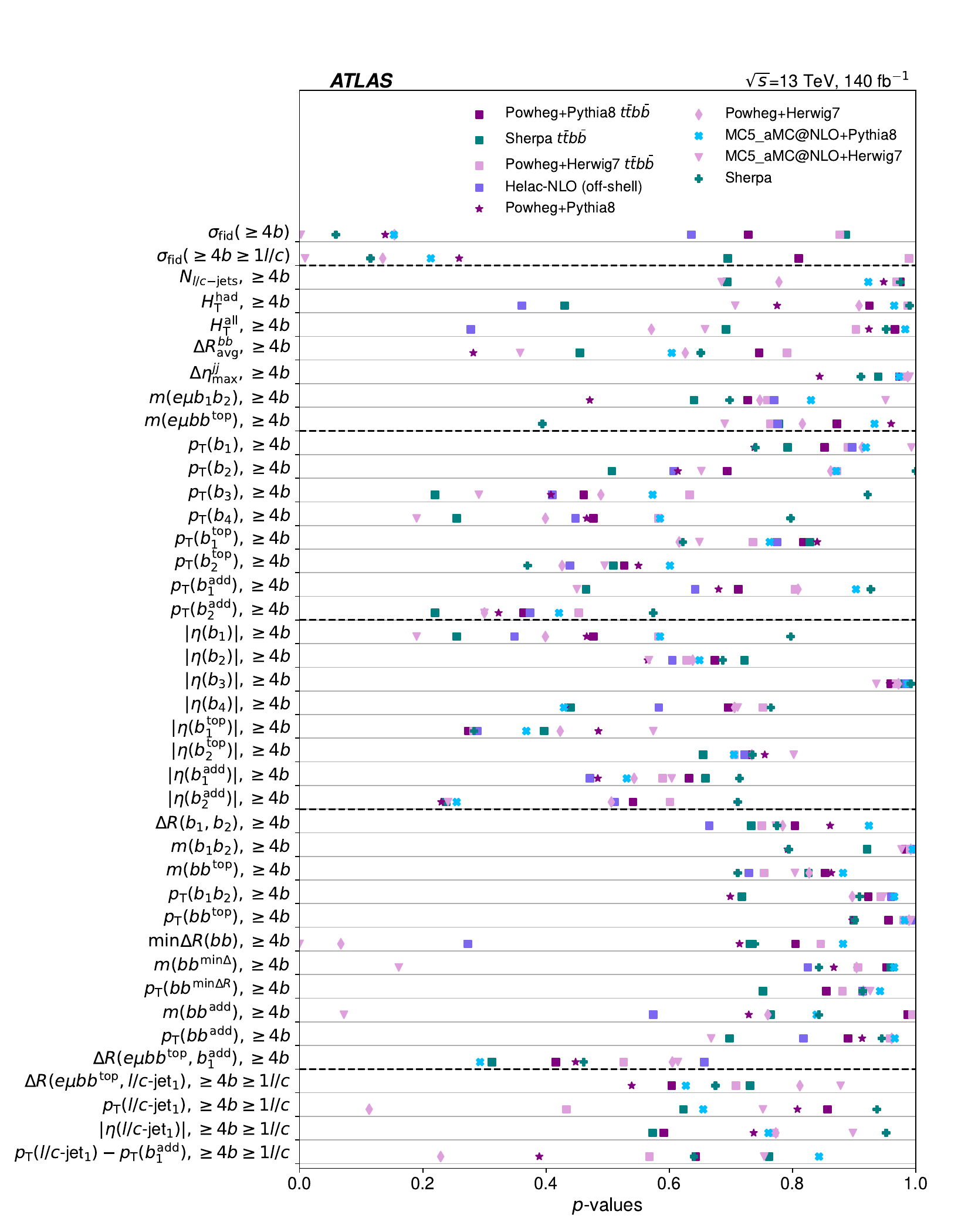}
\caption{Summary of quantitative comparisons for each observable measurement to predictions in form of $p$-values.  This summary refers to observables measured in the phase space with at least four \bjets.  The theoretical uncertainties in the MC predictions are not taken into account when computing the $p$-values.}%
\label{fig:ttbb_pvalue_summary}
\end{figure}

In the case of normalised distributions, one element of $S_{b}$ is discarded in
the calculation along with the corresponding row and column of the covariance
matrix. The resulting $\chi^2$ does not depend on the element of $S_{b}$ or
the row and column of the covariance matrix that is discarded. The $p$-values summarised in Figure~\ref{fig:ttb_pvalue_summary} (Figure \ref{fig:ttbb_pvalue_summary})
for the observables measured in phase space with at least three (four) \bjets give a comprehensive overview of the compatibility of a wide range of measured observables to various predictions.
As discussed above, the quantities measured in the phase space with four or more \bjets generally show good agreement with various predictions, but the measurements suffer from the limited amount of data.
On the other hand, the difference between any two predictions is relatively larger in the phase space with three or more \bjets; however, the comparisons with data are consistent within the measurement uncertainties for most observables.  To facilitate the comparisons with any future MC generator predictions, all measurements together with the correlation matrices will be published in HEPData~\cite{HEPData}.

\clearpage


%

%
%
%

%
\section{Conclusion}
\label{sec:conclusion}

The fiducial and normalised differential cross-sections of \ttbar production in association with \bjets are measured in $pp$ collisions at $\sqrt{s}=13$~\TeV\ using a data sample corresponding to an integrated luminosity of \lumi collected by the ATLAS detector at the LHC.
The results are presented in the $e\mu$ channel within fiducial phase spaces at the stable particle-level.
The fiducial inclusive cross-sections are measured in four phase spaces depending on the number of \bjets and $l/c$-jets.
The measurement precision reaches $8.5\%$ for the $\geq 3b$ phase space, $13\%$ for the $\geq 4b$ phase space, $10\%$ for the $\geq 3b$ $\geq 1l/c$ phase space and $16\%$ for the $\geq 4b$ $\geq 1l/c$ phase space, which are the best to date in the $e\mu$ channel and also better than the theoretical precision of the \ttbb predictions at NLO.
Dominant sources of uncertainties are $b$-tagging, jet energy scale and \ttbar modelling uncertainties.
The measured fiducial cross-sections are compared with various theoretical predictions, and are found to have good compatibility with \ttbb matrix element (4FS) predictions particularly in the regions with at least four \bjets.

The large variety of variables helps in probing the MC simulations in a comprehensive way.
The normalised fiducial differential cross-sections presented as a function of several kinematic variables and global event properties are measured with precisions of 4\%--10\% (10\%--20\%) in the $\geq 3b$ ($\geq 4b$) phase space.
The \bjets have been assigned to the top decay or gluon radiation, with probabilities of correct assignment varying between $40\%$ and $85\%$.
Most observables are generally well described by the majority of MC predictions.
While all models show good agreement with the data for events containing exactly two \bjets, they fail to accurately represent the additional \bjets production in events with three or more.
The \textsc{Sherpa} \ttbb simulation shows good consistency in the full spectrum of the \bjet multiplicity, while \powhwseven \ttbb describes the distributions of \bjet \pt and $H_{\mathrm{T}}^{\mathrm{had}}$ very well and performs better than \powpy \ttbb in modelling \bjets and additional $l/c$-jets multiplicity spectra.  Various predictions where the additional \bjets are generated mainly by the parton shower describe several differential measurements well, however, \mcnlohwseven \ttbar has poor agreement in observables constructed from \bjet pairs originating from gluon emission.
The kinematic distributions of \bjets assigned to top quark or gluons show similar level of compatibility to the data within uncertainties as those of the \bjets that are simply ordered in \pt.

Despite these overall agreements, differences between any two nominal predictions are often smaller than the QCD scale variations of the theory predictions and the uncertainty of the measurement. This highlights the need for further refinement in theoretical calculations and measurement techniques to better discriminate between models and improve our understanding of \ttbar production in association with \bjets.


%

%
\section*{Acknowledgements}
%

%
%

%
%

We thank CERN for the very successful operation of the LHC and its injectors, as well as the support staff at
CERN and at our institutions worldwide without whom ATLAS could not be operated efficiently.

The crucial computing support from all WLCG partners is acknowledged gratefully, in particular from CERN, the ATLAS Tier-1 facilities at TRIUMF/SFU (Canada), NDGF (Denmark, Norway, Sweden), CC-IN2P3 (France), KIT/GridKA (Germany), INFN-CNAF (Italy), NL-T1 (Netherlands), PIC (Spain), RAL (UK) and BNL (USA), the Tier-2 facilities worldwide and large non-WLCG resource providers. Major contributors of computing resources are listed in Ref.~\cite{ATL-SOFT-PUB-2023-001}.

We gratefully acknowledge the support of ANPCyT, Argentina; YerPhI, Armenia; ARC, Australia; BMWFW and FWF, Austria; ANAS, Azerbaijan; CNPq and FAPESP, Brazil; NSERC, NRC and CFI, Canada; CERN; ANID, Chile; CAS, MOST and NSFC, China; Minciencias, Colombia; MEYS CR, Czech Republic; DNRF and DNSRC, Denmark; IN2P3-CNRS and CEA-DRF/IRFU, France; SRNSFG, Georgia; BMBF, HGF and MPG, Germany; GSRI, Greece; RGC and Hong Kong SAR, China; ISF and Benoziyo Center, Israel; INFN, Italy; MEXT and JSPS, Japan; CNRST, Morocco; NWO, Netherlands; RCN, Norway; MNiSW, Poland; FCT, Portugal; MNE/IFA, Romania; MSTDI, Serbia; MSSR, Slovakia; ARIS and MVZI, Slovenia; DSI/NRF, South Africa; MICIU/AEI, Spain; SRC and Wallenberg Foundation, Sweden; SERI, SNSF and Cantons of Bern and Geneva, Switzerland; NSTC, Taipei; TENMAK, T\"urkiye; STFC/UKRI, United Kingdom; DOE and NSF, United States of America.

Individual groups and members have received support from BCKDF, CANARIE, CRC and DRAC, Canada; CERN-CZ, FORTE and PRIMUS, Czech Republic; COST, ERC, ERDF, Horizon 2020, ICSC-NextGenerationEU and Marie Sk{\l}odowska-Curie Actions, European Union; Investissements d'Avenir Labex, Investissements d'Avenir Idex and ANR, France; DFG and AvH Foundation, Germany; Herakleitos, Thales and Aristeia programmes co-financed by EU-ESF and the Greek NSRF, Greece; BSF-NSF and MINERVA, Israel; NCN and NAWA, Poland; La Caixa Banking Foundation, CERCA Programme Generalitat de Catalunya and PROMETEO and GenT Programmes Generalitat Valenciana, Spain; G\"{o}ran Gustafssons Stiftelse, Sweden; The Royal Society and Leverhulme Trust, United Kingdom.

In addition, individual members wish to acknowledge support from Armenia: Yerevan Physics Institute (FAPERJ); CERN: European Organization for Nuclear Research (CERN PJAS); Chile: Agencia Nacional de Investigaci\'on y Desarrollo (FONDECYT 1230812, FONDECYT 1230987, FONDECYT 1240864); China: Chinese Ministry of Science and Technology (MOST-2023YFA1605700), National Natural Science Foundation of China (NSFC - 12175119, NSFC 12275265, NSFC-12075060); Czech Republic: Czech Science Foundation (GACR - 24-11373S), Ministry of Education Youth and Sports (FORTE CZ.02.01.01/00/22\_008/0004632), PRIMUS Research Programme (PRIMUS/21/SCI/017); EU: H2020 European Research Council (ERC - 101002463); European Union: European Research Council (ERC - 948254, ERC 101089007), Horizon 2020 Framework Programme (MUCCA - CHIST-ERA-19-XAI-00), European Union, Future Artificial Intelligence Research (FAIR-NextGenerationEU PE00000013), Italian Center for High Performance Computing, Big Data and Quantum Computing (ICSC, NextGenerationEU); France: Agence Nationale de la Recherche (ANR-20-CE31-0013, ANR-21-CE31-0013, ANR-21-CE31-0022, ANR-22-EDIR-0002), Investissements d'Avenir Labex (ANR-11-LABX-0012); Germany: Baden-Württemberg Stiftung (BW Stiftung-Postdoc Eliteprogramme), Deutsche Forschungsgemeinschaft (DFG - 469666862, DFG - CR 312/5-2); Italy: Istituto Nazionale di Fisica Nucleare (ICSC, NextGenerationEU), Ministero dell'Università e della Ricerca (PRIN - 20223N7F8K - PNRR M4.C2.1.1); Japan: Japan Society for the Promotion of Science (JSPS KAKENHI JP22H01227, JSPS KAKENHI JP22H04944, JSPS KAKENHI JP22KK0227, JSPS KAKENHI JP23KK0245); Netherlands: Netherlands Organisation for Scientific Research (NWO Veni 2020 - VI.Veni.202.179); Norway: Research Council of Norway (RCN-314472); Poland: Ministry of Science and Higher Education (IDUB AGH, POB8, D4 no 9722), Polish National Agency for Academic Exchange (PPN/PPO/2020/1/00002/U/00001), Polish National Science Centre (NCN 2021/42/E/ST2/00350, NCN OPUS nr 2022/47/B/ST2/03059, NCN UMO-2019/34/E/ST2/00393, NCN \& H2020 MSCA 945339, UMO-2020/37/B/ST2/01043, UMO-2021/40/C/ST2/00187, UMO-2022/47/O/ST2/00148, UMO-2023/49/B/ST2/04085, UMO-2023/51/B/ST2/00920); Slovenia: Slovenian Research Agency (ARIS grant J1-3010); Spain: Generalitat Valenciana (Artemisa, FEDER, IDIFEDER/2018/048), Ministry of Science and Innovation (MCIN \& NextGenEU PCI2022-135018-2, MICIN \& FEDER PID2021-125273NB, RYC2019-028510-I, RYC2020-030254-I, RYC2021-031273-I, RYC2022-038164-I), PROMETEO and GenT Programmes Generalitat Valenciana (CIDEGENT/2019/027); Sweden: Carl Trygger Foundation (Carl Trygger Foundation CTS 22:2312), Swedish Research Council (Swedish Research Council 2023-04654, VR 2018-00482, VR 2022-03845, VR 2022-04683, VR 2023-03403, VR grant 2021-03651), Knut and Alice Wallenberg Foundation (KAW 2018.0157, KAW 2018.0458, KAW 2019.0447, KAW 2022.0358); Switzerland: Swiss National Science Foundation (SNSF - PCEFP2\_194658); United Kingdom: Leverhulme Trust (Leverhulme Trust RPG-2020-004), Royal Society (NIF-R1-231091); United States of America: U.S. Department of Energy (ECA DE-AC02-76SF00515), Neubauer Family Foundation.

%
%


%

%
%
%
%
%
%
%

%
\clearpage
\printbibliography
\clearpage
 
\begin{flushleft}
\hypersetup{urlcolor=black}
{\Large The ATLAS Collaboration}

\bigskip

\AtlasOrcid[0000-0002-6665-4934]{G.~Aad}$^\textrm{\scriptsize 104}$,
\AtlasOrcid[0000-0001-7616-1554]{E.~Aakvaag}$^\textrm{\scriptsize 17}$,
\AtlasOrcid[0000-0002-5888-2734]{B.~Abbott}$^\textrm{\scriptsize 123}$,
\AtlasOrcid[0000-0002-0287-5869]{S.~Abdelhameed}$^\textrm{\scriptsize 119a}$,
\AtlasOrcid[0000-0002-1002-1652]{K.~Abeling}$^\textrm{\scriptsize 56}$,
\AtlasOrcid[0000-0001-5763-2760]{N.J.~Abicht}$^\textrm{\scriptsize 50}$,
\AtlasOrcid[0000-0002-8496-9294]{S.H.~Abidi}$^\textrm{\scriptsize 30}$,
\AtlasOrcid[0009-0003-6578-220X]{M.~Aboelela}$^\textrm{\scriptsize 45}$,
\AtlasOrcid[0000-0002-9987-2292]{A.~Aboulhorma}$^\textrm{\scriptsize 36e}$,
\AtlasOrcid[0000-0001-5329-6640]{H.~Abramowicz}$^\textrm{\scriptsize 155}$
% ...
\bigskip
\\

% ...
$^{17}$Department for Physics and Technology, University of Bergen, Bergen; Norway.\\
% ...
$^{30}$Physics Department, Brookhaven National Laboratory, Upton NY; United States of America.\\
% ...
$^{36}$$^{(a)}$Facult\'e des Sciences Ain Chock, Universit\'e Hassan II de Casablanca;$^{(b)}$Facult\'{e} des Sciences, Universit\'{e} Ibn-Tofail, K\'{e}nitra;$^{(c)}$Facult\'e des Sciences Semlalia, Universit\'e Cadi Ayyad, LPHEA-Marrakech;$^{(d)}$LPMR, Facult\'e des Sciences, Universit\'e Mohamed Premier, Oujda;$^{(e)}$Facult\'e des sciences, Universit\'e Mohammed V, Rabat;$^{(f)}$Institute of Applied Physics, Mohammed VI Polytechnic University, Ben Guerir; Morocco.\\
% ...
$^{45}$Physics Department, Southern Methodist University, Dallas TX; United States of America.\\
% ...
$^{50}$Fakult\"{a}t Physik , Technische Universit{\"a}t Dortmund, Dortmund; Germany.\\
% ...
$^{56}$II. Physikalisches Institut, Georg-August-Universit\"{a}t G\"ottingen, G\"ottingen; Germany.\\
% ...
$^{104}$CPPM, Aix-Marseille Universit\'e, CNRS/IN2P3, Marseille; France.\\
% ...
$^{119}$$^{(a)}$New York University Abu Dhabi, Abu Dhabi;$^{(b)}$United Arab Emirates University, Al Ain; United Arab Emirates.\\
% ...
$^{123}$Homer L. Dodge Department of Physics and Astronomy, University of Oklahoma, Norman OK; United States of America.\\
% ...
$^{155}$Raymond and Beverly Sackler School of Physics and Astronomy, Tel Aviv University, Tel Aviv; Israel.\\
% ...
$^{a}$ Also Affiliated with an institute covered by a cooperation agreement with CERN.\\
$^{b}$ Also at An-Najah National University, Nablus; Palestine.\\
$^{c}$ Also at Borough of Manhattan Community College, City University of New York, New York NY; United States of America.\\
$^{d}$ Also at Center for Interdisciplinary Research and Innovation (CIRI-AUTH), Thessaloniki; Greece.\\
$^{e}$ Also at CERN, Geneva; Switzerland.\\
$^{f}$ Also at CMD-AC UNEC Research Center, Azerbaijan State University of Economics (UNEC); Azerbaijan.\\
% ...

\end{flushleft}


%

%
%

%

%
%
%
%

\end{document}